\documentclass[aps,preprint,onecolumn,12pt,nofootinbib]{revtex4}
\usepackage{hyperref}
\usepackage{amsfonts}
\usepackage{bm}
\usepackage{mathrsfs}
\usepackage{soul}
\usepackage{makecell,multirow}
\usepackage{rotating}
\usepackage[utf8]{inputenc}
\usepackage{amsmath}
\usepackage{makecell}

\usepackage{amssymb}
\usepackage{tensor}
\usepackage{graphicx}
\usepackage{bigints}
\usepackage{bbm}
\usepackage{graphicx}
\usepackage{subfigure}
\usepackage{epsf}
\usepackage{bm}
\usepackage{amsmath}
\usepackage{amsfonts}
\usepackage{amssymb}
\usepackage{graphicx}
\usepackage{tabularx}
\usepackage{multirow}
\usepackage{color}%
\providecommand{\U}[1]{\protect\rule{.1in}{.1in}}

\newcommand{\ie}{\begin{equation}}
\newcommand{\fe}{\end{equation}}

\newcommand{\mincir}{\raise
-3.truept\hbox{\rlap{\hbox{$\sim$}}\raise4.truept\hbox{$<$}\ }}
\newcommand{\magcir}{\raise
-3.truept\hbox{\rlap{\hbox{$\sim$}}\raise4.truept\hbox{$>$}\ }}

\providecommand{\U}[1]{\protect\rule{.1in}{.1in}}

\usepackage{tikz,xcolor,hyperref}

\usepackage{hyperref}             
\hypersetup{
    colorlinks=true,              
    breaklinks=true,              
    citecolor=blue,               
    linkcolor=[rgb]{0,0.5,0.9},   
    urlcolor=red,                 
    filecolor=green               
}

\definecolor{lime}{HTML}{A6CE39}
\DeclareRobustCommand{\orcidicon}{%
	\begin{tikzpicture}
	\draw[lime, fill=lime] (0,0) 
	circle [radius=0.16] 
	node[white] {{\fontfamily{qag}\selectfont \tiny ID}};
	\draw[white, fill=white] (-0.0625,0.095) 
	circle [radius=0.007];
	\end{tikzpicture}
	\hspace{-2mm}
}

\foreach \x in {A, ..., Z}{%
	\expandafter\xdef\csname orcid\x\endcsname{\noexpand\href{https://orcid.org/\csname orcidauthor\x\endcsname}{\noexpand\orcidicon}}
}

\begin{document}

\title{\Large{A Non-Commutative Kalb-Ramond Black Hole}}


\author{A. A. Ara\'{u}jo Filho\orcidB{}}
\email{dilto@fisica.ufc.br (The corresponding author)}

\affiliation{Departamento de Física, Universidade Federal da Paraíba, Caixa Postal 5008, 58051-970, João Pessoa, Paraíba, Brazil}
\affiliation{Departamento de Física, Universidade Federal de Campina Grande Caixa Postal 10071, 58429-900 Campina Grande, Paraíba, Brazil.}

\author{N. Heidari\orcidA{}}
\email{heidari.n@gmail.com}

\affiliation{Center for Theoretical Physics, Khazar University, 41 Mehseti Street, Baku, AZ-1096, Azerbaijan.}
\affiliation{School of Physics, Damghan University, Damghan, 3671641167, Iran.}


\author{Iarley P. Lobo\orcidD{}}
\email{lobofisica@gmail.com}

\affiliation{Department of Chemistry and Physics, Federal University of Para\'iba, Rodovia BR 079 - km 12, 58397-000 Areia-PB,  Brazil.}
\affiliation{Departamento de Física, Universidade Federal de Campina Grande Caixa Postal 10071, 58429-900 Campina Grande, Paraíba, Brazil.}


\begin{abstract}

This work presents a new black hole solution within the framework of a non--commutative gauge theory applied to Kalb--Ramond gravity. Using the method recently proposed in the literature [Nucl.Phys.B 1017 (2025) 116950], we employ the Moyal twist $\partial_r \wedge \partial_\theta$ to implement non--commutativity, being encoded by parameter $\Theta$. We begin by verifying that the resulting black hole no longer possesses spherical symmetry, while the event horizon remains unaffected by non--commutative corrections. The Kretschmann scalar is computed to assess the corresponding regularity. It turns out that the solution is regular, provided that the Christoffel symbols and related quantities are not expanded to second order in $\Theta$. We derive the thermodynamic quantities, including the Hawking temperature $T^{(\Theta,\ell)}$, entropy $S^{(\Theta,\ell)}$, and heat capacity $C_V^{(\Theta,\ell)}$. The remnant mass $M_{\text{rem}}$ is estimated by imposing $T^{(\Theta,\ell)} \to 0$, although the absence of a physical remnant indicates complete evaporation. Quantum radiation for bosons and fermions is analyzed via the tunneling method, where divergent integrals are treated using the residue theorem. Notably, in the low--frequency regime, the particle number density for bosons surpasses that of fermions (at least within the scope of the methods considered here). The evaporation lifetime and emission rate are also examined. It turns out that non--commutativity accelerates the black hole evaporation process, making it faster compared to the Kalb--Ramond and Schwarzschild black holes. The effective potential for a massless scalar field is obtained perturbatively, enabling the computation of quasinormal modes and the time--domain profiles. Finally, further bounds on $\Theta$ and $\ell$ (Lorentz--violating paramter) are derived from solar system tests, including the perihelion precession of Mercury, light deflection, and the Shapiro time delay.

\end{abstract}
\maketitle

\tableofcontents


\pagebreak
    
\section{Introduction}

Gravitational phenomena in general relativity are described through the curvature of spacetime, where the nonlinearity of Einstein’s field equations poses substantial difficulties for obtaining {analytical} solutions. Even when symmetries or simplifying assumptions are introduced, the task of solving these equations exactly remains largely intractable \cite{misner1973gravitation,wald2010general}. A common strategy to address these challenges involves linearizing the equations under the assumption of a weak gravitational field. This perturbative approach facilitates the analysis of spacetime fluctuations, particularly gravitational waves. These perturbations are crucial in the context of black hole dynamics, as they affect processes such as Hawking radiation, the system’s stability, and its interaction with the surrounding medium.

While general relativity does not prescribe an intrinsic limitation on the accuracy of spatial measurements, numerous theoretical frameworks suggest the existence of a fundamental minimal length—commonly associated with the Planck scale. In light of this, several models have been developed that treat spacetime as inherently non--commutative, a concept inspired largely by string--theoretic considerations \cite{3,szabo2003quantum,szabo2006symmetry}. These non--commutative geometries have also gained prominence in the analysis of supersymmetric Yang--Mills theories, where they have been extensively applied and studied \cite{ferrari2003finiteness}. In gravitational contexts, one frequently employs the Seiberg--Witten map to introduce non--commutative effects, thereby altering the symmetry structure of spacetime \cite{chamseddine2001deforming}.

Non--commutative geometry has emerged as a powerful framework for probing the physics of black holes from new perspectives \cite{karimabadi2020non,heidari2023gravitational,zhao2024quasinormal,modesto2010charged,araujo2024effects,Anacleto:2019tdj,heidari2024exploring,1,lopez2006towards,mann2011cosmological,Heidari:2025sku,nicolini2009noncommutative,campos2022quasinormal,2,araujo202as5properties}. A substantial body of research has investigated how such deformations of spacetime structure influence black hole dynamics, particularly in relation to evaporation processes \cite{myung2007thermodynamics,23araujo2023thermodynamics}. The thermodynamic properties of these systems are also significantly altered under non--commutative corrections, prompting detailed examinations of entropy, temperature, and heat capacity across various models \cite{nozari2006reissner,banerjee2008noncommutative,nozari2007thermodynamics,lopez2006towards,sharif2011thermodynamics}.

The concept of a non--commutative spacetime stems from redefining the fundamental relations between coordinates, replacing their classical behavior with the commutation rule $[x^\mu, x^\nu] = \mathbbm{i} \Theta^{\mu \nu}$, where the antisymmetric matrix $\Theta^{\mu \nu}$ quantifies the deformation. This shift away from traditional geometrical assumptions has led to several methods for embedding non--commutativity within gravitational theories. Among these, a notable route consists in modifying the underlying gauge structure—extending the SO(4,1) de Sitter symmetry and blending it with the Poincaré group ISO(3,1)—by employing the Seiberg--Witten map. Within this context, Chaichian and collaborators \cite{chaichian2008corrections} introduced a version of the Schwarzschild solution that incorporates non--commutative corrections to the spacetime geometry.

An alternative strategy for incorporating non--commutativity into general relativity was advanced by Nicolini and collaborators \cite{nicolini2006noncommutative}, who opted to modify the matter source rather than the geometric sector of the Einstein equations. In this formulation, the classical notion of a point--like mass was replaced by a smeared distribution, effectively encoding non--commutative effects in the energy--momentum content. Two specific functional forms have been adopted to describe this extended mass configuration: a Gaussian profile given by $\rho_\Theta = M (4\pi \Theta)^{-3/2} e^{-r^2/4\Theta}$ and a Lorentzian distribution expressed as $\rho_\Theta = M \sqrt{\Theta} \pi^{-3/2} (r^2 + \pi \Theta)^{-2}$.

In a more recent investigation, Jurić et al. \cite{Juric:2025kjl} proposed novel strategies for deriving black hole configurations in the setting of non--commutative gauge gravity. Their analysis revisited the earlier work of Chaichian et al. \cite{chaichian2008corrections}, highlighting that the original formulation lacked a crucial component in its treatment of non--commutative effects. Specifically, the authors identified an additional term that had been omitted from the initial set of corrections, thereby offering a more complete and consistent extension of the model
\ie
 - \frac{1}{16} \Theta^{\nu \rho} \Theta^{\lambda\tau} \Big[ \tilde{\omega}^{ac}_{\nu} \Tilde{\omega}^{cd}_{\lambda} \Big(  D_{\tau}R^{d5}_{\rho\mu} + \partial_{\tau}R^{d5}_{\rho\mu}   \Big)   \Big].
\nonumber
\fe
The additional contributions uncovered in the revised formulation significantly modified the tetrad structure, which in turn alter the associated metric components. Based on the new formulation of \cite{Juric:2025kjl}, a recent study introduced a Kalb--Ramond--inspired model by implementing the Moyal deformation characterized by the twist $\partial_r \wedge \partial_\theta$ \cite{heidari2025non}.

Two distinct Kalb--Ramond black hole configurations have emerged in recent theoretical investigations. The first, originally introduced in \cite{Yang:2023wtu}, has become the subject of a broad array of studies encompassing multiple physical aspects. Among them are the analysis of quasinormal spectra \cite{araujo2024exploring}, lensing signatures in strong field regimes \cite{junior2024gravitational}{,} and greybody emissions \cite{guo2024quasinormal}. Researchers have also examined how this geometry accommodated spontaneous symmetry--breaking effects \cite{junior2024spontaneous} (by addressing bounds), kinetic theory scenarios involving Vlasov gas accretion \cite{jiang2024accretion} and particle dynamics in circular orbits and quasi--periodic oscillations \cite{jumaniyozov2024circular}. Investigations have further extended to mechanisms of particle production \cite{12araujo2024particle}, global monopole structures \cite{belchior2025global}, and slowly rotating extensions of the solution \cite{Liu:2024lve}. Building upon this framework, an electrically charged version has been constructed \cite{duan2024electrically}, and additional features concerning thermal aspects, lensing phenomena, radiation profiles, and topological remarks were also taken into account \cite{al-Badawi:2024pdx,Pantig:2025eda,heidari2024impact,aa2024antisymmetric,chen2024thermal,Zahid:2024ohn, hosseinifar2024shadows}. A second solution, proposed more recently in \cite{Liu:2024oas}, has initiated interest in the degradation of entanglement under Lorentz symmetry violation \cite{liu2024lorentz}, as well as in quantum field processes such as particle emission and greybody corrections \cite{12araujo2024particle, 12araujo2025does}.

Hawking’s pioneering work bridged black hole physics and quantum field theory by revealing that black holes was not entirely black, but rather emit radiation with a thermal spectrum \cite{o11,o1,o111}. This radiation—now known as Hawking radiation—gradually reduces a black hole’s mass over time, pointing to a fundamental link between quantum mechanics and curved spacetime geometry \cite{gibbons1977cosmological}. His approach inspired extensive research into quantum processes occurring in gravitational backgrounds \cite{araujo2023analysis,o9,o8,araujo2024dark,o6,aa2024implications,o3,o4,o7,sedaghatnia2023thermodynamical}. Later developments \cite{o13,o10,011,o12}, reframed the emission mechanism as a quantum tunneling phenomenon. The treatment of this process (a semi--analytical one) has since been adapted to a wide array of black hole spacetimes \cite{vanzo2011tunnelling,hollands2015quantum,senjaya2024bocharova,zhang2005new,giavoni2020quantum,mitra2007hawking,del2024tunneling,calmet2023quantum,johnson2020hawking,mirekhtiary2024tunneling,medved2002radiation}.

Gravitational waves serve as a powerful probe for exploring astrophysical systems and cosmological phenomena, including events such as stellar pulsations and the coalescence of compact binaries \cite{heuvel2011compact, dziembowski1992effects,pretorius2005evolution,unno1979nonradial,kjeldsen1994amplitudes,hurley2002evolution,yakut2005evolution}. The frequency content of these signals is strongly influenced by the dynamics and composition of their sources. In the case of perturbed black holes, the emitted gravitational radiation exhibits a discrete spectrum characterized by so--called quasinormal modes—damped oscillations that possess information about the black hole’s geometry and parameters \cite{blazquez2018scalar,konoplya2011quasinormal,heidari2024impact,roy2020revisiting,Berti:2022xfj,horowitz2000quasinormal,london2014modeling,flachi2013quasinormal,kokkotas1999quasi,nollert1999quasinormal,ferrari1984new,jusufi2024charged,araujo2024dark,ovgun2018quasinormal,maggiore2008physical,berti2009quasinormal,herceg2025noncommutative}.

This paper presents a new class of black hole solutions derived within a non--commutative gauge framework embedded in Kalb--Ramond gravity. Based on the method recently proposed in \cite{Juric:2025kjl}, the analysis incorporates non--commutativity via the Moyal twist $\partial_r \wedge \partial_\theta$, where the deformation is controlled by the parameter $\Theta$. Although the geometry no longer exhibits spherical symmetry, the event horizon remains unaffected by these non--commutative modifications. Regularity of the spacetime is examined through the Kretschmann scalar, which confirms the absence of singularities as long as second--order expansions in $\Theta$ are avoided in computing Christoffel symbols and related terms.

Thermodynamic aspects are systematically addressed, including the derivation of Hawking temperature $T^{(\Theta,\ell)}$, entropy $S^{(\Theta,\ell)}$, and heat capacity $C_V^{(\Theta,\ell)}$. While the condition $T^{(\Theta,\ell)} \to 0$ leads to a remnant mass $M_{\text{rem}}$, no evidence is found for the existence of a physical remnant, pointing instead to full evaporation. The emission of particles—both bosonic and fermionic—is analyzed through the quantum tunneling formalism. To handle divergences in the integral expressions, the residue theorem is employed. Results show that, under the low--frequency limit and using the methods applied, the number density of bosons surpasses that of fermions.

A perturbative expansion is used to derive the effective potential for massless scalar fields, which provides the basis for computing quasinormal frequencies and analyzing time--domain behavior of perturbations. Additionally, the study examines evaporation dynamics by calculating the emission rates and evaporation timescales in the high--frequency regime for both energy and particle channels. Lastly, complementary bounds on the deformation parameters are extracted from classical tests of general relativity in the solar system, including Mercury’s perihelion shift, the Sun’s light bending, and the Shapiro time delay.

This paper is organized as follows: In Sec. \ref{Sec2}, we present the general framework of the non--commutative gauge theory and outline the steps for constructing a new black hole solution within Kalb--Ramond gravity. We also examine the effects of non--commutativity on the event horizon and the Kretschmann scalar. In Sec. \ref{Sec3}, we investigate the thermodynamic behavior, focusing on the Hawking temperature, entropy, and heat capacity, and estimate the remnant mass. In Sec. \ref{Sec4}, we analyze the quantum radiation for bosons and fermions, where divergent integrals are handled using the residue method. In Sec. \ref{Sec9}, we evaluate the evaporation lifetime as the black hole approaches its final stage, and compute the energy and particle emission rates in the high--frequency regime. In Sec. \ref{Sec5}, we derive the effective potential for scalar perturbations using a perturbative approach. In Sec. \ref{Sec6}, based on the effective potential, we apply the WKB method to compute the quasinormal modes. In Sec. \ref{Sec7}, we study the time evolution of perturbations by solving the time--domain profile. In Sec. \ref{Sec12}, we establish bounds on the parameters $\Theta$ and $\ell$ using solar system tests such as Mercury’s perihelion precession, light deflection, and the Shapiro time delay. Finally, in Sec. \ref{Sec13}, we present our concluding remarks.


\section{\label{Sec2}The new non--commutative black hole}

Initially, the analysis begins by outlining the underlying framework for treating gravity through the lens of a non--commutative gauge theory. The core of this construction lies the $\mathrm{SO}(4,1)$ group, associated with the de Sitter symmetry, which serves as the foundational gauge group. As a starting point, the formalism is developed in a standard $(3+1)$--dimensional Minkowski spacetime, still within a commutative setting. Within this background, the line element in spherical coordinates takes the familiar form:
\ie
\mathrm{d}s^{2}=\mathrm{d}r^{2}+r^{2}\mathrm{d}\Omega^{2} - c^{2}\mathrm{d}t^{2}.
\fe

The metric’s angular sector is described by $\mathrm{d}\Omega^{2} = \mathrm{d}\theta^{2} + \sin^{2}\theta \,\mathrm{d}\varphi^{2}$, as one should expect. As mentioned before, the underlying symmetry is governed by the $\mathrm{SO}(4,1)$ group, whose algebra is generated by ten antisymmetric elements, labeled $\mathcal{M}_{ab}$ and satisfying the condition $\mathcal{M}_{ab} = -\mathcal{M}_{ba}$. Here, the indices $a$ and $b$ span the range ${0,1,2,3,5}$, with the subset ${0,1,2,3}$ denoted by lowercase Latin indices from the beginning of the alphabet. These generators can be separated into two distinct classes: $\mathcal{M}_{ab}$ (with $a,b = 0,1,2,3$), associated with local Lorentz rotations, and $P_a = \mathcal{M}_{a5}$, which fundamentally are generators of spacetime translations.

In the commutative setting, the gauge potentials $\Tilde{\omega}^{ab}_{\mu}(x)$ exhibit antisymmetry in their Lorentz indices, satisfying $\Tilde{\omega}^{ab}_{\mu}(x) = -\Tilde{\omega}^{ba}_{\mu}(x)$. These quantities are distinct from the standard spin connection and the tetrad fields $e^a_\mu(x)$, although they relate to each other under certain conditions. Specifically, the components $\hat{\Tilde{\omega}}^{a,5}_{\mu}(x)$ are proportional to the tetrads through the relation $\hat{\Tilde{\omega}}^{a,5}_{\mu}(x) = K\, \hat{e}^a_\mu(x)$, where $K$ denotes a contraction parameter that governs the symmetry reduction. In addition, a supplementary gauge field arises, identified as $\hat{\Tilde{\omega}}^{55}_{\mu}(x) = K \hat{\phi}_\mu(x, \Theta)$, where the function $\hat{\phi}_\mu(x, \Theta)$ vanishes in the limit $K \to 0$. This limiting process effectively reduces the full de Sitter group $\mathrm{SO}(4,1)$ to the Poincaré group $\mathrm{ISO}(3,1)$ \cite{2,1}. The corresponding field strength tensor constructed from the gauge connection $\Tilde{\omega}^{ab}_{\mu}(x)$ is given by 
\ie
F^{ab}_{\mu} = \partial_{\mu}\omega^{ab}_{\nu} - \partial_{\nu}\omega^{ab}_{\mu} + \left(\Tilde{\omega}^{ac}_{\mu}\Tilde{\omega}^{db}_{\nu}-\Tilde{\omega}^{ac}_{\nu}\Tilde{\omega}^{db}_{\mu}\right)\eta_{cd}.
\fe 
Here, the indices labeling spacetime coordinates are taken as $\mu, \nu = 0,1,2,3$, while the internal metric of the gauge group is defined by $\eta_{ab} = \mathrm{diag}(+,+,+,-,+)$. This setup permits an equivalent reformulation of the governing expression in an alternative mathematical form as follows
\begin{subequations}
	\begin{align}
&F^{{a}5}_{\mu\nu}=K\left[\partial_{\mu}e^{{a}}_{\nu}-\partial_{\nu}e^{{a}}_{\mu}+\left(\Tilde{\omega}^{{ab}}_{\mu}e^{a}_{\nu}-\Tilde{\omega}^{{ab}}_{\nu}e^{{c}}_{\mu}\right)\eta_{{bc}}\right]=KT^{{a}}_{\mu\nu},\label{torsion}\\
&F^{{ab}}_{\mu\nu} = \partial_{\mu} \Tilde{\omega}^{{ab}}_{\nu}-\partial_{\nu}\Tilde{\omega}^{{ab}}_{\mu}+\left(\Tilde{\omega}^{{ac}}_{\mu}\Tilde{\omega}^{{db}}_{\nu}-\Tilde{\omega}^{{ac}}_{\nu}\Tilde{\omega}^{{db}}_{\mu}\right)\eta_{{cd}}+K\left(e^{{a}}_{\mu}e^{{b}}_{\nu}-e^{{a}}_{\nu}e^{{b}}_{\mu}\right)=R^{{ab}}_{\mu\nu}.
\end{align}
\end{subequations}

It is important to notice that the metric structure is specified by $\eta_{ab} = \mathrm{diag}(+,+,+,-)$, consistent with the Minkowski signature in four dimensions. In this context, the Poincaré gauge group is naturally correlated to the Riemann--Cartan geometry, a framework that extends general relativity by incorporating both curvature and torsion effect features \cite{6,1,Juric:2025kjl}. Within this formalism, torsion arises through the definition $T^a_{\mu\nu} = F^{a5}_{\mu\nu}/K$, whereas the spacetime curvature relies on the expression $R^{ab}_{\mu\nu} = F^{ab}_{\mu\nu}$. These field strengths are constructed using the spin connection $\Tilde{\omega}^{ab}_{\mu}(x)$ and the tetrad fields $e^a_\mu(x)$. When torsion is absent—corresponding to the condition stated in Eq. \eqref{torsion}—the spin connection ceases to be an independent variable and becomes fully determined by the tetrad configuration.

Our attention is now directed toward exploring a viable configuration of gauge fields that respects spherical symmetry, formulated under the symmetry group $\mathrm{SO}(4,1)$ \cite{6,1,Juric:2025kjl}
\ie
	e^{1}_{\mu} = \left(\frac{1}{\Tilde{\mathcal{A}}}, 0,0,0\right), \quad e^{2}_{\mu} = \left(0, r,0,0\right), \quad e^{3}_{\mu} = \left(0,0,r\, \mathrm{sin}\theta,0\right), \quad e^{0}_{\mu} = \left(0,0,0, \Tilde{\mathcal{A}}\right),
\fe
and 
\ie
	\begin{split}
		& \Tilde{\omega}^{12}_{\mu} = \left(0, \Tilde{\mathcal{W}},0,0\right), \quad 
		\Tilde{\omega}^{13}_{\mu} = \left(0,0, \Tilde{\mathcal{Z}}\, \sin \theta,0\right),  \quad \Tilde{\omega}^{10}_{\mu} = \left(0,0,0,\Tilde{\mathcal{U}}\right),\\
		& \Tilde{\omega} ^{23}_{\mu} = \left(0,0,-\cos \theta, \Tilde{\mathcal{V}}\right), \quad \Tilde{\omega}^{20}_{\mu} = \Tilde{\omega}^{30}_{\mu} = \left(0,0,0,0\right).
	\end{split}
\fe

The functions $\Tilde{\mathcal{A}}$, $\Tilde{\mathcal{U}}$, $\Tilde{\mathcal{V}}$, $\Tilde{\mathcal{W}}$, and $\Tilde{\mathcal{Z}}$ are assumed to depend solely on the radial coordinate (within $3D$ space). Moreover, the torsion tensor is characterized by nonzero components, which are expressed in the following form \cite{2,Juric:2025kjl}
\ie
\label{torsion2}
	\begin{split}
		&T^{0}_{01} = -\frac{\Tilde{\mathcal{A}}\Tilde{\mathcal{A}}'+\Tilde{\mathcal{U}}}{\mathcal{A}}, \qquad 
		T^{2}_{03} = r\, \Tilde{\mathcal{V}} \sin \theta \,T^{2}_{12} = \frac{\Tilde{\mathcal{A}} + \Tilde{\mathcal{W}}}{\Tilde{\mathcal{A}}},\\
		& T^{3}_{02} = -r\, \Tilde{\mathcal{V}}, \,\,\,\qquad\qquad T^{3}_{13} = \frac{\left(\Tilde{\mathcal{A}} + \Tilde{\mathcal{Z}}\right)\sin \theta}{\Tilde{\mathcal{A}}}.
	\end{split}
\fe
As a result, the curvature tensor takes the form \cite{2,Juric:2025kjl}
\ie
\label{curvature}
\begin{split}
&R^{01}_{01} = \Tilde{\mathcal{U}}', \quad R^{23}_{01} = - \Tilde{\mathcal{V}}', \quad R^{13}_{23} = \left(\Tilde{\mathcal{Z}} -\Tilde{\mathcal{W}}\right) \cos\theta,\quad R^{01}_{01} = -\Tilde{\mathcal{U}}\Tilde{\mathcal{W}}, \quad R^{13}_{01} = - \Tilde{\mathcal{V}}\Tilde{\mathcal{W}},\\
& R^{03}_{03} = -\Tilde{\mathcal{U}} \Tilde{\mathcal{Z}} \sin \theta, \quad R^{12}_{03} = \Tilde{\mathcal{V}} \Tilde{\mathcal{Z}} \sin \theta \, R^{12}_{12} = \Tilde{\mathcal{W}}',\quad R^{23}_{23} = \left(1-\Tilde{\mathcal{Z}} \Tilde{\mathcal{W}}\right)\sin\theta, \quad R^{13}_{13} = \Tilde{\mathcal{Z}}' \sin\theta.	\end{split}
\fe

Derivatives with respect to the radial variable are denoted by primes in the functions $\Tilde{\mathcal{A}}'$, $\Tilde{\mathcal{U}}'$, $\Tilde{\mathcal{V}}'$, $\Tilde{\mathcal{W}}'$, and $\Tilde{\mathcal{Z}}'$. To ensure the torsion vanishes—consistent with the condition outlined in Eq. \eqref{torsion2}—a specific set of constraints must be imposed on these functions
\ie
\Tilde{\mathcal{V}} = 0, \qquad \Tilde{\mathcal{U}} = - \Tilde{\mathcal{A}}\Tilde{\mathcal{A}}', \qquad \Tilde{\mathcal{W}} =  - \Tilde{\mathcal{A}} = \Tilde{\mathcal{Z}},
\fe
where, accordingly, by taking into account the associated field equations, the system must satisfy:
\ie
R^{{a}}_{\mu} - \frac{1}{2} R\, e^{{a}}_{\mu} = 0.
\fe

They are expressed through the tetrad fields $e^a_\mu(x)$, the curvature is characterized by components $R^a_\mu = R^{ab}_{\mu\nu} e^\nu_b$, while the scalar curvature takes the form $R = R^{ab}_{\mu\nu} e^\mu_a e^\nu_b$. Under these definitions, the resulting solution can be written as
\ie
\Tilde{\mathcal{A}}(r) = \sqrt{\frac{1}{1-\ell}-\frac{2 M}{r}}.
\fe

Within this framework, the parameter $M$ identifies the mass of the black hole, while $\ell$ serves as the deformation parameter associated with the Kalb--Ramond sector \cite{Yang:2023wtu}. The modified line element is written as $\mathrm{d}s^{2} = g^{(\Theta,\ell)}_{\mu\nu}(x,\Theta)\mathrm{d}x^{\mu}\mathrm{d}x^{\nu}$, where the coordinates $x^{\mu} = (t, r, \theta, \varphi)$ describe a $(3+1)$--dimensional spacetime endowed with non--commutative structure. Constructing this geometry involves determining the deformed tetrad fields $\hat{e}^{a}_{\mu}(x,\Theta)$, which arise through a contraction mechanism linking the non--commutative gauge group $\mathrm{SO}(4,1)$ to the Poincaré group $\mathrm{ISO}(3,1)$, following the Seiberg--Witten map formalism \cite{5,4,3}. The resulting spacetime, modified by non--commutativity, is subject to the following structural constraints:
\ie
\left[x^{\mu},x^{\nu}\right]=i\Theta^{\mu\nu}.
\fe

Furthermore, the components $\Theta^{\mu\nu}$ are assumed to be real--valued and satisfy the antisymmetric relation $\Theta^{\mu\nu} = -\Theta^{\nu\mu}$. In this non--commutative gravitational setting, both the deformed tetrads $\hat{e}^a_\mu(x,\Theta)$ and the gauge connection $\hat{\Tilde{\omega}}^{ab}_\mu(x,\Theta)$ are treated as functions that can be expanded perturbatively in powers of the deformation parameter $\Theta$ \cite{4,2,3,1,Juric:2025kjl}
\ie
\begin{split}
		&\hat{e}^{{a}}_{\mu}(x,\Theta) = e^{{a}}_{\mu}(x) - i \Theta^{\nu\rho}e^{{a}}_{\mu\nu\rho}(x) + \Theta^{\nu\rho}\Theta^{\lambda\tau}e^{{a}}_{\mu\nu\rho\lambda\tau}(x)\dots,\\	
		&\hat{\Tilde{\omega}}^{ab}_{\mu}(x,\Theta) = \Tilde{\omega}^{ab}_{\mu}(x)-i \Theta^{\nu\rho} \Tilde{\omega}^{ab}_{\mu\nu\rho}(x) + \Theta^{\nu\rho}\Theta^{\lambda\tau}\Tilde{\omega}^{ab}_{\mu\nu\rho\lambda\tau}(x)\dots  \,\,. \label{o12megan23123oncom}
	\end{split}
\fe

The non--commutative corrections applied to the gauge connection $\hat{\Tilde{\omega}}^{ab}_{\mu}(x,\Theta)$ give rise to modified tetrad fields $\hat{e}^{a}_{\mu}(x,\Theta)$. These ones are obtained through an expansion derived from the formal expression in Eq. \eqref{o12megan23123oncom}, retaining terms up to second order in the deformation parameter $\Theta$ as outlined in Ref. \cite{Juric:2025kjl}
\ie
\omega^{ab}_{\mu\nu\rho} (x) = \frac{1}{4} \left\{\Tilde{\omega}_{\nu},\partial_{\rho}\Tilde{\omega}_{\mu}+R_{\rho\mu}\right\}^{ab},\label{naoancaomamacorr-ateatraaad}
\fe
{
\begin{equation}
	\begin{split}
		\Tilde{\omega}_{\mu\nu\rho\lambda\tau}^{ab} = &\frac{1}{16}\biggl[-\left\{\left\{\Tilde{\omega}_{\lambda},\left(\partial_\tau \Tilde{\omega}_{\nu} + R_{\tau\nu}\right)\right\},\left(\partial_\rho  \Tilde{\omega}_{\mu} + R_{\rho\mu}\right)\right\}^{ab} \\
  & - \left\{\Tilde{\omega}_{\nu}, \partial_\rho \left\{\Tilde{\omega}_{\lambda},\left(\partial_{\tau}\Tilde{\omega}_{\mu} + R_{\tau\mu}\right)\right\} \right\}^{ab}  +2 \left[\partial_{\lambda} \Tilde{\omega}_{\nu}, \partial_{\tau} \left(\partial_{\rho} \Tilde{\omega}_{\mu} + R_{\rho\mu}\right)\right]^{ab}  \\
  & + \left\{\Tilde{\omega}_{\nu},2\left\{R_{\rho\lambda},R_{\mu\tau}\right\}\right\}^{ab} - \left\{\Tilde{\omega}_{\nu},\left\{\Tilde{\omega}_{\lambda}, D_\tau R_{\rho\mu} + \partial_\tau R_{\rho \mu}\right\}\right\}^{ab}   \biggr].
	\end{split}
\label{n1o2n2com3mco3rr-om3eg3a}
\end{equation}
}
Eqs. \eqref{naoancaomamacorr-ateatraaad} and \eqref{n1o2n2com3mco3rr-om3eg3a}, obtained through the application of the Seiberg--Witten map, are required to fulfill the following consistency conditions
\ie
\left[\alpha,\beta\right]^{ab} =  \alpha^{ac}\beta^{b}_{c}-\beta^{ac}\alpha^{b}_{c},\qquad
\left\{\alpha,\beta\right\}^{ab} = \alpha^{ac}\beta^{b}_{c}+\beta^{ac}\alpha^{b}_{c},
\fe
and
\ie
	D_{\mu}R^{ab}_{\rho\sigma} = \partial_{\mu}R^{ab}_{\rho\sigma} +\left(\Tilde{\omega}^{ac}_{\mu}R^{db}_{\rho\sigma}+\Tilde{\omega}^{bc}_{\mu}R^{da}_{\rho\sigma}\right)\eta_{cd}.
\fe

It is important to emphasize that the gauge connection $\hat{\Tilde{\omega}}^{ab}_{\mu}(x,\Theta)$ is subject to specific conditions that impose the following restrictions
\ie
\label{CondDefOmega}
\hat{\Tilde{\omega}}^{ab\star}_{\mu}(x,\Theta) = -\hat{\Tilde{\omega}}^{ab}_{\mu}(x,\Theta), \quad 
\hat{\Tilde{\omega}}^{ab}_{\mu}(x,\Theta) ^{r} \equiv \hat{\Tilde{\omega}}^{ab}_{\mu}(x,-\Theta) = -\hat{\omega}^{ba}_{\mu}(x,\Theta).
\fe

Note that the symbol ${}^\star$ is used to denote complex conjugation. In addition, the corrections introduced by non--commutativity, as constrained by the conditions in Eq. \eqref{CondDefOmega}, take the following explicit form
\ie
\Tilde{\omega}^{ab}_{\mu} (x) = - \Tilde{\omega}^{ba}_{\mu} (x), \quad \Tilde{\omega}^{ab}_{\mu\nu\rho} (x) = \Tilde{\omega}^{ba}_{\mu\nu\rho} (x), \quad \Tilde{\omega}^{ab}_{\mu\nu\rho\lambda\tau} (x) = -\Tilde{\omega}^{ba}_{\mu\nu\rho\lambda\tau} (x).
\fe

These results arise by evaluating Eqs. \eqref{naoancaomamacorr-ateatraaad} and \eqref{n1o2n2com3mco3rr-om3eg3a} under the assumptions that the torsion tensor $T^a_{\mu\nu}$ vanishes and the contraction parameter approaches zero, i.e., $K \to 0$. With these conditions in place, one obtains the following relations
\ie
\label{ComConjDefTetrads}
\hat{e}^{{a}\star}_{\mu}(x,\Theta) = e^{{a}}_{\mu}(x)+i \Theta^{\nu\rho}e^{{a}}_{\mu\nu\rho}(x)+\Theta^{\nu\rho}\Theta^{\lambda\tau}e^{{a}}_{\mu\nu\rho\lambda\tau}(x)\dots,
\fe
with
\ie
\begin{split}
e^{{a}}_{\mu\nu\rho} &= \frac14	\left[\Tilde{\omega}^{{ac}}_{\nu}\partial_{\rho} e^{{d}}_{\mu} + \left(\partial_{\rho}\Tilde{\omega}^{{ac}}_{\mu} + R^{{ac}}_{\rho\mu}\right)e^{{d}}_{\nu}\right]\eta_{{cd}},
\end{split}
\fe
where \cite{Juric:2025kjl}
\begin{align}
&e_{\mu \nu \rho \lambda \tau }^{a} = \\
&= \frac{1}{16}\Bigl[
   2\,\Bigl\{R_{\tau \nu},\,R_{\mu \rho}\Bigr\}^{ab}\,e_{\lambda}^{c}
   \;-\;\Tilde{\omega}_{\lambda}^{a\,b}\,\Bigl(D_{\rho}\,R_{\tau \mu}^{c\,d}
     \;+\;\partial_{\rho}\,R_{\tau \mu}^{c\,d}\Bigr)\,e_{\nu}^{m}\,\eta_{d\,m}
   -\,\Bigl\{\Tilde{\omega}_{\nu},\,\bigl(D_{\rho}\,R_{\tau \mu}
     + \partial_{\rho}\,R_{\tau \mu}\bigr)\Bigr\}^{ab}\,e_{\lambda}^{c}
     \nonumber \\
&\quad
   \;-\;\partial_{\tau}\,\Bigl\{\Tilde{\omega}_{\nu},\,\bigl(\partial_{\rho}\,\Tilde{\omega}_{\mu}
     + R_{\rho \mu}\bigr)\Bigr\}^{a\,b}\,e_{\lambda}^{c}
   -\,\Tilde{\omega}_{\lambda}^{a\,b}\,\partial_{\tau}\Bigl(
       \Tilde{\omega}_{\nu}^{c\,d}\,\partial_{\rho}\,e_{\mu}^{m}
       + \bigl(\partial_{\rho}\,\Tilde{\omega}_{\mu}^{c\,d}
         + R_{\rho \mu}^{c\,d}\bigr)\,e_{\nu}^{m}
     \Bigr)\,\eta_{d\,m}
     \nonumber \\
&\quad
   \;+\;2\,\partial_{\nu}\,\Tilde{\omega}_{\lambda}^{a\,b}\,
        \partial_{\rho}\partial_{\tau}\,e_{\mu}^{c}
   -\,2\,\partial_{\rho}\Bigl(\partial_{\tau}\,\Tilde{\omega}_{\mu}^{a\,b}
     + R_{\tau \mu}^{a\,b}\Bigr)\,\partial_{\nu}\,e_{\lambda}^{c}
   \;-\;\Bigl\{\Tilde{\omega}_{\nu},\,\bigl(\partial_{\rho}\,\Tilde{\omega}_{\lambda}
     + R_{\rho \lambda}\bigr)\Bigr\}^{a\,b}\,\partial_{\tau}\,e_{\mu}^{c}
\nonumber \\
&\quad
   \;-\,\Bigl(\partial_{\tau}\,\Tilde{\omega}_{\mu}^{a\,b}
     + R_{\tau \mu}^{a\,b}\Bigr)\,\Bigl(
       \Tilde{\omega}_{\nu}^{c\,d}\,\partial_{\rho}\,e_{\lambda}^{m}
       + \bigl(\partial_{\rho}\,\Tilde{\omega}_{\lambda}^{c\,d}
         + R_{\rho \lambda}^{c\,d}\bigr)\,e_{\nu}^{m}\,\eta_{d\,m}
     \Bigr)
\Bigr]\;\eta_{b\,c}
\nonumber \\
&\quad
\; - \frac{1}{16}\,\Tilde{\omega}_{\lambda}^{a\,c}\,\Tilde{\omega}_{\nu}^{d\,b}\,e_{\rho}^{f}\,
   R_{\tau \mu}^{g\,m}\,\eta_{c\,d}\,\eta_{f\,g}\,\eta_{b\,m}\,.
\label{3.12}
\end{align}

It should be emphasized that this last contribution was absent from the original analysis in Ref. \cite{chaichian2008corrections}, as well as in later developments that followed its approach. Incorporating this previously neglected term, the corrected metric tensor is therefore
\ie
\label{DefMetTensor}
g^{(\Theta)}_{\mu\nu}\left(x,\Theta\right) = \frac{1}{2} \eta_{{a}{b}}\Bigg[\hat{e}^{{a}}_{\mu}(x,\Theta)\ast\hat{e}^{{b}\star}_{\nu}(x,\Theta)+\hat{e}^{{b}}_{\mu}(x,\Theta)\ast\hat{e}^{{a}\star}_{\nu}(x,\Theta)\Bigg].
\fe
Here, the symbol $\ast$ stands for the conventional Moyal (star) product used in non--commutative geometry. {To obtain the black hole solution, we adopt the Moyal twist 
$\partial_r \wedge \partial_\theta$, from which the corresponding matrix deformation is derived as
($\Theta=\Theta^{23}=-\Theta^{32}$)
\begin{equation}
	\Theta^{\mu\nu}=\left(\begin{matrix}
		0	& 0 & 0 & 0 \\
		0	& 0 & \Theta & 0 \\
		0	& -\Theta & 0 & 0 \\
		0	& 0 & 0 & 0
	\end{matrix}
	\right), \qquad \mu,\nu=0,1,2,3.
\end{equation}
In other words, this prescription gives rise to the following commutation relation:
\ie
 [\,r \,\overset{\star}{,}\, \theta \,] = i \Theta.
\fe

}

Throughout the remainder of this analysis, we adopt natural units by setting $\hbar = c = G = 1$. The corresponding line element can now be written explicitly as follows:
\ie
\label{metrictensorss}
\mathrm{d}s^{2} = g^{(\Theta)}_{\mu\nu}\left(x,\Theta\right) \mathrm{d}x^{\mu} \mathrm{d}x^{\nu}   = - A^{(\Theta,\ell)} \mathrm{d}t^{2} +  B^{(\Theta,\ell)} \mathrm{d}r^{2} + C^{(\Theta,\ell)} \mathrm{d}\theta^{2} + D^{(\Theta,\ell)} \mathrm{d}\varphi^{2},
\fe
in which
\ie \label{gtt}
A^{(\Theta,\ell)} = \frac{1}{1-\ell} - \frac{2 M}{r} - \frac{\Theta ^2 M (11 (\ell-1) M+4 r)}{2 (\ell-1) r^4},
\fe
\ie \label{grr}
B^{(\Theta,\ell)} = \frac{1}{\frac{1}{1-\ell}-\frac{2 M}{r}} + \frac{\Theta ^2 (\ell-1) M (3 (\ell-1) M+2 r)}{2 r^2 (2 (\ell-1) M+r)^2},
\fe
\ie \label{gtheta}
C^{(\Theta,\ell)} = r^2 -\frac{\Theta ^2 \left(64 (\ell-1)^2 M^2+32 (\ell-1) M r+r^2\right)}{16 (\ell-1) r (2 (\ell-1) M+r)},
\fe
\ie \label{gphi}
D^{(\Theta,\ell)} = r^2 \sin ^2(\theta ) +\frac{1}{16} \Theta ^2 \left[5 \cos ^2(\theta )+\frac{4 \sin ^2(\theta ) \left(-2 (\ell-1) M^2+4 (\ell-1) M r+r^2\right)}{r (2 (\ell-1) M+r)}\right].
\fe

Moreover, a direct analysis of the expression $1/B^{(\Theta,\ell)}$ reveals that the introduction of non--commutative corrections leaves the position of the event horizon unchanged relative to that of the Kalb--Ramond black hole. Specifically, this implies that:
\ie
\begin{split}
\label{eventhorizonhay}
r_{h} = 2 (1-\ell) M.
\end{split}
\fe
Conversely, by isolating $M$ in the equation, the resulting expression reads:
\ie
\label{masss}
M = \frac{r_{h}}{2-2 \ell} \approx \, \, \frac{\ell\, r_{h}}{2}+\frac{r_{h}}{2}.
\fe

\begin{figure}
    \centering
    \includegraphics[scale=0.7]{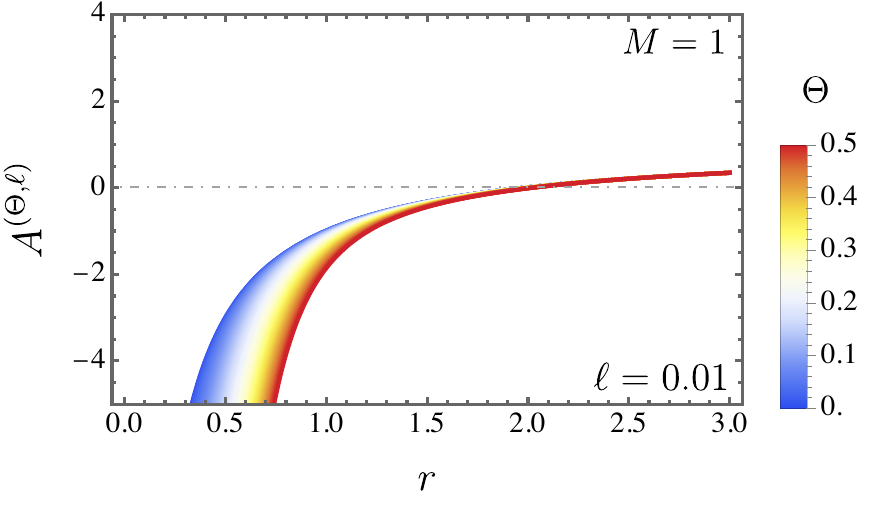}
    \caption{The temporal component of the metric, $A^{(\Theta,\ell)}$, is plotted against the radial coordinate $r$ for several choices of the non--commutative parameter $\Theta$, while keeping the Lorentz--violating parameter fixed at $\ell = 0.01$.}
    \label{metriccompg00}
\end{figure}

\begin{figure}
    \centering
    \includegraphics[scale=0.7]{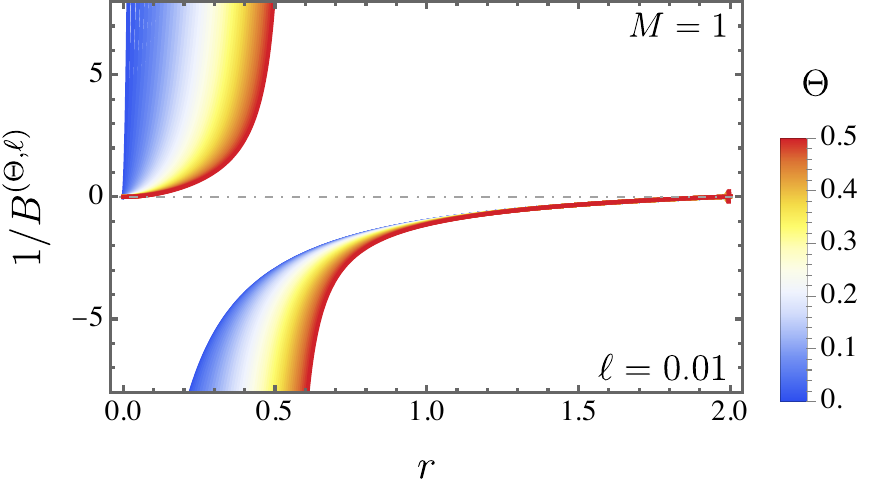}
    \caption{The radial metric component $1/B^{(\Theta,\ell)}$ is shown as a function of $r$, considering various values of the non--commutative deformation parameter $\Theta$ while holding $\ell$ constant at $0.01$.}
    \label{metriccompog11}
\end{figure}

Figs. \ref{metriccompg00} and \ref{metriccompog11} display the profiles of the metric functions $A^{(\Theta,\ell)}$ and $1/B^{(\Theta,\ell)}$, respectively. Due to the excessive length of the resulting expression—with 21 lengthy terms—the Kretschmann scalar will not be written explicitly. However, in the limit $r \to 0$, its behavior reduces to:
\ie
\label{ktozero}
\begin{split}
\tilde{\mathcal{K}}_{r\to 0} = & \frac{1552}{3 \Theta ^4},
\end{split}
\fe
where no dependence on the Lorentz--violating parameter $\ell$ arises. In contrast to the black hole discussed in Ref.\cite{Yang:2023wtu}, our solution seems to be regular. Furthermore, in the limit $r \to 0$, as shown in Eq.(\ref{ktozero}), our results coincide with those found in the literature for the Schwarzschild case under the same Moyal twist, namely $\partial_r \wedge \partial_\theta$. One additional remark is in order: to compute the Kretschmann scalar, we did not expand the Christoffel symbols and related geometric quantities up to second order, as was done in Ref.~\cite{Juric:2025kjl}. Had we done so, the Kretschmann scalar would have signaled a non--regular black hole, similarly to the behavior observed for the Schwarzschild case in that work.

\section{\label{Sec3}Thermodynamics}

This part is dedicated to exploring the system's thermodynamic behavior, starting with the Hawking temperature $T^{(\Theta,\ell)}$, which is determined via the surface gravity. The discussion proceeds with the evaluation of entropy $S^{(\Theta,\ell)}$ and heat capacity $C_V^{(\Theta,\ell)}$. Given the complexity of the full expressions, each quantity will be expanded, retaining terms up to second order in the non--commutative parameter $\Theta$.

In the subsequent analysis, the Hawking temperature is presented as functions of the black hole mass $M$. As will become clear from the respective plot, it indicates the absence of a remnant mass.


\subsection{Hawking temperature}

Employing the surface gravity method, one arrives at the following expression for the Hawking temperature:
\ie
\begin{split}
\label{temppppe}
T^{(\Theta,\ell)} & =   \frac{1}{{4\pi \sqrt {{{A^{(\Theta,\ell)}}}{{B^{(\Theta,\ell)}}}} }}{\left. {\frac{{\mathrm{d}{{A^{(\Theta,\ell)}}}}}{{\mathrm{d}r}}} \right|_{r = {r_{h}}}}  \\
  & = \frac{(\ell-1) M r_{h}^3+\Theta ^2 M (11 (\ell-1) M+3 r_{h})}{\pi  (\ell-1) r_{h}^5 \sqrt{\frac{\left(2 r_{h}^3 (2 (\ell-1) M+r_{h})+\Theta ^2 M (-3 (\ell-1) M-2 r_{h})\right) \left(2 r_{h}^3 (2 (\ell-1) M+r_{h})+\Theta ^2 M (11 (\ell-1) M+4 r_{h})\right)}{r_{h}^6 (2 (\ell-1) M+r_{h})^2}}}\\
& \approx  \, \, \frac{M}{2 \pi  r_{h}^2} -\frac{13 \Theta ^2 \ell M^3}{2 \pi  r_{h}^4 (r-2 M)^2}+\frac{6 \Theta ^2 \ell M^2}{\pi  r_{h}^3 (r_{h}-2 M)^2}+\frac{10 \Theta ^2 M^3}{\pi  r_{h}^5 (2 M-r_{h})}\\
  & -\frac{33 \Theta ^2 M^2}{4 \pi  r_{h}^4 (2 M-r_{h})}+\frac{3 \Theta ^2 M}{2 \pi  r_{h}^3 (2 M-r_{h})} -\frac{3 \Theta ^2 \ell M}{2 \pi  r_{h}^2 (r_{h}-2 M)^2},
\end{split}
\fe
where we have employed the series expansion, retaining terms up to second order in the non--commutative parameter $\Theta$ and first order in the Lorentz--violating parameter $\ell$. Unlike the standard Schwarzschild scenario—based on the same Moyal twist—the surface gravity remains regular and finite in the present case. Substituting the mass expression from Eq.(\ref{masss}) into Eq.(\ref{temppppe}) yields:
\ie
\begin{split}
\label{refMhawking}
T^{(\Theta,\ell)} \approx & \, \frac{1}{4 \pi  r_{h}} +\frac{\ell}{4 \pi  r_{h}}+\frac{3 \Theta ^2}{16 \pi  r_{h}^3} .
\end{split}
\fe
Given that both $\Theta$ and $\ell$ are treated as small parameters—an assumption that will be validated through observational bounds in subsequent sections—we restrict the expansion to the dominant contributions already presented. In this expression, the first two terms on the right-hand side correspond, respectively, to the standard Schwarzschild and Kalb--Ramond black holes. The third contribution, as expected, reflects the non--commutative modifications introduced in this analysis.

\begin{figure}
    \centering
    \includegraphics[scale=0.7]{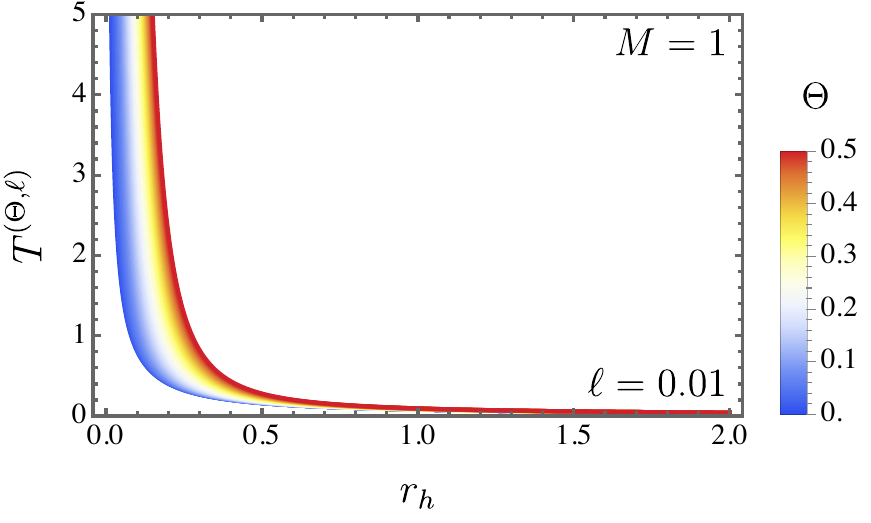}
    \caption{It displays the variation of the Hawking temperature $T^{(\Theta,\ell)}$ with respect to the event horizon radius $r_{h}$, considering several values of the non--commutative parameter $\Theta$ while keeping the Lorentz--violating parameter fixed at $\ell = 0.01$.}
    \label{h1a2w3ki3ngt3em3p3pr}
\end{figure}

To study the dependence of the Hawking temperature on the event horizon radius, its behavior is plotted in Fig. \ref{h1a2w3ki3ngt3em3p3pr}. The graph illustrates that higher values of $\Theta$ lead to an increase in the temperature $T^{(\Theta,\ell)}$. Alternatively, one may express the temperature as a function of the black hole mass $M$. This is done by substituting Eq.(\ref{eventhorizonhay}) into Eq.~(\ref{refMhawking}), resulting in the following relation:
\ie
\label{masshawww}
T^{(\Theta,\ell)} \approx \,  \frac{1}{8 \pi  (1-\ell) M} + \frac{\ell}{8 \pi  (1-\ell) M} + \frac{3 \Theta ^2}{128 \pi  (1-\ell)^3 M^3}.
\fe

Notice that by taking into account the temperature in terms of the mass is essential for probing whether a remnant mass ($M_{\text{rem}}$) emerges and for evaluating the black hole’s evaporation dynamics within the framework of the Stefan--Boltzmann law. While the detailed thermodynamic evolution—including aspects such as greybody factors, and absorption cross sections—is of considerable interest, these topics extend beyond the objectives of the present work. It is important to emphasize that, as the effective potential will be constructed through a perturbative approach (as discussed later), determining these quantities—particularly the last one—is far from straightforward. The complexity involved, therefore, makes the calculation highly nontrivial.

Imposing the condition $T^{(\Theta,\ell)} \to 0$ in Eq.~(\ref{masshawww}) yields two possible values for $M$. Nevertheless, both solutions fail to produce physically meaningful or real values, specifically:
\ie
M^{1}_{\text{rem}} = \frac{i \sqrt{3} \Theta }{{4(1- \ell)} \sqrt{\ell+1}},
\fe
\ie
M^{2}_{\text{rem}} = \frac{i \sqrt{3} \Theta }{4 (\ell-1) \sqrt{\ell+1}},
\fe
{or, to simplify the notation, one may also express it as:
\ie
M^{(1,2)}_{\text{rem}} =  \pm \frac{i \sqrt{3} \Theta }{4 (\ell-1) \sqrt{\ell+1}}.
\fe
}

\begin{figure}
    \centering
    \includegraphics[scale=0.7]{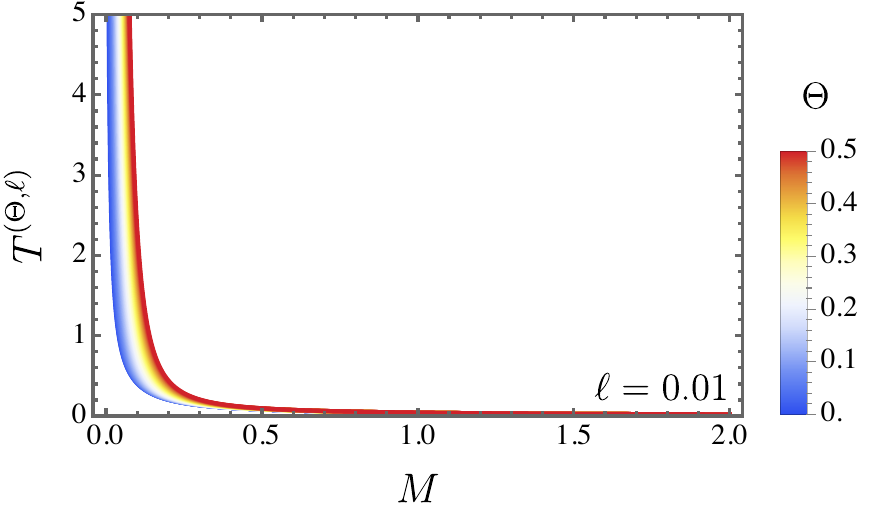}
    \caption{The behavior of the Hawking temperature $T^{(\Theta,\ell)}$ as a function of the black hole mass $M$ is illustrated for various choices of the non--commutative parameter $\Theta$, while maintaining the Lorentz--violating parameter fixed at $\ell = 0.01$.}
    \label{h2aw2kin2gte2mp2p}
\end{figure}

It can be readily confirmed that both solutions yield imaginary values, indicating the lack of physical validity. This outcome suggests that the black hole under consideration does not retain a stable remnant but instead evaporates entirely. This interpretation is further corroborated by the temperature curve in Fig.~\ref{h2aw2kin2gte2mp2p}, which indicates that nonzero remnant mass remains at the end of the evaporation process.

As it is commonly considered in the literature, the entropy itself is a essential quantity to be taken into account within the context of thermodynamic investigations. Nevertheless, in this particular scenario, using the conventional expression $S = \pi\, r_{h}^{2}$ \cite{furtado2023thermal} leads to no dependence on the non--commutative parameter $\Theta$, thereby excluding any deformation effects. As a result, we do not present graphical representations of the entropy in this analysis. Nevertheless, it will be used to accomplish the analysis of the heat capacity in the next subsection.


\subsection{Heat capacity}

To conclude the examination of the system’s thermodynamic properties, we now turn to the analysis of the heat capacity. This quantity is expressed as follows:
\ie
\begin{split}
C^{(\Theta,\ell)}_{V} & =  T \frac{\partial S}{\partial T} =  -2 \left(\pi  r_{h}^2\right) -\frac{2 \pi  r^2 \left(3 \Theta ^2+4 (\ell+1) r_{h}^2\right)}{9 \Theta ^2+4 (\ell+1) r_{h}^2} \\
&  \approx \, -2 \left(\pi  r_{h}^2\right) +   {3\pi(1 - \ell)}\Theta^2. 
\end{split}
\fe

\begin{figure}
    \centering
    \includegraphics[scale=0.7]{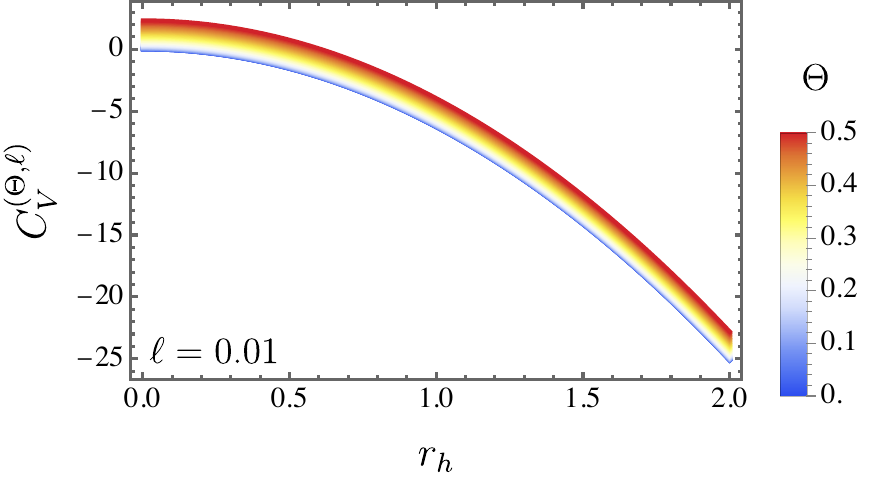}
    \caption{The behavior of the heat capacity as a function of the black hole mass $M$ is illustrated for various choices of $\Theta$, while keeping fixed $\ell = 0.01$.}
    \label{1h1eat1t1t}
\end{figure}

In this manner, Fig. \ref{1h1eat1t1t} depicts how the heat capacity responds to changes in the non--commutative parameter $\Theta$, with the Lorentz--violating parameter set to $\ell = 0.01$. 

\section{\label{Sec4}Quantum radiation}

Having established in the previous section that the black hole radiates thermally via the Hawking mechanism, we now turn to a detailed investigation of the quantum particle emission itself. This analysis includes both bosonic and fermionic modes. The number densities associated with each type of particle are calculated separately, enabling a direct comparison between their respective emission rates. As will be demonstrated, for a fixed frequency $\omega$ (in the low frequency regime scenario), bosonic particles are produced in greater quantities than their fermionic counterparts.


\subsection{Bosonic particle modes}

To ensure that energy conservation is properly incorporated into the analysis of black hole radiation, we employ the tunneling method developed in \cite{011,vanzo2011tunnelling,parikh2004energy,calmet2023quantum}. This approach begins by reformulating the spacetime geometry using the Painlevé--Gullstrand coordinate system, under which the metric is expressed as
\ie
\mathrm{d}s^2 = -A^{(\Theta,\ell)}\,\mathrm{d}t^2 + 2 H^{(\Theta,\ell)}\,\mathrm{d}t\,\mathrm{d}r + \mathrm{d}r^2 + Z^{(\Theta,\ell,\theta)}\,\mathrm{d}\theta^2 + Y^{(\Theta,\ell,\theta)}\,\mathrm{d}\varphi^2,
\fe
where the function $H^{(\Theta,\ell)}$ is defined by $H^{(\Theta,\ell)} = \sqrt{A^{(\Theta,\ell)}\left(B^{(\Theta,\ell)} - 1\right)}$, following the formulation proposed in \cite{calmet2023quantum}. Within this framework, the emission probability of a quantum particle is linked to the imaginary component of its classical action, as discussed in \cite{parikh2004energy,vanzo2011tunnelling,calmet2023quantum}. For a massless particle, the action is written as
\ie
\mathcal{S} = \int p_\mu\, \mathrm{d}x^\mu,
\fe
{where $p_{\mu}$ is the conjugate momentum.}

When extracting the corresponding imaginary part, $\text{Im} \, \mathcal{S}$, it is the radial contribution that governs the result. The temporal term $p_{t}\, \mathrm{d}t = -\omega \, \mathrm{d}t$ remains purely real and, therefore, does not contribute to the tunneling amplitude. As such, only the radial integration yields a nonzero imaginary component, which enable us to evaluate the tunneling probability as follows
\ie
\text{Im}\,\mathcal{S}=\text{Im}\,\int_{r_i}^{r_f} \,p_r\,\mathrm{d}r=\text{Im}\,\int_{r_i}^{r_f}\int_{0}^{p_r} \,\mathrm{d}p_r'\,\mathrm{d}r.
\fe

Let us define the Hamiltonian as $H = M - \omega'$, with $\omega'$ denoting the instantaneous energy of the particle being emitted. According to Hamilton’s equations, this leads to the differential relation $\mathrm{d}H = -\mathrm{d}\omega'$. Integrating over the entire energy range of the emitted particle, from $0$ to the total energy $\omega$, the imaginary part of the action becomes:
\ie
\begin{split}
\text{Im}\, \mathcal{S} & = \text{Im}\,\int_{r_i}^{r_f}\int_{M}^{M-\omega} \,\frac{\mathrm{d}H}{\mathrm{d}r/\mathrm{d}t}\,\mathrm{d} r  =\text{Im}\,\int_{r_i}^{r_f}\,\mathrm{d}r\int_{0}^{\omega} \,-\frac{\mathrm{d}\omega'}{\mathrm{d}r/\mathrm{d}t}\,.
\end{split}
\fe
Now, let us rearranging the integration order and applying a suitable change of variables, so that
\ie
\begin{split}
 \frac{\mathrm{d}r}{\mathrm{d}t} = &  -h(r)+\sqrt{f(r)+h(r)^2} \\
& = \frac{1}{2} \left(\sqrt{\frac{\gamma(\Theta,\ell)  \left(2 r^3 (2 (\ell-1) M+r)+\Theta ^2 M (-3 (\ell-1) M-2 r)\right)}{r^6 (2 (\ell-1) M+r)^2}} \right.\\
& \left.  -\sqrt{\frac{\gamma(\Theta,\ell)  \left(2 r^2 (2 (\ell-1) M+r) (2 (\ell-1) M + \ell r)-\Theta ^2 (\ell-1) M (3 (\ell-1) M+2 r)\right)}{(\ell-1) r^6 (2 (\ell-1) M+r)^2}}\right) \\
& \approx \,  1 - \frac{ \sqrt{\frac{2 \ell M+\ell r -2 M}{\ell-1}}}{\sqrt{r}} + \frac{(4 \ell M^2-4 M^2+M r)\Theta^{2}}{2\,r^3 (2 \ell M-2 M+r)}\\
& -\frac{\sqrt{\frac{2 \ell M+\ell r -2 M}{\ell-1}} \left(22 \ell^2 M^3+8 \ell^2 M^2 r-44 \ell M^3+3 \ell M^2 r+2 \ell M r^2+22 M^3-11 M^2 r+2 M r^2\right)\Theta^{2}}{4 r^{7/2} (2 \ell M-2 M+r) (2 \ell M+\ell r -2 M)},
\end{split}
\fe
and, expanding the expression up to second order in the non--commutative parameter $\Theta$, we introduce the auxiliary function $\gamma(\Theta,\ell) \equiv 2 r^3 \big[2(\ell - 1) M + r\big] + \Theta^2 M \big[11(\ell - 1) M + 4r\big]$. To account for the particle's energy loss during emission, the substitution $M \rightarrow (M - \omega')$ is applied to the metric components. This results in a modified form of the function, now expressed as:
\ie
\begin{split}
\label{ims}
&\text{Im}\, \mathcal{S} =\text{Im}\,\int_{0}^{\omega} -\mathrm{d}\omega'\\
& \times \int_{r_i}^{r_f}\,\frac{\mathrm{d}r}{1 - \frac{ \sqrt{\frac{2 \ell(M - \omega') + \ell r -2 (M - \omega')}{\ell-1}}}{\sqrt{r}} + \frac{(4 \ell(M - \omega')^2-4 (M - \omega')^2+(M - \omega') r)\Theta^{2}}{2\,r^3 (2\ell(M - \omega')-2 (M - \omega')+r)} + \Xi(\Theta,\ell,\omega')},
\end{split}
\fe
where
\ie
\Xi(\Theta,\ell,\omega') \equiv \, - \frac{\sqrt{\frac{2\ell(M - \omega')+\ell r -2 (M - \omega')}{\ell-1}} \gamma(\Theta,\ell,\omega') }{4 r^{7/2} (2\ell(M - \omega')-2 (M - \omega')+r) (2\ell(M - \omega')+\ell r -2 (M - \omega'))}
\fe
and
\ie
\begin{split}
\gamma(\Theta,\ell,\omega') & \equiv  \,  \left[22 \ell^2 (M - \omega')^3+8 \ell^2 (M - \omega')^2 r-44 \ell(M - \omega')^3+3 \ell(M - \omega')^2 r \right.\\
& \left.  +2\ell(M - \omega') r^2+22 (M - \omega')^3-11 (M - \omega')^2 r+2 (M - \omega') r^2\right] \Theta^{2}.
\end{split}
\fe

To illustrate, define the auxiliary quantity $\Delta(r,\omega') \equiv 2 (M - \omega')(1 - \ell)$. Introducing the energy-dependent shift $M \rightarrow (M - \omega')$ alters the position of the pole, which now appears at the adjusted horizon radius $r = 2(M - \omega')(1 - \ell)$. Before proceeding with the integration, the integrand is expanded in order to us obtain analytical outcomes, retaining terms up to second order in the non--commutative parameter $\Theta$. Therefore, it reads
\ie
\begin{split}
\label{imspertubated}
&\text{Im}\, \mathcal{S} \approx \, \text{Im}\,\int_{0}^{\omega} -\mathrm{d}\omega'\\
& \times \int_{r_i}^{r_f}\, 
\left\{ \frac{1}{1-\frac{\sqrt{\frac{\ell r}{\ell-1}+2 M-2 \omega'}}{\sqrt{r}}}-\frac{\zeta(\Theta,\ell,\omega')  (M-\omega')}{r^{7/2} \left(2-\frac{2 \sqrt{\frac{\ell r}{\ell-1}+2 M-2 \omega'}}{\sqrt{r}}\right)^2 (2 (\ell-1) M-2 (\ell-1) \omega'+r)} \right\} \mathrm{d}r,
\end{split}
\fe
where 
\ie
\zeta(\Theta,\ell,\omega') \equiv \,  \left(\frac{\beta(\Theta,\ell,\omega') }{(\ell-1) \sqrt{\frac{\ell r}{\ell-1}+2 M-2 \omega'}}+2 \sqrt{r} (4 (\ell-1) M-4 (\ell-1) \omega'+r)\right)\Theta^2 ,
\fe
with
\ie
\begin{split}
 \beta(\Theta,\ell,\omega') \equiv & \, -22 (\ell-1)^2 M^2-(\ell-1) M ((8 \ell+11) r-44 (\ell-1) \omega') \\
& -2 (\ell+1) r^2+(\ell-1) (8 \ell+11) r \omega'-22 (\ell-1)^2 \omega'^2.
\end{split}
\fe

Performing the contour integral around the shifted pole in the complex $r$-plane, oriented counterclockwise, results in the following expression:
\begin{eqnarray}
    \text{Im}\, \mathcal{S}  =  2 \pi  (1-\ell)^2 \omega  (2 M-\omega ) +\frac{\pi   (16 \ell-17) \Big[\ln (M)-\ln (M-\omega )\Big]\Theta^2}{16 (1-\ell)^2}.
\end{eqnarray}
A few remarks are in order here. First, the leading term on the right--hand side corresponds to the imaginary part of the action associated with the Kalb--Ramond black hole, as previously established in the literature \cite{Yang:2023wtu,12araujo2024particle}. In contrast, the second term encodes the contribution stemming from the non--commutative gauge framework, directly reflecting the deformation effects discussed in earlier sections of this study.

Accordingly, once non--commutative contributions are incorporated, the modified expression for the particle emission rate is given by:
\ie
\begin{split}
\Gamma (\Theta,\ell,\omega) & \sim e^{-2 \, \text{Im}\, S} = e^{ - 4 \pi  (1-\ell)^2 \omega  (2 M-\omega ) -\frac{\pi  (16 \ell-17) \big[\ln M - \ln (M-\omega )\big]\Theta^2}{8 (1-\ell)^2} } \\
&  { \approx \, \, 8 \pi  \ell \omega  e^{-4 \pi  \omega  (2 M-\omega )} (2 M-\omega )+e^{-4 \pi  \omega  (2 M-\omega )}    }  \\
& { \Bigg[  \frac{1}{4} \pi  \ell e^{-4 \pi  \omega  (2 M-\omega )} \left(136 \pi  M \omega -68 \pi  \omega ^2+9\right) (\ln M - \ln (M-\omega )) } \\
& {+\frac{17}{8} \pi  e^{-4 \pi  \omega  (2 M-\omega )} (\ln M -\ln (M-\omega )) \Bigg]\Theta^{2}  }.
\end{split}
\fe
Notice that the first term on the right--hand side, $e^{- 4 \pi (1 - \ell)^2 \omega (2M - \omega)}$, aligns with recent findings in the context of Kalb--Ramond gravity, as reported in \cite{12araujo2024particle}. As anticipated, the second term, $e^{- \frac{\pi (16 \ell - 17)\left[\ln M - \ln (M - \omega)\right] \Theta^2}{8 (1 - \ell)^2}}$, arises solely due to the presence of non--commutative corrections.

Moreover, the particle number density is directly related to the tunneling probability and can be written as:
\ie
\label{bonsfsdfdsf}
\begin{split}
n(\Theta,\ell,\omega) & = \frac{\Gamma(\Theta,\ell,\omega)}{1 - \Gamma(\Theta,\ell,\omega)} = \frac{1}{\exp \left\{  4 \pi  (1-\ell)^2 \omega  (2 M-\omega )+ \frac{\pi  (16 \ell-17) \big[\ln M-\ln (M-\omega )\big]\Theta^2}{8 (1-\ell)^2}\right\}-1} \\
&  { \approx \, \, \frac{8 \pi  \ell \omega  e^{4 \pi  \omega  (2 M-\omega )} (2 M-\omega )}{\left(e^{4 \pi  \omega  (2 M-\omega )}-1\right)^2}+\frac{1}{e^{4 \pi  \omega  (2 M-\omega )}-1}   }   \\
& { + \Bigg[ \frac{17 \pi  e^{4 \pi  \omega  (2 M-\omega )} (\ln M - \ln (M-\omega ))}{8 \left(e^{4 \pi  \omega  (2 M-\omega )}-1\right)^2}    }  \\
&   { + \frac{\pi  e^{4 \pi  \omega  (2 M-\omega )}}{4 \left(e^{4 \pi  \omega  (2 M-\omega )}-1\right)^3} \times \Bigg(  -68 \pi  \omega ^2 e^{4 \pi  \omega  (2 M-\omega )}+136 \pi  M \omega  e^{4 \pi  \omega  (2 M-\omega )}+136 \pi  M \omega }\\
& {+9 e^{4 \pi  \omega  (2 M-\omega )}-68 \pi  \omega ^2-9\Bigg) \times \Bigg( \ell (\ln M - \ln (M-\omega )) \Bigg)    \Bigg]\Theta^{2}}.
\end{split}
\fe

\begin{figure}
    \centering
    \includegraphics[scale=0.67]{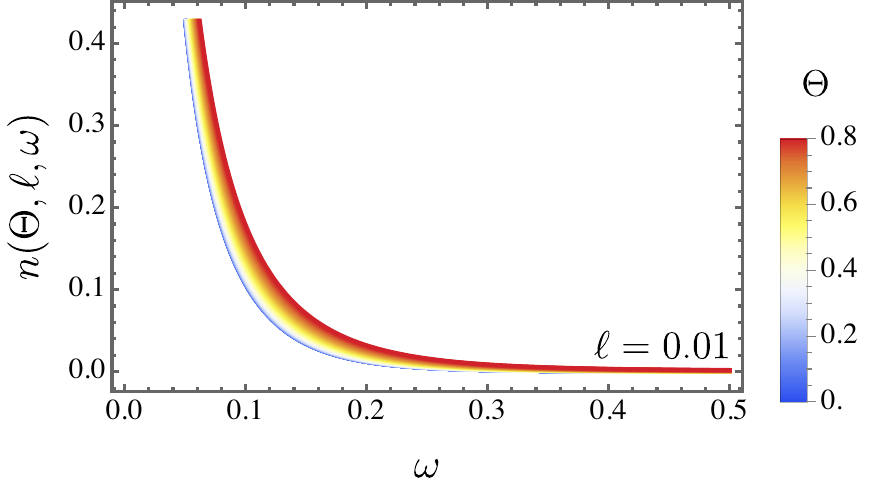}
    \includegraphics[scale=0.67]{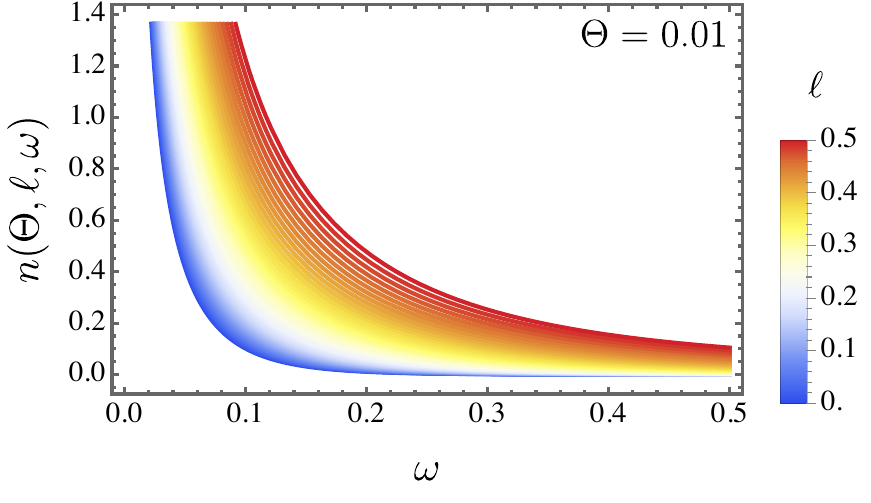}
    \caption{The particle creation density $n(\Theta,\ell,\omega)$ is plotted against the frequency $\omega$, with two distinct parameter settings: in the upper panel, $\ell$ is fixed at $0.01$ while $\Theta$ varies; in the lower panel, $\Theta$ is held constant at $0.01$ and different values of $\ell$ are considered.}
    \label{bosparticleonsss}
\end{figure}

Fig. \ref{bosparticleonsss} presents the distribution of bosonic particle production, denoted by {$n(\Theta,\ell,\omega)$}\footnote{{  It should be emphasized that, while the expansion in the second line of Eq. (\ref{bonsfsdfdsf}) is written only up to second order in $\Theta$ and first order in $\ell$, the plots in Fig. \ref{bosparticleonsss} were produced using the complete expression. This was done so that the effects of non--commutativity and Lorentz violation appear more clearly and can be more readily distinguished in the figures.     }}. As observed, both the non--commutative parameter $\Theta$ and the Lorentz--violating parameter $\ell$ increase the particle creation density. However, it is worth emphasizing that the system exhibits a stronger response to variations in $\ell$ than to changes in $\Theta$.


\subsection{Fermionic particle modes}


This section investigates the quantum tunneling of fermions across the event horizon associated with the black hole geometry under consideration. While various approaches—such as those using generalized Painlevé--Gullstrand or Kruskal--Szekeres coordinates—have been explored in earlier foundational studies \cite{o69}, our analysis starts from the general metric form provided in Eq.~(\ref{metrictensorss}). In a curved spacetime background, the evolution of spin--$\tfrac{1}{2}$ particles is governed by the Dirac equation, which takes the following form:
\ie
\Big(\gamma^\mu \nabla_\mu + m \Big) \Psi^{\pm}_{\uparrow\downarrow}(x) = 0
\fe
in which
\ie
\nabla_\mu = \partial_\mu + \frac{\mathbbm{i}}{2} {\Gamma^\alpha_{\;\mu}}^{\;\beta} \,\Sigma_{\alpha\beta}\fe
where 
\ie
\Sigma_{\alpha\beta} = \frac{\mathbbm{i}}{4} [\gamma_\alpha,  \gamma_\beta].
\fe

It should be emphasized that the coordinate system is defined by $x \equiv (t, r, \theta, \varphi)$. The $\gamma^\mu$ matrices represent the generators of the Clifford algebra in curved spacetime and satisfy the defining anticommutation relation:
\ie
\{\gamma_\alpha,\gamma_\beta\} = 2 g_{\alpha\beta} \mathbbm{1}.
\fe
Here, $\mathbbm{1}$ refers to the $4 \times 4$ identity matrix. In this framework, the explicit representation of the $\gamma$ matrices adopted in the analysis is given as follows: 
\begin{eqnarray*}
 \gamma^{t}(\Theta,\ell) &=&\frac{\mathbbm{i}}{\sqrt{{A^{(\Theta,\ell)}}}}\left( \begin{array}{cc}
\bf{1}& \bf{ 0} \\ 
\bf{ 0} & -\bf{ 1}%
\end{array}%
\right), \;\;
\gamma^{r}(\Theta,\ell) =\sqrt{\frac{1}{B^{(\Theta,\ell)}{}}}\left( 
\begin{array}{cc}
\bf{0} &  \sigma_{3} \\ 
 \sigma_{3} & \bf{0}%
\end{array}%
\right), \\
\gamma^{\theta }(\Theta,\ell) &=&\frac{1}{\sqrt{{C^{(\Theta,\ell)}}}}\left( 
\begin{array}{cc}
\bf{0} &  \sigma_{1} \\ 
 \sigma_{1} & \bf{0}%
\end{array}%
\right), \;\;
\gamma^{\varphi }(\Theta,\ell) =\frac{1}{{\sqrt{D^{(\Theta,\ell)}} }}\left( 
\begin{array}{cc}
\bf{0} &  \sigma_{2} \\ 
 \sigma_{2} & \bf{0}%
\end{array}%
\right).
\end{eqnarray*}%
Within this setup, the symbols $\sigma$ denote the standard Pauli matrices, which satisfy the well--known algebraic relation:
$\sigma_i \sigma_j = \mathbbm{1} \delta_{ij} + \mathbbm{i} \varepsilon_{ijk} \sigma_k, \quad \text{for} \quad i,j,k = 1,2,3.$
Furthermore, the matrix corresponding to $\gamma^5(\Theta,\ell)$—accounting for non--commutative and Lorentz--violating effects—can be identified with the following structure:
\begin{equation*}
\gamma^{5}(\Theta,\ell) = \mathbbm{i} \gamma^{t}(\Theta,\ell)\gamma^{r}(\Theta,\ell)\gamma^{\theta }(\Theta,\ell)\gamma^{\varphi}(\Theta,\ell) = \mathbbm{i}\sqrt{\frac{1}{{A^{(\Theta,\ell)} \, B^{(\Theta,\ell)} \, C^{(\Theta,\ell)} \, D^{(\Theta,\ell)} }}}\left( 
\begin{array}{cc}
\bf{ 0} & - \bf{ 1} \\ 
\bf{ 1} & \bf{ 0}%
\end{array}%
\right)\:.
\end{equation*}

To represent a Dirac field whose spin is oriented upward along the positive $r$--direction, the following ansatz is employed \cite{vanzo2011tunnelling}:
\begin{equation}
\psi^{(+)}_{\uparrow}(x) = \left( \begin{array}{c}
H^{(+)}_{\uparrow}(x) \\ 
0 \\ 
Y^{(+)}_{\uparrow}(x) \\ 
0
\end{array}
\right) \exp \left[ \mathbbm{i} \, \psi^{(+)}_{\uparrow}(x)\right]\;.
\label{spinupbh} 
\end{equation}

The analysis centers on the spin--up $(+)$ configuration, corresponding to alignment along the positive $r$--axis. The spin-down $(-)$ case, which represents orientation in the opposite direction, can be handled through an analogous procedure. Inserting the spinor ansatz from Eq.~(\ref{spinupbh}) into the Dirac equation yields the following set of expressions, in line with the method outlined by Vanzo et al. \cite{vanzo2011tunnelling}
\begin{align}
&-\frac{\mathbbm{i}}{\sqrt{A^{(\Theta,\ell)}}}\left( H^{(+)}_{\uparrow}(x)\,\partial_{t}\psi^{(+)}_{\uparrow}\right)
-\sqrt{\frac{1}{B^{(\Theta,\ell)}}}\left( Y^{(+)}_{\uparrow}(x)\,\partial_{r}\psi^{(+)}_{\uparrow}\right)
+m \mathbbm{i} H^{(+)}_{\uparrow}(x) = 0, \label{eq1} \\[10pt]
& - \frac{1}{\sqrt{C^{(\Theta,\ell)}}}\left(Y^{(+)}_{\uparrow}(x)\,\partial_{\theta}\psi^{(+)}_{\uparrow}\right)
-\frac{1}{\sqrt{D^{(\Theta,\ell)}}}\left( \mathbbm{i}Y^{(+)}_{\uparrow}(x)\,\partial_{\varphi}\psi^{(+)}_{\uparrow}\right) = 0, \label{eq2} \\[10pt]
&    \frac{\mathbbm{i}}{\sqrt{A^{(\Theta,\ell)}}}\left(Y^{(+)}_{\uparrow}(x)\,\partial_{t}\psi^{(+)}_{\uparrow}\right)
-\sqrt{\frac{1}{B^{(\Theta,\ell)}}}\left(H^{(+)}_{\uparrow}(x)\,\partial_{r}\psi^{(+)}_{\uparrow}\right)
+ m \mathbbm{i} Y^{(+)}_{\uparrow}(x) = 0, \label{eq3} \\[10pt]
& - \frac{1}{\sqrt{C^{(\Theta,\ell)}}}\left( H^{(+)}_{\uparrow}(x)\,\partial_{\theta}\psi^{(+)}_{\uparrow}\right)
-\frac{\mathbbm{i}}{\sqrt{D^{(\Theta,\ell)}}}\left( H^{(+)}_{\uparrow}(x)\,\partial_{\varphi}\psi^{(+)}_{\uparrow}\right) = 0. \label{eq4}
\end{align}
Here, only the leading contributions in $\hbar$ have been retained in the expansion. We now proceed by introducing the following form for the action:
\ie
\psi^{(+)}_{\uparrow}= - \omega\, t + \chi(r) + L(\theta ,\varphi ) 
\fe
so that
\cite{vanzo2011tunnelling} 
\begin{align}
&+\frac{\mathbbm{i}\, \omega H^{(+)}_{\uparrow}(x)}{\sqrt{A^{(\Theta,\ell)}}} 
-\sqrt{\frac{1}{B^{(\Theta,\ell)}}} Y^{(+)}_{\uparrow}(x)\,\chi^{\prime}(r)
+m \mathbbm{i} H^{(+)}_{\uparrow}(x) = 0, \label{eq11} \\[10pt]
& - Y^{(+)}_{\uparrow}(x) \left(  \frac{\partial_{\theta}L(\theta,\varphi)}{\sqrt{C^{(\Theta,\ell)}}}
+\frac{\mathbbm{i}\,\partial_{\varphi} L(\theta,\varphi)}{\sqrt{D^{(\Theta,\ell)}}} \right) = 0, \label{eq21} \\[10pt]
&    -\frac{\mathbbm{i} \, \omega Y^{(+)}_{\uparrow}(x)}{\sqrt{A^{(\Theta,\ell)}}}
-\sqrt{\frac{1}{B^{(\Theta,\ell)}}}H^{(+)}_{\uparrow}(x)\chi^{\prime}(r)
+ m \mathbbm{i} Y^{(+)}_{\uparrow}(x) = 0, \label{eq31} \\[10pt]
& - H^{(+)}_{\uparrow}(x) \left(  \frac{\partial_{\theta}L(\theta,\varphi)}{\sqrt{C^{(\Theta,\ell)}}}
+\frac{\mathbbm{i}\,\partial_{\varphi} L(\theta,\varphi)}{\sqrt{D^{(\Theta,\ell)}}} \right) = 0. \label{eq41}
\end{align}

The particular expressions of $H^{(+)}_{\uparrow}(x)$ and $Y^{(+)}_{\uparrow}(x)$ are not essential to the outcome, as the key point is that Eqs.~(\ref{eq21}) and (\ref{eq41}) together enforce the following constraint: 
$\frac{\partial_{\theta}L(\theta,\varphi)}{\sqrt{C^{(\Theta,\ell)}}}
+\frac{\mathbbm{i}\,\partial_{\varphi} L(\theta,\varphi)}{\sqrt{D^{(\Theta,\ell)}}} = 0.$
In other words, this feature indicates that $L(\theta, \varphi)$ must be treated as a complex--valued function, a condition that applies equally to both emission and absorption processes. As a result, when evaluating the ratio of outgoing to ingoing tunneling probabilities, the contributions involving $L(\theta, \varphi)$ cancel out, making it unnecessary for the remainder of the analysis—an observation also noted in Ref. \cite{vanzo2011tunnelling}. For the massless case ($m = 0$), Eqs. (\ref{eq11}) and (\ref{eq31}) yield two distinct and independent solutions:
\ie
H^{(+)}_{\uparrow}(x) = - \mathbbm{i} Y^{(+)}_{\uparrow}(x), \qquad \chi^{\prime }(r) = \chi_{\text{out}}' {(r)} = \frac{\omega}{\sqrt{\frac{A^{(\Theta,\ell)}}{B^{(\Theta,\ell)}}}},
\fe
\ie
H^{(+)}_{\uparrow}(x) = \mathbbm{i} Y^{(+)}_{\uparrow}(x), \qquad \chi^{\prime }(r) = \chi_{\text{in}}' {(r)} = - \frac{\omega}{\sqrt{\frac{A^{(\Theta,\ell)}}{B^{(\Theta,\ell)}}}}.
\fe
Within this framework, the functions $\chi_{\text{out}}(r)$ and $\chi_{\text{in}}(r)$ correspond to the radial components of the wavefunction for particles tunneling outward and inward, respectively \cite{vanzo2011tunnelling}. As a result, the tunneling probability for fermionic particles is expressed as
$\Gamma_{\psi}(\Theta,\ell,\omega) \sim \exp\left[-2\, \text{Im} \left(\chi_{\text{out}}(r) - \chi_{\text{in}}(r)\right)\right].$
Accordingly, the final expression becomes:
\ie
 \chi_{ \text{out}}(r)= -  \chi_{ \text{in}} (r) = \int \mathrm{d} r \,\frac{\omega}{\sqrt{\frac{A^{(\Theta,\ell)}}{B^{(\Theta,\ell)}}}}\:.
\fe

An additional point deserves attention: given that both $\Theta$ and $\ell$ are assumed to be very small—an assumption that will be quantitatively supported later using solar system constraints—the behavior of the metric functions in the vicinity of the event horizon $r = r_h$ can be approximated as linear. This approximation justifies treating the singularity as a simple pole in the relevant integrals. Under this consideration, we can express the following:
\ie
A^{(\Theta,\ell)} \frac{1}{B^{(\Theta,\ell)}} \approx \, A^{(\Theta,\ell)\prime} \frac{1}{B^{(\Theta,\ell)\prime}}(r - r_{h})^{2} + ... \, .
\fe

Applying Feynman’s prescription for handling the pole structure, one arrives at the following result:
\ie
2\mbox{Im}\;\left[  \chi_{ \text{out}}{(r)} -  \chi_{ \text{in}} {(r)}\right] =\mbox{Im}\int \mathrm{d} r \,\frac{4\omega}{\sqrt{\frac{A^{(\Theta,\ell)}}{B^{(\Theta,\ell)}}}}=\frac{2\pi\omega}{\kappa(\Theta,\ell)},
\fe
with $\kappa$ is given by
\ie
\kappa(\Theta,\ell) \approx \, \, \frac{M}{r_{h}^2}+ \left[ \frac{M (28 M-9 r_{h})}{4 r_{h}^5}-\frac{9 \ell M}{4 r_{h}^4} \right]\Theta^{2}.
\fe

The fermionic particle number density $n_{\psi}$ associated with the black hole background obeys the relation $\Gamma_{\psi} (\Theta,\ell,\omega) \sim \exp\left(-\frac{2\pi \omega}{\kappa(\Theta,\ell)}\right)$. By substituting the expression for the event horizon $r_{h}$ obtained earlier from Eq.~(\ref{eventhorizonhay}), the result becomes:
\ie
\begin{split}
\label{fermmiiiipartii}
n_{\psi}(\Theta,\ell) & = {\frac{{\Gamma}_{\psi}(\Theta,\ell)}{1 + {\Gamma}_{\psi}(\Theta,\ell)}} \\
&  = \, \, \frac{1}{\exp \left(\frac{128 \pi  (\ell-1)^5 M^3 \omega }{16 (\ell-1)^3 M^2-\Theta ^2 \left(9 \ell^2+5\right)}\right)+1} \\
& {  \approx  \, \,  \frac{16 \pi  \ell M \omega  e^{8 \pi  M \omega }}{\left(e^{8 \pi  M \omega }+1\right)^2}+\frac{1}{e^{8 \pi  M \omega }+1} }\\
& {+ \left[ \frac{5 \pi  \ell \omega  e^{8 \pi  M \omega } \left(16 \pi  M \omega  e^{8 \pi  M \omega }-16 \pi  M \omega +e^{8 \pi  M \omega }+1\right)}{2 M \left(e^{8 \pi  M \omega }+1\right)^3}+\frac{5 \pi  \omega  e^{8 \pi  M \omega }}{2 M \left(e^{8 \pi  M \omega }+1\right)^2} \right]\Theta^{2} }.
\end{split}
\fe

\begin{figure}
    \centering
    \includegraphics[scale=0.66]{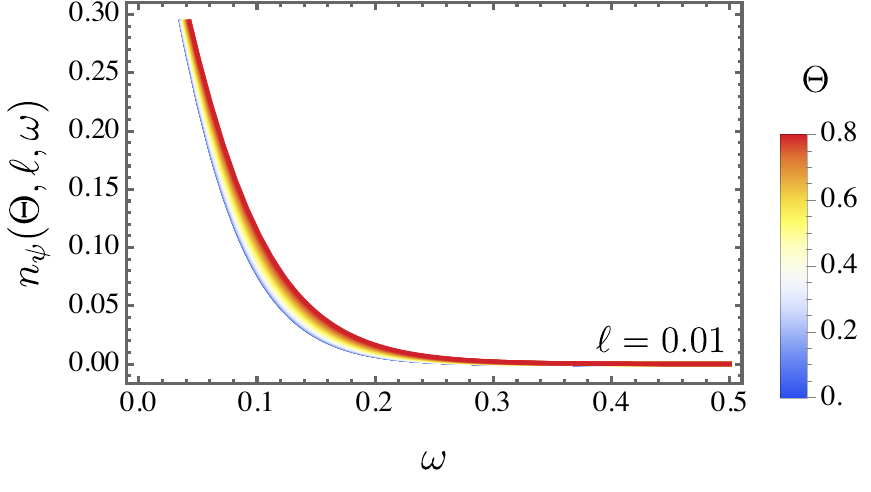}
    \includegraphics[scale=0.66]{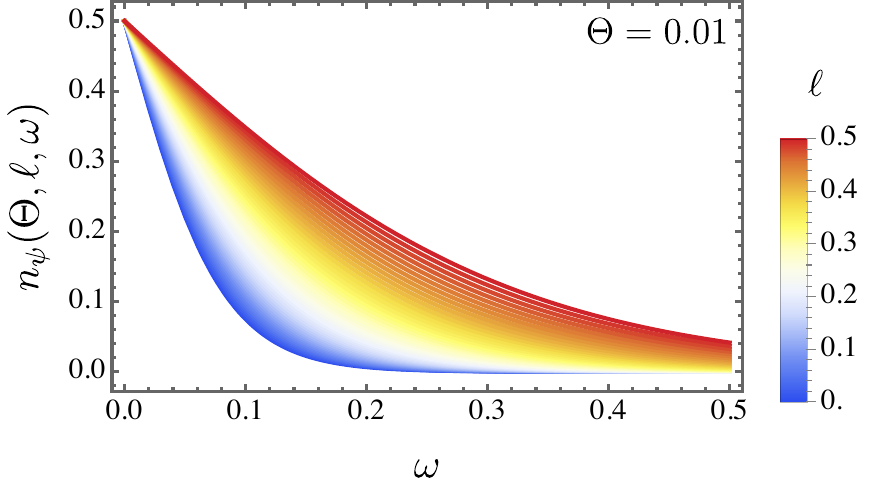}
    \caption{The particle creation density $n_{\psi}(\Theta,\ell,\omega)$ is plotted against the frequency $\omega$, with the upper panel showing its behavior for various values of the non--commutative parameter $\Theta$ at a fixed Lorentz--violating parameter $\ell = 0.01$, and the lower panel showing the effect of varying $\ell$ while keeping $\Theta$ constant at $0.01$.}
    \label{f1e1rr1mm1ss1s1s}
\end{figure}

Fig. \ref{f1e1rr1mm1ss1s1s} displays the fermionic particle creation density, $n_{\psi}$\footnote{{  Here, it is worth noting that although the expansion in the third line of Eq. (\ref{fermmiiiipartii}) is presented up to second order in $\Theta$ and first order in $\ell$, the plots in Fig. \ref{f1e1rr1mm1ss1s1s} were generated using the full expression. This choice was made to render the effects of non--commutativity and Lorentz violation more pronounced and easier to distinguish in the figures.     }}. As observed in the bosonic case, both the non--commutative parameter $\Theta$ and the Lorentz--violating parameter $\ell$ enhance the overall particle production rate, leading to a noticeable increase in the density’s amplitude therefore.

\begin{figure}
    \centering
    \includegraphics[scale=0.73]{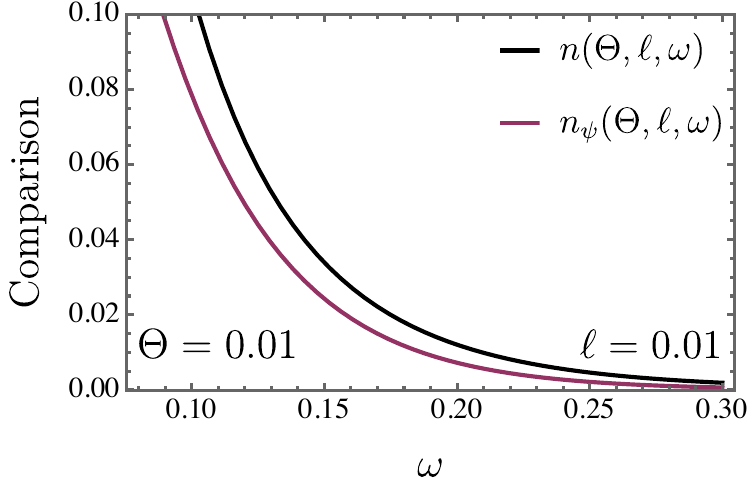}
    \caption{A comparison between the bosonic and fermionic particle creation densities is shown for fixed parameter values $\Theta = 0.01$ and $\ell = 0.01$.}
    \label{comp}
\end{figure}

Additionally, as the particle densities for bosons and fermions were obtained using distinct methodologies, a comparative analysis is presented in Fig.~\ref{comp} to highlight the differences between the two approaches. In general lines, it is observed that bosonic particle modes exhibit higher particle creation densities than their fermionic counterparts, particularly in the low--frequency regime. The outcomes (in the comparison) considered here were based on the following fixed parameters: $\Theta = 0.01$, $\ell = 0.01$ and $M=1$.

\section{\label{Sec9}Evaporation and Emission rate}

In this section, we provide a qualitative analysis of the black hole's evaporation behavior. The investigation is primarily based on the application of the \textit{Stefan--Boltzmann} law, which serves as the foundational framework for estimating energy loss through thermal radiation, as commoly reported in the literature \cite{ong2018effective,12aa2025particle}
\ie
\label{sflaw}
\frac{\mathrm{d}M}{\mathrm{d}t}  =  - \alpha a {\Tilde{\sigma}} T^{4},
\fe
with ${\Tilde{\sigma}}$ corresponds to the effective radiating (cross--section) area, $a$ is the radiation constant, $\alpha$ accounts for the greybody factor, and $T$ denotes the Hawking temperature. It is important to highlight that the emitted radiation predominantly consists of massless particles, such as photons and neutrinos \cite{hiscock1990evolution,page1976particle}. The cross--sectional area $\sigma$ is approximated by $\pi R_{sh}^2$, where $R_{sh}$ denotes the shadow radius, serving as the effective emission radius in the high--frequency (geometric optics) limit—often referred to as ${\Tilde{\sigma}}_{\text{lim}}$. Under these conditions, the greybody factor simplifies to $\alpha \to 1$, as noted in \cite{liang2025einstein}. Moreover, after some algebraic manipulations, we can write $\sigma_{lim}$ as follows:
\ie
{\Tilde{\sigma}}_{lim} \approx \, 27 \pi  M^2 -81 \pi  \ell M^2 -\frac{3 \pi  \Theta ^2}{4}+\frac{3}{4} \pi  \Theta ^2 \ell,
\fe
where we have neglected superior order than $\Theta^{2}$ in above expression.

Hawking’s groundbreaking discovery in Ref.\cite{hawking1975particle} revealed that black holes radiate thermally, with the emission governed by a characteristic temperature now known as the Hawking temperature. Using Eq.(\ref{masshawww}), this temperature can be re--expressed as a function of the black hole mass, i.e., 
\ie
\label{haah}
T^{(\Theta,\ell)} \approx \,  \frac{1}{8 \pi  (1-\ell) M} + \frac{l}{8 \pi  (1-\ell) M} + \frac{3 \Theta ^2}{128 \pi  (1-\ell)^3 M^3},
\nonumber
\fe 
we now move forward with the calculation. Although the Hawking temperature can alternatively be derived using Bogoliubov transformations, such an approach falls outside the scope of this work. Given the considerations outlined above, Eq.~(\ref{sflaw}) takes the following form:
\ie
\begin{split}
\label{dmmm}
\frac{\mathrm{d}M}{\mathrm{d}t} = &  { \, \frac{3 \xi \left(3 \Theta ^2+16 (\ell-1)^2 (\ell+1) M^2\right)^4 \left(36 (3 \ell-1) M^2-\Theta ^2 (\ell-1)\right)}{1073741824 \pi ^3 (\ell-1)^{12} M^{12}}} \\
\approx & \, \,  -\frac{27 \xi}{4096 \left(\pi ^3 M^2\right)} -\frac{135 \xi \ell}{4096 \left(\pi ^3 M^2\right)} + \left(-\frac{465 \xi \ell}{16384 \pi ^3 M^4}-\frac{39 \xi}{8192 \left(\pi ^3 M^4\right)}\right)\Theta ^2   \, ,
\end{split}
\fe
{where it is expanded up to second order in $\Theta$ and first order in $\ell$ so that the expression takes a simplified form; here, we have set $\xi = a\alpha$.} The next step involves evaluating the following integral:
\ie
\begin{split}
\label{dmmeee}
 \int_{0}^{t_{\text{evap}}} \xi \mathrm{d}\tau  =  &\int_{M_{i}}^{M_{f}} 
\left[ -\frac{27 }{4096 \left(\pi ^3 M^2\right)} -\frac{135  l}{4096 \left(\pi ^3 M^2\right)} + \left(-\frac{465  l}{16384 \pi ^3 M^4}-\frac{39 \xi}{8192 \left(\pi ^3 M^4\right)}\right)\Theta ^2    \right]^{-1} \mathrm{d}M, \\
& \approx \, \, \frac{20480 \pi ^3 \ell \left(M_{f}^3-M_{i}^3\right)}{81 } -\frac{4096 \left(\pi ^3 \left(M_{f}^3-M_{i}^3\right)\right)}{81 } \\
& \, \, \, \, \, \, \, + \left(\frac{26624 \pi ^3 (M_{f}-M_{i})}{243 }-\frac{35840 \ell \left(\pi ^3 (M_{f}-M_{i})\right)}{81 }\right)\Theta^2.
\end{split}
\fe
Here, $t_{\text{evap}}$ denotes the total evaporation time of the black hole. Under the given assumptions, it can be written as:
\ie
\begin{split}
t_{\text{evap}} = & \, \, \frac{20480 \pi ^3 \ell \left(M_{f}^3-M_{i}^3\right)}{81 \xi} -\frac{4096 \left(\pi ^3 \left(M_{f}^3-M_{i}^3\right)\right)}{81 \xi} \\
& \, \, \, \, \, \, \, + \left(\frac{26624 \pi ^3 (M_{f}-M_{i})}{243 \xi}-\frac{35840 \ell \left(\pi ^3 (M_{f}-M_{i})\right)}{81 \xi}\right)\Theta^2.
\end{split}
\fe

It is important to note that $t_{\text{evap}}$ is obtained through an explicit analytical evaluation. Given that the black hole under consideration does not yield a physical remnant, we assume complete evaporation occurs. In this context, the final mass approaches zero, i.e., $M_{f} \to 0$, leading to:
\ie
t_{\text{evap}} = \, \, \frac{4096 \pi ^3 M_{i}^3}{81 \xi} -\frac{20480 \pi ^3 \ell M_{i}^3}{81 \xi}+\frac{35840 \pi ^3 \Theta ^2 \ell M_{i}}{81 \xi}-\frac{26624 \pi ^3 \Theta ^2 M_{i}}{243 \xi}.
\fe

Observe that the leading contribution in the expression corresponds to the standard Schwarzschild evaporation time. The second term, however, deviates slightly from the result recently presented in Ref.\cite{12araujo2024particle}, due to our expansion being carried out only to first order in $\ell$ in Eqs.(\ref{dmmm}) and (\ref{dmmeee}).

\begin{figure}
    \centering
     \includegraphics[scale=0.51]{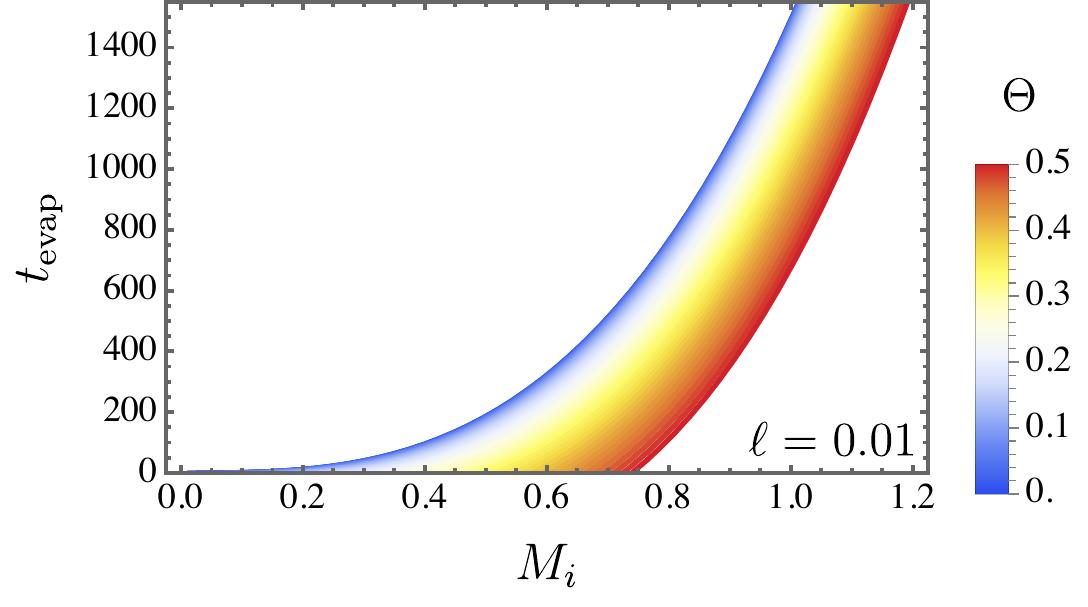}
     \includegraphics[scale=0.51]{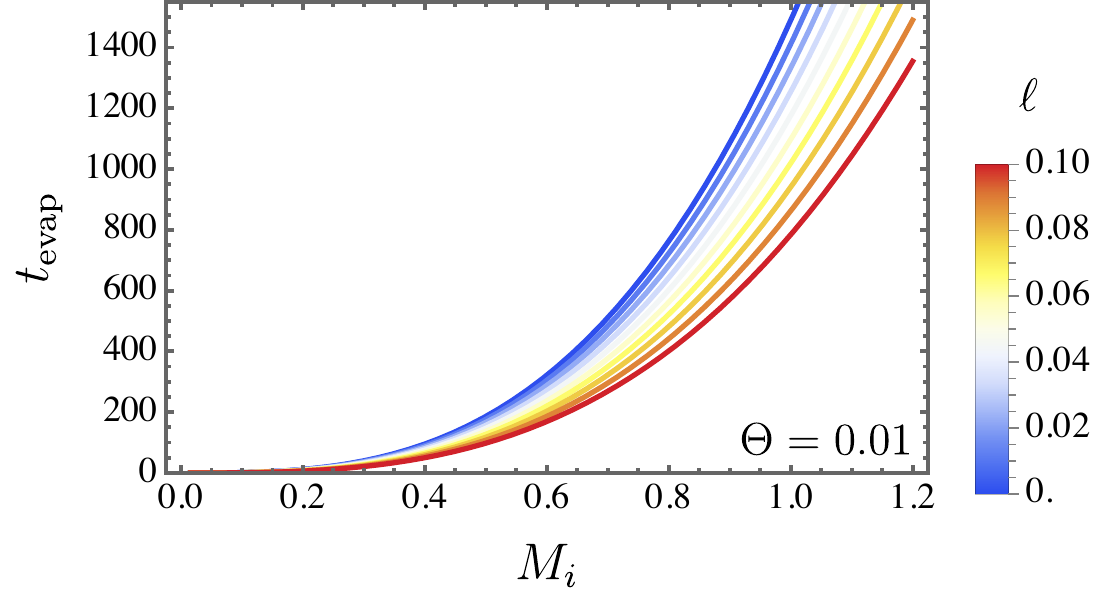}
    \caption{The evaporation lifetime at the its final stage $M_{f} \to 0$ for different values of initial mass $M_{i}$ {by considering different values of $\Theta$} (up panel) and different values of $\ell$ (bottom panel).}
    \label{tempevap}
\end{figure}

To better visualize the behavior of the evaporation process, we depict in Fig.~\ref{tempevap} the evaporation time $t_{\text{evap}}$ at its final stage ($M_f \to 0$) as a function of the initial mass $M_i$, considering varying values of $\Theta$ (upper panel) and $\ell$ (lower panel). A few remarks are in order. For fixed $\ell$, increasing the non--commutative parameter $\Theta$ accelerates the evaporation compared to the Schwarzschild case. A similar trend is found for fixed $\Theta = 0.01$, where larger values of $\ell$ also lead to faster evaporation. These behaviors are in line with prior results in the literature \cite{12araujo2024particle,12araujo2025does}. Finally, consistent with previous analyses in the high--frequency limit, we adopt $\xi \sim 1$ when constructing the plots, which corresponds to the limit in which greybody factors approach unity.

A natural question emerges at this point: how does non--commutativity influence the black hole evaporation time in the scenario under consideration? To address this properly, we present a comparative analysis involving different black hole solutions available in the literature—namely, the Schwarzschild solution, the non--commutative Schwarzschild case \cite{Juric:2025kjl}, the Kalb--Ramond black hole \cite{Yang:2023wtu}, and the non--commutative Kalb--Ramond black hole investigated in this work.

\begin{figure}
    \centering
     \includegraphics[scale=0.6]{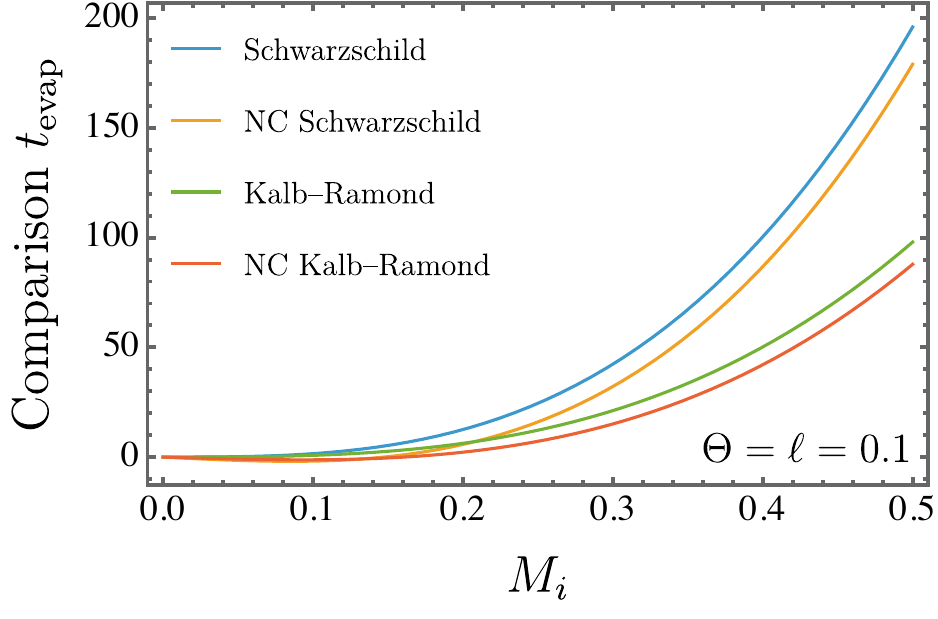}
    \caption{Comparison of the final evaporation states for different black hole configurations. The plot presents the evaporation agains the initial black hole mass $M_{i}$, taking into account four different cases: the Schwarzschild black hole, its non--commutative counterpart incorporating the same Moyal twist adopted in this work \cite{Juric:2025kjl}, the Kalb--Ramond black hole \cite{Yang:2023wtu}, and the non--commutative Kalb--Ramond solution developed in this study. Overall, the hierarchy observed is:
$t_{\text{evap}}^{\text{Schwarzschild}} > t_{\text{evap}}^{\text{NC Schwarzschild}} > t_{\text{evap}}^{\text{Kalb--Ramond}} > t_{\text{evap}}^{\text{NC Kalb--Ramond}}$.
 }
    \label{tempevapcomp}
\end{figure}

Fig.~\ref{tempevapcomp} displays the corresponding evaporation lifetimes for all these cases, providing a direct comparison to clarify the question posed above. From this analysis, it becomes evident that the introduction of non--commutativity consistently leads to a shorter evaporation time across all models considered. Furthermore, the hierarchy among the evaporation times follows the pattern:
$t_{\text{evap}}^{\text{Schwarzschild}} > t_{\text{evap}}^{\text{NC Schwarzschild}} > t_{\text{evap}}^{\text{Kalb--Ramond}} > t_{\text{evap}}^{\text{NC Kalb--Ramond}}.$ In other words, this clearly shows that both the non--commutative deformation and the presence of the Kalb--Ramond field accelerate the black hole evaporation process when compared to the standard Schwarzschild scenario.

Quantum effects manifest prominently in the vicinity of a black hole's event horizon, where transient particle–antiparticle pairs emerge due to vacuum fluctuations. Occasionally, one of these particles—carrying positive energy—may traverse the potential barrier and escape via quantum tunneling. This escape mechanism leads to a net energy loss from the black hole, contributing to its gradual evaporation, a process commonly known as Hawking radiation. For an observer located far from the black hole, this evaporation is indirectly associated with the absorption profile at high energies. In such regimes, the absorption cross--section tends toward a limiting value, denoted by ${\Tilde{\sigma}}_{\text{lim}}$. As reported in Refs. \cite{decanini2011universality,papnoi2022rotating}, this behavior governs the spectral energy emission, which can be formulated as follows:
\ie
\label{emission}
	\frac{{{\mathrm{d}^2}E}}{{\mathrm{d}\omega \mathrm{d}t}} = \frac{{2{\pi ^2}{\Tilde{\sigma}}_{lim}}}{{{e^{\frac{\omega }{T^{(\Theta,\ell)}}}} - 1}} {\omega ^3},
\fe
in which the photon frequency is denoted by $\omega$. At high frequencies, the absorption cross--section approaches a constant value, $\sigma_{\text{lim}}$, which is approximately given by the geometric relation $\sigma_{\text{lim}} \approx \pi R_{\text{sh}}^2$, where $R_{\text{sh}}$ stands for the radius associated with the black hole’s shadow. By incorporating the corresponding expressions for both the shadow radius and the black hole's Hawking temperature into the formalism, one arrives at the modified expression for the rate at which energy is radiated:
\ie
\frac{\mathrm{d}^{2}E}{\mathrm{d}\omega \mathrm{d} t} = \frac{\pi ^3 \omega ^3 \left(\Theta ^2 (\ell+2)+72 (3 \ell-2) M^2\right)^2}{384 M^2 \left( e^{-\frac{128 \pi  (\ell-1)^3 M^3 \omega }{3 \Theta ^2+16 (\ell-1)^2 (\ell+1) M^2}}-1\right)}.
\fe

\begin{figure}
    \centering
     \includegraphics[scale=0.51]{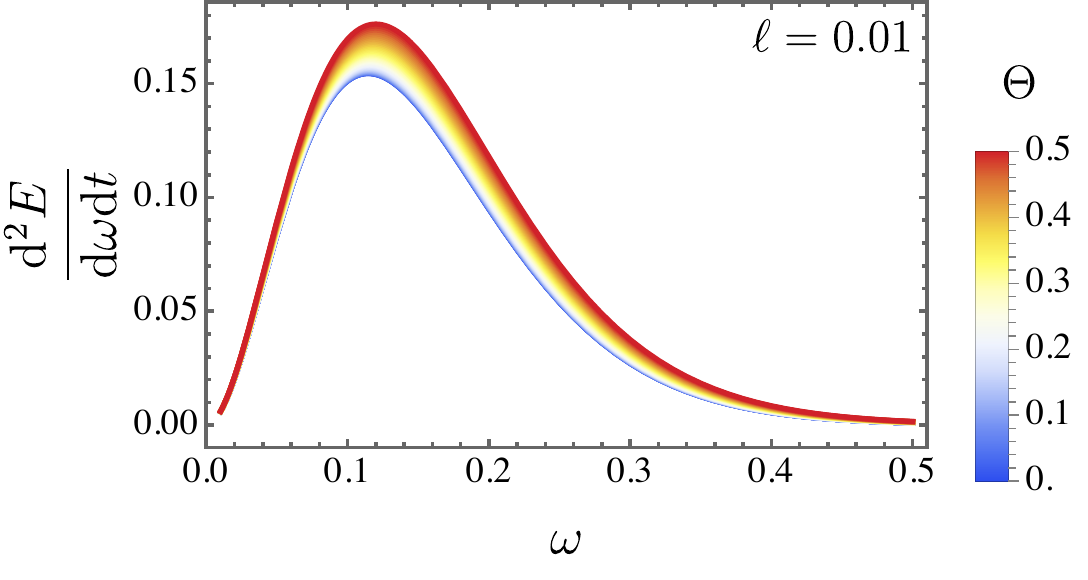}
     \includegraphics[scale=0.51]{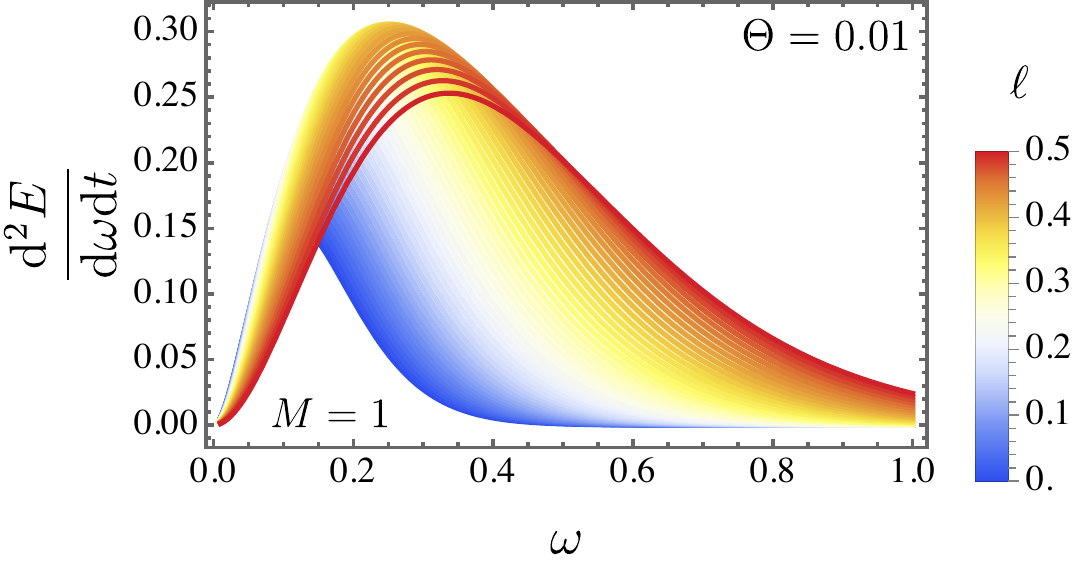}
    \caption{The energy emission rate of the radiated particles is presented for varying values of $\Theta$ (top panel) and distinct values of $\ell$ (bottom panel).}
    \label{emissionrateenergy}
\end{figure}

To clarify the physical implications of our results, Fig. \ref{emissionrateenergy} displays the spectral energy emission rate, $\frac{\mathrm{d}^{2}E}{\mathrm{d}\omega \mathrm{d} t}$. The upper panel illustrates how this quantity varies with different values of $\Theta$, while the lower panel presents its dependence on various choices of the Lorentz--violating parameter $\ell$.
In general lines, the emission rate is found to increase with higher values of both $\Theta$ and $\ell$. This behavior is consistent with the earlier results concerning the evaporation timescale. In essence, a greater energy flux—driven by larger values of $\Theta$ and $\ell$—implies a more rapid loss of mass, thereby accelerating the black hole's evaporation. This trend is precisely reflected in our previous analyses, as shown in Figs. \ref{tempevap} and \ref{tempevapcomp}.

Moreover, the emission rate for the radiated particles can also be readily determined
\begin{equation}
\frac{\mathrm{d}^{2}N}{\mathrm{d}\omega \mathrm{d}t}
= \frac{2\pi^{2}\,{\Tilde{\sigma}}_{lim}\,\omega^{2}}
       {e^{\,\omega / T^{(\Theta,\ell)}} - 1} = \frac{\pi ^3 \omega ^2 \left(\Theta ^2 (\ell+2)+72 (3\ell-2) M^2\right)^2}{384 M^2 \left(e^{-\frac{128 \pi  (\ell-1)^3 M^3 \omega }{3 \Theta ^2+16 (\ell-1)^2 (\ell+1) M^2}}-1\right)}.
\end{equation}

\begin{figure}
    \centering
     \includegraphics[scale=0.51]{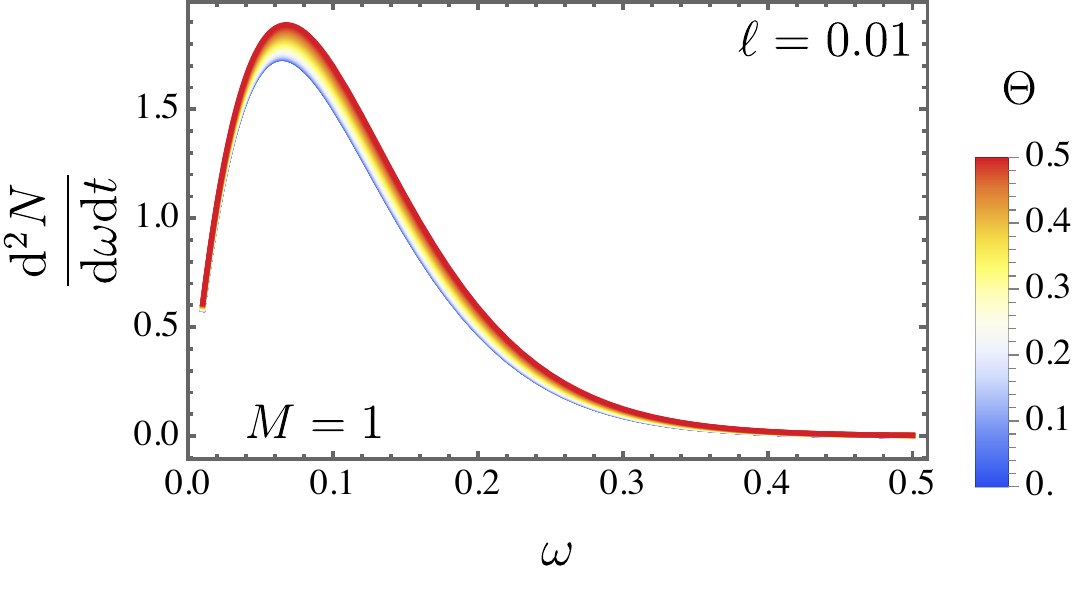}
     \includegraphics[scale=0.51]{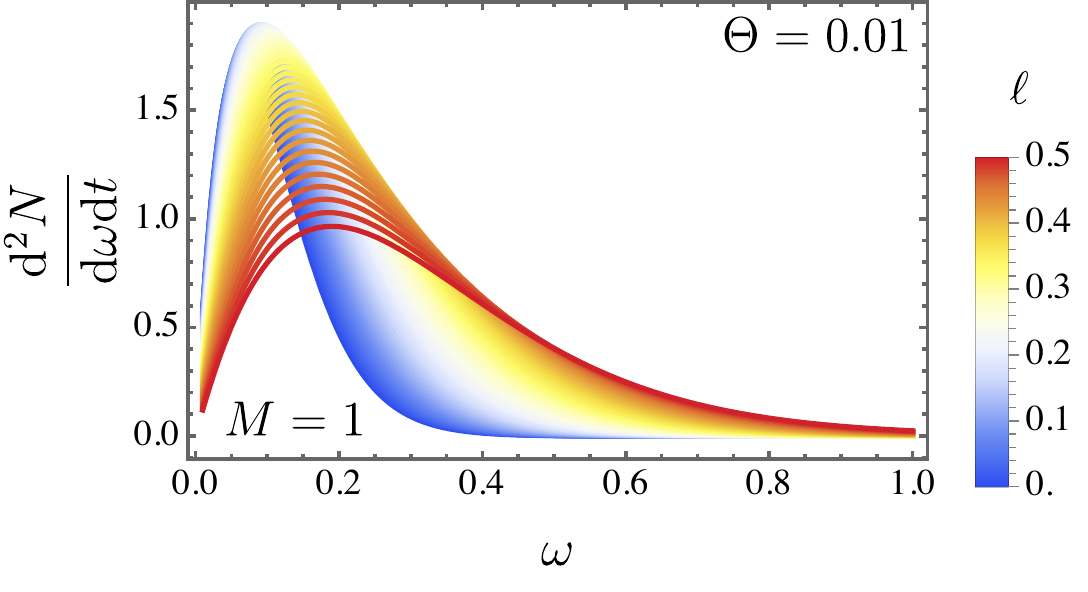}
    \caption{The particle emission rate is plotted as a function of $\omega$, with the top panel illustrating the effect of varying $\Theta$ for a fixed $\ell = 0.01$, and the bottom panel showing the influence of different $\ell$ values while keeping $\Theta = 0.01$ constant.}
    \label{ndensity}
\end{figure}

In Fig. \ref{ndensity}, the particle emission rate is plotted as a function of $\omega$. The top panel displays the results for different values of $\Theta$ with $\ell$ fixed at $0.01$, while the bottom panel shows the behavior for varying $\ell$ and a fixed $\Theta = 0.01$. In the first case, increasing $\Theta$ leads to a clear enhancement of the particle emission. In the second case, however, the effect of $\ell$ depends on the frequency range: in the low--frequency regime, larger values of $\ell$ reduce the emission rate $\frac{\mathrm{d}^{2}N}{\mathrm{d}\omega \mathrm{d}t}$, whereas in higher--frequency intervals, the emission rate increases with $\ell$.

\section{\label{Sec5}Massless Scalar Field in a Non-Commutative Background}

This section focuses on analyzing the behavior of a massless scalar field within a modified gravitational background. To facilitate this investigation, we first introduce corrections to the Kalb--Ramond black hole geometry by integrating effects derived from non--commutative geometry, as outlined in Eqs.~(\ref{gtt}--\ref{gphi}). These corrections yield a revised spacetime metric described by 
\[
g_{\mu\nu}^{\text{NC}} = g_{\mu\nu} + \Theta^2 h_{\mu\nu}^{\text{NC}},\]
where $g_{\mu\nu}$ denotes the original Kalb--Ramond black hole metric, and $h_{\mu\nu}^{\text{NC}}$ encapsulates the terms introduced by non--commutative effects, as explored in previous studies~\cite{chaichian2008corrections, chamseddine2001deforming, zet2003desitter}.

To proceed, we adopt the formalism developed in Ref.~\cite{chen2022eikonal}, which presents an innovative method for incorporating metric deformations. The deformation is characterized by a small, dimensionless expansion parameter $\epsilon$. This perturbative scheme allows for a modified representation of black hole spacetime, starting from the base metric given in Eqs.~(\ref{gtt}--\ref{gphi}) reformed in a $\cos\theta$ series as
	\begin{align}\label{g00}
		A^{(\Theta,\ell)} &= - f(r)(1 + \epsilon{A_j}{{\mathop{\rm cos}\nolimits} ^j}\theta ),\\
		B^{(\Theta,\ell)} &= { f(r)}^{-1}(1 + \epsilon{B_j}{{\mathop{\rm cos}\nolimits} ^j}\theta ),\\
		C^{(\Theta,\ell)}& = {r^2}(1 + \epsilon \, {C_j}{{\mathop{\rm cos}\nolimits} ^j}\theta ),\\ \label{g33}
		D^{(\Theta,\ell)}&= {r^2}{{\mathop{\rm sin}\nolimits} ^2}\theta (1 + \epsilon{D_j}{{\mathop{\rm cos}\nolimits} ^j}\theta ),
\end{align}
where the other non--diagonal components of the spacetime are zero before and after deformation and the function $f(r) = \frac{1}{1 - \ell} - \frac{2M}{r}$ represents the basic metric function associated with the Kalb--Ramond black hole spacetime. The deformation parameter $\epsilon$, introduced in the previous section, corresponds to the square of the non--commutative parameter, i.e., $\epsilon = \Theta^{2}$. Building on the methodology outlined in Refs.~\cite{chen2022eikonal, zhao2024quasinormal}, we proceed to derive the expressions for the metric correction terms resulting from the deformation
\begin{align}\label{A}
		&{A_0} =\frac{M (11 M (\ell -1)+4 r)}{2 r^3 (2 M (\ell -1)+r)},\\
        &{B_0} =-\frac{M (3 M (\ell -1)+2 r)}{2 r^3 (2 M (\ell -1)+r)},\\ 
		&{C_0} =-\frac{64 M^2 (\ell-1)^2+32 M r (\ell -1)+r^2}{16 r^3 (\ell-1) (2 M (\ell-1)+r)},\\
		&{D_0} =\frac{-2 M^2 (\ell -1)+4 M r (\ell -1)+r^2}{4 r^3 (2 M (\ell -1)+r)},\\
		&{A_j}={B_j}={C_j}=0, \quad \text{and} \quad
		{D_j}= {\frac{5({1 + {{\left( { - 1} \right)}^j}})}{{16{r^2}}}} \quad \text{for}\quad j>0.
\end{align}

We initiate our study by considering the Klein--Gordon equation for a massless scalar field propagating in a spacetime, which is given by
\begin{equation}
    \frac{1}{\sqrt{-g}} \, \partial_\mu \left( \sqrt{-g} \, g^{\mu \nu} \, \partial_\nu \psi \right) = 0,
\end{equation}
where $\psi$ denotes the scalar field. Exploiting the stationarity and axisymmetry of the background, we expand the field $\psi$ in terms of eigenmodes associated with the Killing vectors $\partial_{t}$ and $\partial_\varphi$. The scalar field can thus be decomposed as:
\begin{equation}
    \psi = \int_{-\infty}^{\infty} \! \mathrm{d} \omega \sum_{m = -\infty}^{\infty} e^{im\varphi} \, D_{m, \omega}^2 \, \psi_{m,\omega}(r,\theta) \, e^{-i\omega t},
\end{equation}
where the operator $D_{m,\omega}^2$ acts on the radial and polar coordinates. The parameters $m$ and $\omega$ correspond to the azimuthal quantum number and frequency, respectively, and each mode satisfies the relation $D_{m,\omega}^2 \psi_{m,\omega}(r,\theta) = 0$.

To include first--order corrections arising from non--commutative geometry, we express the operator $D_{m,\omega}^2$ as a perturbative expansion in  $\Theta$, such that~\cite{chen2022eikonal, zhao2024quasinormal}
\begin{equation}
    D_{m,\omega}^2 = D_{0m,\omega}^2 + \Theta^2 \, D_{1m,\omega}^2,
\end{equation}
where $D_{0m,\omega}^2$ represents the operator in the commutative (classical) case, and $D_{1m,\omega}^2$ takes into account the leading--order non--commutative corrections.

Substituting the metric components defined in Eqs.~(\ref{g00})--(\ref{g33}) into the Klein--Gordon equation, we can explicitly determine the structure of these operators in the deformed background
\begin{align}
    D_{0m,\omega}^2 &= - \left( \omega^2 - \frac{m^2 f(r)}{r^2 \sin^2 \theta} \right) - \frac{f(r)}{r^2} \partial_r \left( r^2 f(r) \partial_r \right) - \cos \theta \, \partial_r \left( r^2 f(r) \partial_r \right) \\
    &- \frac{f(r)}{r^2 \sin^2 \theta} \partial_\theta \left( \sin \theta \, \partial_\theta \right), \\
    D_{1m,\omega}^2 &= \frac{m^2 f(r)}{r^2 \sin^2 \theta} (A_j - D_j) \cos \theta - \frac{f(r)}{r^2} (A_j - B_j) \cos \theta \, \partial_r \left( r^2 f(r) \partial_r \right) \\
    &- \frac{f(r)^2}{r^2} (A'_j - B'_j + C'_j + D'_j) \cos \theta \, \partial_r - \frac{f(r)}{r^2} (A_j - C_j) \cos \theta \left( \cot \theta \, \partial_\theta + \partial_\theta^2 \right) \\
    &- \frac{f(r)}{2r^2} (A_j + B_j - C_j + D_j) \partial_\theta \cos \theta \, \partial_\theta - \frac{2i\omega f(r)}{r} a_j \cos \theta (r \partial_r + 1).
\end{align}

To proceed further, we introduce the tortoise coordinate $r^{*}$, which is defined through the differential relation
\begin{equation}
    r^* = \int\frac{{\mathrm{d}r}}{f(r) \left( 1 + \frac{1}{2} \Theta^2 b_{lm}^j (A_j - B_j) \right)},
\end{equation}
where the term involving $\Theta^2$ accounts for leading--order non--commutative corrections. The scalar field $\psi_{m,\omega}$ is then expanded in terms of associated Legendre polynomials  $P_{l'}^m(\cos\theta)$ and radial functions $R_{l',m}(r)$, which yields the mode decomposition
\begin{equation}\label{psilm}
    \psi_{m,\omega}(r,\theta) = \sum_{l' = |m|}^\infty P_{l'}^m (\cos\theta) \, R_{l',m}(r).
\end{equation}
A new radial wave function is introduced as $\Psi_{l,m}$, which satisfies the following expression
\begin{equation}\label{Rlm}
R_{l, m} = \frac{\Psi_{l, m} }{r} \left( 1 +  \frac{\epsilon}{4} \, b_j^{l m} (A_j - B_j) - \epsilon \int \mathrm{d} r \, \frac{\mathcal{Z}_{l m}(r)}{4r^2} \right),
\end{equation}
where
\begin{equation}\label{Zlm}
\mathcal{Z}_{l m}(r) \equiv b^j_{l m} r^2 \left( A_j' - B_j' + C_j' + D_j' \right) + 4i g_{l m}^j d_j + \frac{4i \omega}{f(r)} r^2 b^j_{l m} a_j.
\end{equation}

{Substituting the Eqs.} \eqref{Rlm}-\eqref{Zlm} into the Eq. \eqref{psilm} and considering the new redefined radial function  $\Psi_{lm}(r^*)$ in the Klein--Gordon equation, a one--dimensional Schr\"{o}dinger--like wave equation will be obtained as 
\begin{equation}
    \frac{\mathrm{d}^2 \Psi_{lm}}{\mathrm{d}r^{*2}} + \omega^2 \Psi_{lm} = V_{\rm eff}(r) \, \Psi_{lm},
\end{equation}
where  $V_{\rm eff}(r)$ is the effective potential experienced by the scalar field. This potential incorporates both the Kalb--Ramond features and non--commutative contributions and takes the form
\begin{equation}
    V_{\rm eff}(r) = V_{\rm KR}(r) + \Theta^2 V_{\rm NC}(r),
\end{equation}
with $V_{\rm KR}(r)$ representing the effective potential in the standard Kalb--Ramond black hole background, given by
\begin{equation}
    V_{\rm KR}(r) = f(r) \left( \frac{l(l + 1)}{r^2} + \frac{1}{r} \frac{\mathrm{d}f(r)}{\mathrm{d}r} \right).
\end{equation}

The correction term $V_{\rm NC}(r)$ encodes the modifications arising from non--commutative geometry and can be explicitly determined through direct algebraic computation based on the deformed metric structure
\ie
    \begin{split}
		& V_{\rm {NC}}(r)= \frac{f(r)}{r}\frac{{\mathrm{d}f(r)}}{{\mathrm{d}r}}b_{lm}^0\left( {{A_0} - {B_0}} \right) +  \left[\frac{f(r)}{{{r^2}}}(a_{lm}^0\left( {{A_0} - {D_0}} \right) - c_{lm}^0\left( {{A_0} - {C_0}}\right) \right. \\ 
		& \left. -\frac{{d_{lm}^0}}{2}({A_0} + {B_0} - {C_0} + {D_0})
		+ \frac{1}{{4{r^2}}}\frac{\mathrm{d}}{{\mathrm{d}{r^*}}}\left(b_{lm}^0{r^2}\frac{\mathrm{d}}{{\mathrm{d}{r^*}}}({A_0} - {B_0} + {C_0} + {D_0})\right) - \frac{{b_{lm}^0}}{4}\frac{{{\mathrm{d}^2}}}{{\mathrm{d}{r^*}^2}}({A_0} - {B_0})) \right]\\ 
		& - \frac{f(r)}{{{r^2}}}\sum\limits_{j = 1}^\infty  {\left(a_{lm}^j + \frac{1}{2}d_{lm}^j \right){D_j}}  + \sum\limits_{j = 1}^\infty  {\frac{1}{{4{r^2}}}\frac{\mathrm{d}}{{\mathrm{d}{r^*}}}\left(b_{lm}^j{r^2}\frac{\mathrm{d}}{{\mathrm{d}{r^*}}}\right){D_j}}. \nonumber
	\end{split}
    \fe

The coefficients $a_{lm}^j, b_{lm}^j, c_{lm}^j, d_{lm}^j$ are determined by the specific value of $\Theta$ and the associated multipole number $l$, as outlined in Ref.~\cite{zhao2024quasinormal}. Importantly, the introduction of non--commutative effects leads to an effective potential that exhibits explicit dependence on both  $l$ and the azimuthal index $m$, marking a departure from the commutative scenario where only $l$ typically plays a role.

Moreover, it is observed that the coefficients remain invariant under the transformation $m \rightarrow -m$. As an illustrative case, the explicit expression for the effective potential $V_{\rm eff}(r)$ when  $l = 1$ and $m = \pm 1$ is given by
\begin{align}\nonumber
    &V_\text{eff}(r)=f(r) \left(\frac{2}{r^2}+\frac{f'(r)}{r}\right)+\frac{\Theta^2}{1024 r^4 \sqrt{f(r)}}\Big[+2 r \sqrt{f(r)} \left(\frac{128 r^2 f'(r)^2}{f(r)}+192\right) f'(r)\\ \nonumber
    &+32 f(r)^{7/2}+2 f(r)^{3/2} \left(\frac{r^2 f'(r) \left(\frac{96 r^2 f'(r)^3}{f(r)^{3/2}}-\frac{32 f'(r)}{\sqrt{f(r)}}+2624 r \Pi\right)}{2 \sqrt{f(r)}}-312\right)\\ \nonumber
    &+32rf(r)^2\left[-\frac{81 f'(r)}{2 \sqrt{f(r)}}+\frac{29 r^2 f'(r)^3}{2 f(r)^{3/2}}-16 r \Pi\frac{82 r^3 f'(r)^2 \Pi}{f(r)}+\right.\\ \nonumber
    &+\left.28 r^2 \left(\frac{f^{3}(r)}{2 \sqrt{f(r)}}+\frac{3 f'(r)^3}{8 f(r)^{5/2}}-\frac{3 f'(r) f''(r)}{4 f(r)^{3/2}}\right)\right]+f(r)^{5/2}\Big[176-\frac{264 r^2 f'(r)^2}{f(r)}\\ \nonumber
    &+3200 r^4 \Pi^2+\frac{r f'(r) \left(9024 r^2 \Pi+6016 r^3 \left(\frac{f^{3}(r)}{2 \sqrt{f(r)}}+\frac{3 f'(r)^3}{8 f(r)^{5/2}}-\frac{3 f'(r) f''(r)}{4 f(r)^{3/2}}\right)\right)}{2 \sqrt{f(r)}}\Big]\\ \nonumber
    &+rf(r)^3\left(-\frac{64 f'(r)}{\sqrt{f(r)}}+32 r \Pi+1856 r^2 \left(\frac{f^{3}(r)}{2 \sqrt{f(r)}}+\frac{3 f'(r)^3}{8 f(r)^{5/2}}-\frac{3 f'(r) f''(r)}{4 f(r)^{3/2}}\right)\right.\\ \label{VeffL1M1}
    &\left.\left.+768 r^3 \left(\frac{f^{4}(r)}{2 \sqrt{f(r)}}-\frac{3 f''(r)^2}{4 f(r)^{3/2}}-\frac{15 f'(r)^4}{16 f(r)^{7/2}}-\frac{f^{3}(r) f'(r)}{f(r)^{3/2}}+\frac{9 f'(r)^2 f''(r)}{4 f(r)^{5/2}}\right)\right)\right)\Bigg].
\end{align}

Here, we define the quantity $\Pi$ as
\begin{equation}
    \Pi = \frac{f''(r)}{2 \sqrt{f(r)}} - \frac{f'(r)^2}{4 f(r)^{3/2}},
\end{equation}
where the primes denote derivatives with respect to the radial coordinate $r$. {It is worth noting that the effective potential $V_{\text{eff}}(r)$ is well defined only for $l > 0$. Accordingly, since Eq. (\ref{VeffL1M1}) was obtained for $l = 1$, taking the limit $\Theta \to 0$ clearly recovers the standard effective potential for a scalar field in the Kalb--Ramond black hole background with $l = 1$ \cite{araujo2024exploring}, as expected.  }

\begin{figure}[ht]
		\centering
		\includegraphics[width=90mm]{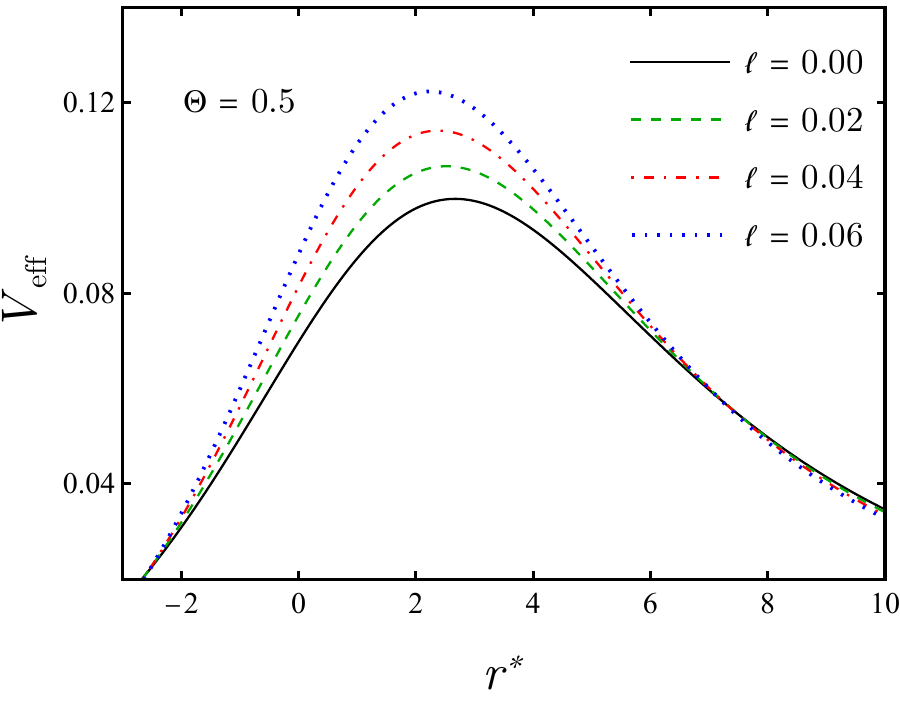} 
		\hfill
	   \caption{The scalar field effective potential is analyzed for $M = 1$ and $l = 1$ ($m = \pm 1 $), considering $\ell = 0$ corresponds to Schwarzschild geometry and $ \ell = 0.02$, $0.04$, and $0.06$ to the Kalb--Ramond spacetime, within the non--commutative geometry framework defined by the parameter $\Theta = 0.5$.
}
		\label{fig:VeffSH}
	\end{figure}

Corroborating our previous results, Fig.~\ref{fig:VeffSH} presents a comparative visualization of the effective potentials for Schwarzschild and Kalb--Ramond black holes. Notably, in the context of non--commutative geometry, the Kalb--Ramond metric yields a higher potential barrier than its Schwarzschild counterpart.

\begin{figure}[ht]
		\centering
		\includegraphics[width=80mm]{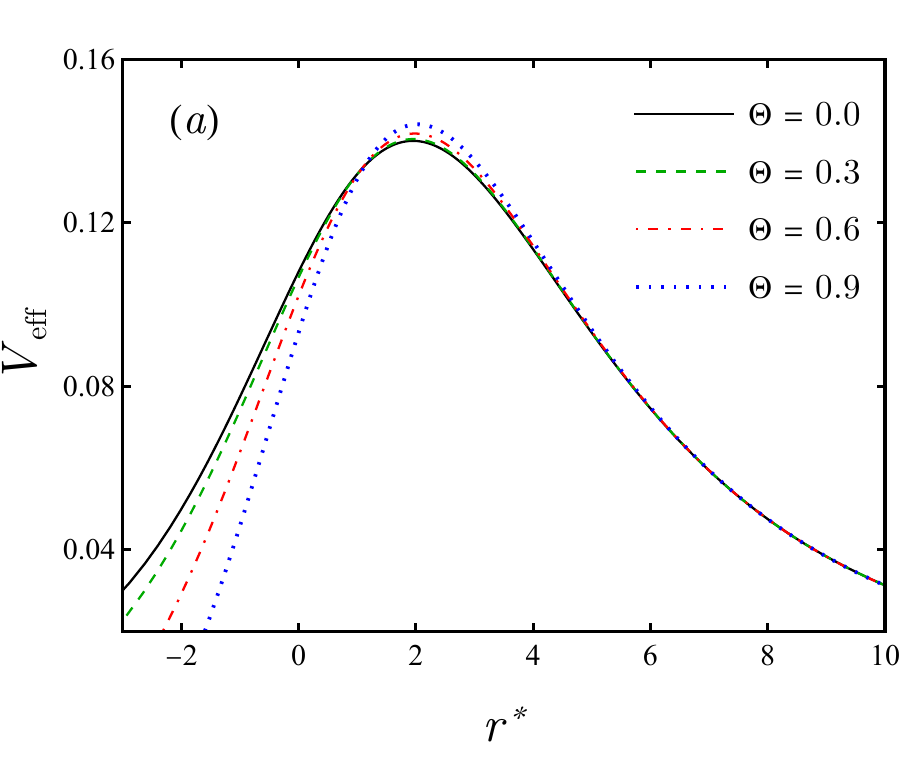} 
		\includegraphics[width=80mm]{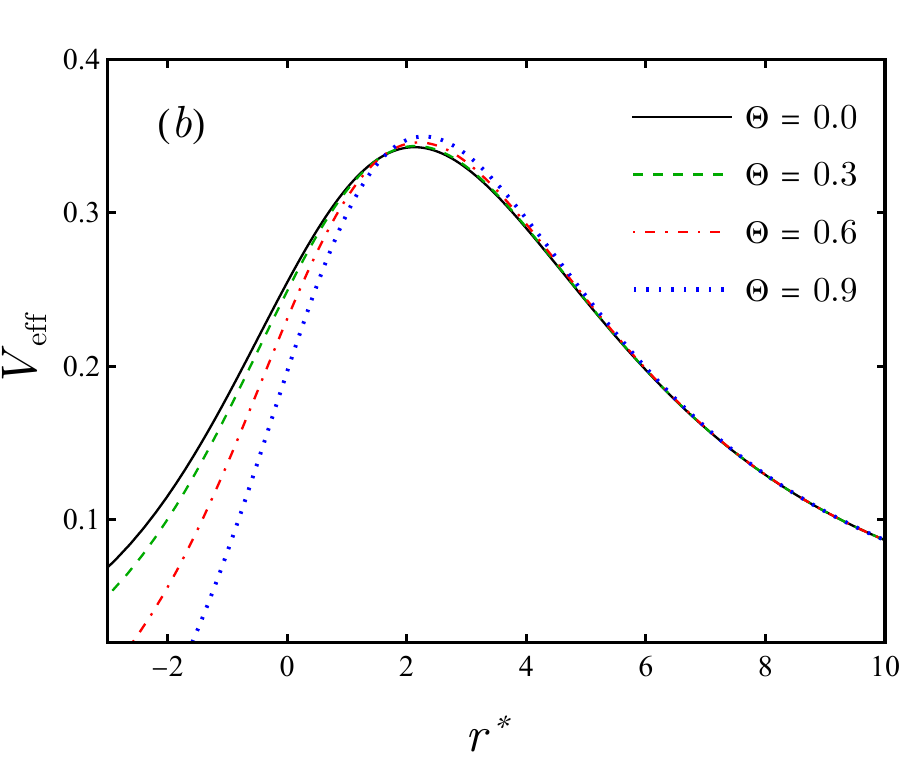}
		\includegraphics[width=80mm]{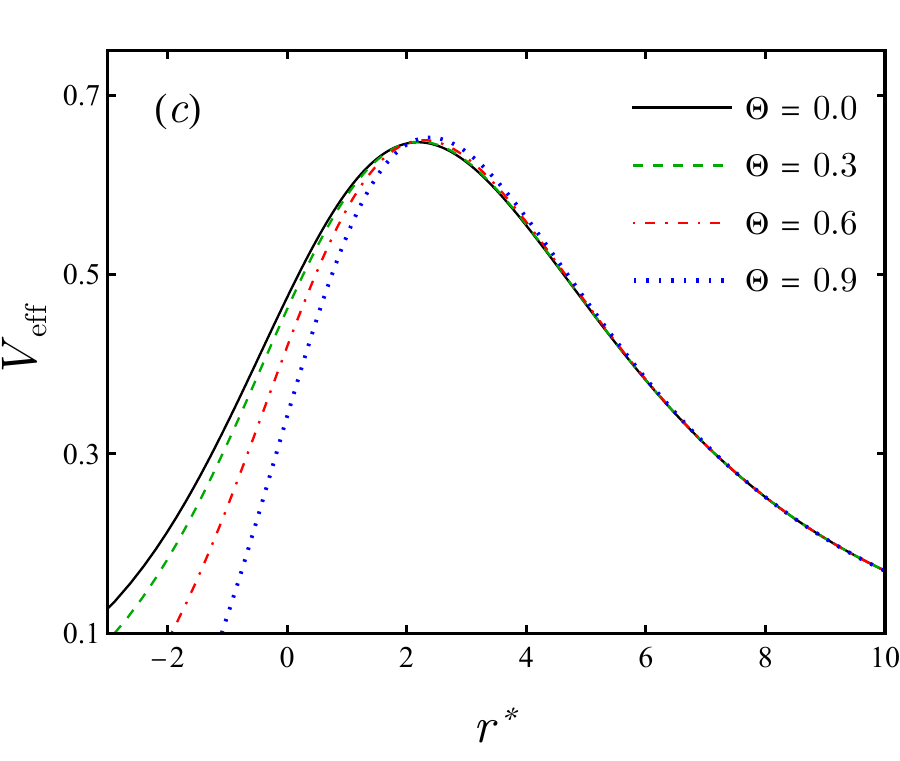}\\
		\caption{The effective potentials for a scalar field with $M = 1$ and $\ell = 0.1$ are shown as follows: panel (a) corresponds to $l = 1$ ($m = \pm 1$); panel (b) represents the case $l = 2$ ($m = \pm 1$); and panel~(c) illustrates $ l = 3$ ($m = \pm 1$), each plotted for different values of $\Theta$.}
		\label{fig:Veffrstar}
	\end{figure}

To further investigate how $\Theta$ influences the effective potential (for a fixed value of $\ell =0.1$). In this manner, we display the corresponding behavior for fixed values of mass  $M$, $l$, and $m$, as illustrated in Fig.~\ref{fig:Veffrstar}. The left panel $(a)$ shows the variation of the potential profile for different values of $\Theta$ when $l = 1$ and $m = \pm 1$. The center panel $(b)$ and right panel $(c)$ highlights similar behavior for $l = 2 ~( m = \pm 1)$ and $l= 3~( m = \pm 1 )$, respectively. In other words, while the impact of the non--commutative parameter $\Theta$ may appear modest, a closer inspection reveals that by increasing it, we have a noticeable enhancement in the height of the potential barrier.

	
\section{\label{Sec6}Quasinormal modes}

Quasinormal modes (QNMs) are the response of a black hole to external perturbations. These complex frequencies are described by the real part, which indicates the oscillation frequency, and the imaginary component, which determines the rate at which the signal decays. They encapsulate how perturbations evolve near the event horizon and encode information about the geometry of the surrounding spacetime. When considered within the framework of non--commutative geometry, QNMs turn out to be an important aspect for highlighting the influence of $\Theta$ on scalar fields and their interaction with the associated effective potential.

Their spectrum is obtained by analyzing the wave dynamics governed by the perturbation equation, subject to boundary conditions that enforce wave absorption at the black hole horizon and emission to spatial infinity. Given the inherent complexity of the governing equations, exact analytical expressions are rarely obtainable. As a result, various approximation techniques—both analytical and numerical—have been formulated to compute QNM frequencies, being essential in characterizing the response of the black hole to external perturbations \cite{leaver1986solutions, ferrari1984new, blome1984quasi, cardoso2001quasinormal, heidari2023investigation}. In the present analysis, we adopt the WKB approximation \cite{iyer1987black, konoplya2011quasinormal} to examine how the introduction of non--commutativity affects the structure of the QNM spectrum.

The WKB approximation provides an efficient means of estimating QNM frequencies through the following expression
\begin{equation}\label{omegawkb}
\frac{i(\omega^2 - V_0)}{\sqrt{-2 V_0''}} + \sum_{j=2}^{3} \Omega_j = n + \frac{1}{2},
\end{equation}
where $V_0$ denotes the peak of the effective potential, $V_0''$ is the second derivative of the potential with respect to the tortoise coordinate $r^{*}$, and $\Omega_j$ are higher--order correction terms.

\begin{table}[]
\caption{Quasinormal modes for the Kalb--Ramond black hole computed under non--commutative corrections via the WKB approach, assuming $M = 1$.}
\label{tab:allQNM}
\begin{tabular}{|cc|c|c|c|c|c|}
\hline
\multicolumn{2}{|c|}{$\ell = 0.1$}                            & $\Theta=0.000$            & $\Theta=0.025$      & $\Theta=0.050$          & $\Theta=0.075$          & $\Theta=0.100$          \\ \hline
\multicolumn{1}{|c|}{\multirow{2}{*}{\begin{tabular}[c]{@{}c@{}}$l=1$,\\$m=\pm 1$\end{tabular}}} &
  $n = 0$ & {0.34317-0.12124i} & {0.34318-0.12125i} & {0.34320-0.12129i} & {0.34325-0.12136i} & {0.34332-0.12146i} \\ \cline{2-7} 
\multicolumn{1}{|c|}{}                      & $n = 1$ & {0.30668-0.38153i} & {0.30671-0.38156i} & {0.30679-0.38166i} & {0.30692-0.38182i} & {0.30711-0.38204i} \\ \hline
\multicolumn{1}{|c|}{\multirow{2}{*}{\begin{tabular}[c]{@{}c@{}}$l=2$,\\$m=\pm 1$\end{tabular}}} &
  $n = 0$ & {0.56741-0.11959i} & {0.56741-0.11961i} & {0.56743-0.11965i} & {0.56747-0.11973i} & {0.56752-0.11983i} \\ \cline{2-7} 
\multicolumn{1}{|c|}{}                      & $n = 1$ & {0.54161-0.36611i} & {0.54163-0.36615i} & {0.54168-0.36628i} & {0.54177-0.36649i} & {0.54190-0.36678i} \\ \cline{2-7} 
\multicolumn{1}{|c|}{}                      & $n = 2$ & {0.50163-0.62371i} & {0.50167-0.62377i} & {0.50177-0.62396i} & {0.50195-0.62428i} & {0.50220-0.62473i} \\ \hline
\multicolumn{1}{|c|}{\multirow{2}{*}{\begin{tabular}[c]{@{}c@{}}$l=2$,\\$m=\pm 2$\end{tabular}}} &
  $n = 0$ & {0.56741-0.11959i} & {0.56741-0.11961i} & {0.56744-0.11964i} & {0.56748-0.11971i} & {0.56754-0.11980i} \\ \cline{2-7} 
\multicolumn{1}{|c|}{}                      & $n = 1$ & {0.54161-0.36611i} & {0.54163-0.36615i} & {0.54168-0.36626i} & {0.54176-0.36645i} & {0.54188-0.36671i} \\ \cline{2-7} 
\multicolumn{1}{|c|}{}                      & $n = 2$ & {0.50163-0.62371i} & {0.50166-0.62377i} & {0.50176-0.62394i} & {0.50192-0.62422i} & {0.50213-0.62462i} \\ \hline
\multicolumn{1}{|c|}{\multirow{2}{*}{\begin{tabular}[c]{@{}c@{}}\\$l=3$,\\$m=\pm 1$\end{tabular}}} &
  $n = 0$ & {0.79189-0.11919i} & {0.79190-0.11921i} & {0.79191-0.11925i} & {0.79193-0.11932i} & {0.79196-0.11943i} \\ \cline{2-7} 
\multicolumn{1}{|c|}{}                      & $n = 1$ & {0.77274-0.36141i} & {0.77275-0.36145i} & {0.77278-0.36158i} & {0.77284-0.36180i} & {0.77292-0.36211i} \\ \cline{2-7} 
\multicolumn{1}{|c|}{}                      & $n = 2$ & {0.74002-0.61147i} & {0.74005-0.61154i} & {0.74012-0.61175i} & {0.74025-0.61210i} & {0.74042-0.61259i} \\ \cline{2-7} 
\multicolumn{1}{|c|}{}                      & $n = 3$ & {0.69853-0.86796i} & {0.69858-0.86805i} & {0.69871-0.86833i} & {0.69892-0.86879i} & {0.69921-0.86944i} \\ \hline
\multicolumn{1}{|c|}{\multirow{2}{*}{\begin{tabular}[c]{@{}c@{}}\\$l=3$,\\$m=\pm 2$\end{tabular}}} &
  $n = 0$ & {0.79189-0.11919i} & {0.79190-0.11920i} & {0.79193-0.11925i} & {0.79199-0.11932i} & {0.79206-0.11941i} \\ \cline{2-7} 
\multicolumn{1}{|c|}{}                      & $n = 1$ & {0.77274-0.36141i} & {0.77275-0.36145i} & {0.77280-0.36157i} & {0.77288-0.36178i} & {0.77300-0.36206i} \\ \cline{2-7} 
\multicolumn{1}{|c|}{}                      & $n = 2$ & {0.74002-0.61147i} & {0.74005-0.61153i} & {0.74014-0.61173i} & {0.74028-0.61206i} & {0.74047-0.61252i} \\ \cline{2-7} 
\multicolumn{1}{|c|}{}                      & $n = 3$ & {0.69853-0.86796i} & {0.69858-0.86805i} & {0.69871-0.86831i} & {0.69893-0.86875i} & {0.69923-0.86936i} \\ \hline
\multicolumn{1}{|c|}{\multirow{2}{*}{\begin{tabular}[c]{@{}c@{}}\\$l=3$,\\$m=\pm 3$\end{tabular}}} &
  $n = 0$ & {0.79189-0.11919i} & {0.79190-0.11920i} & {0.79193-0.11924i} & {0.79198-0.11930i} & {0.79205-0.11939i} \\ \cline{2-7} 
\multicolumn{1}{|c|}{}                      & $n = 1$ & {0.77274-0.36141i} & {0.77275-0.36144i} & {0.77280-0.36156i} & {0.77287-0.36174i} & {0.77297-0.36200i} \\ \cline{2-7} 
\multicolumn{1}{|c|}{}                      & $n = 2$ & {0.74002-0.61147i} & {0.74005-0.61153i} & {0.74012-0.61171i} & {0.74024-0.61201i} & {0.74041-0.61243i} \\ \cline{2-7} 
\multicolumn{1}{|c|}{}                      & $n = 3$ & {0.69853-0.86796i} & {0.69857-0.86804i} & {0.69868-0.86828i} & {0.69887-0.86869i} & {0.69912-0.86925i} \\ \hline
\end{tabular}
\end{table}

\begin{figure}
	\centering
	\includegraphics[width=81mm]{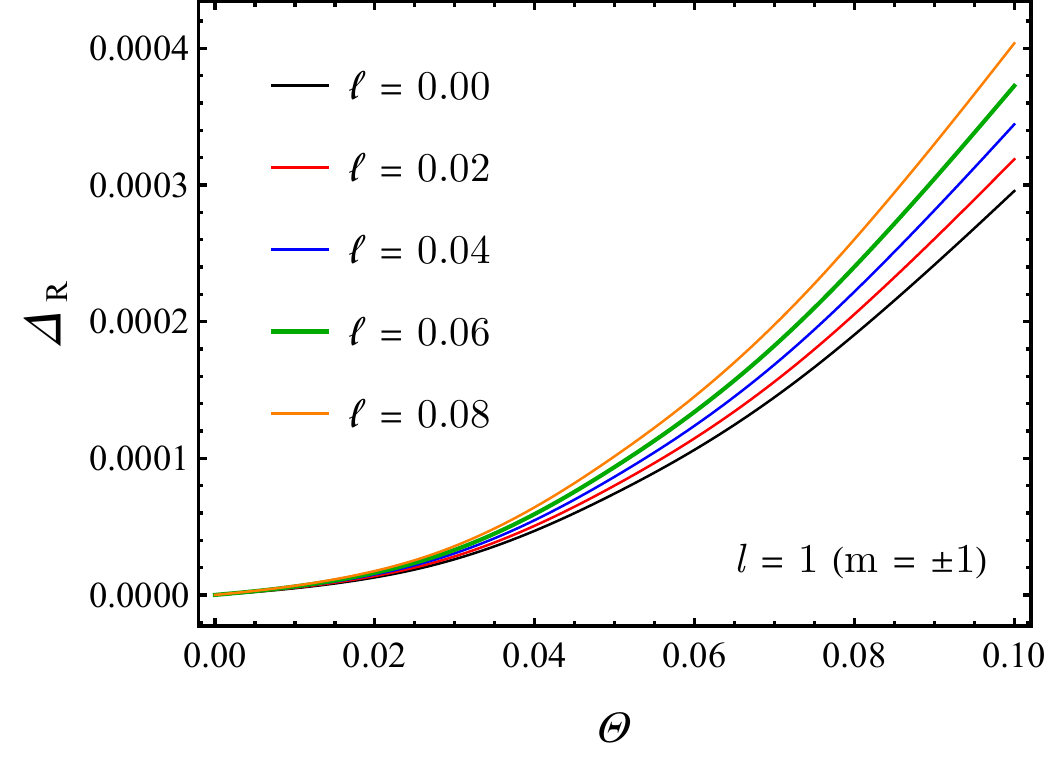}
	\hfil
    \includegraphics[width=81mm]{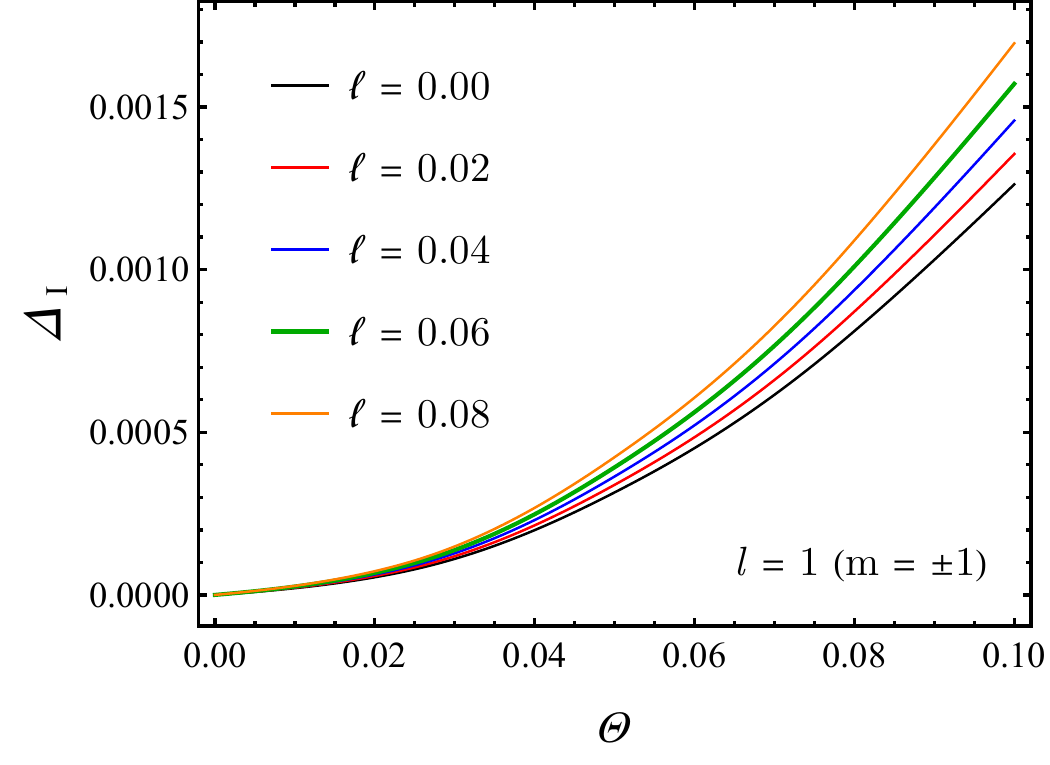}\\
	
	\caption{The deviations in the real and imaginary parts of the quasinormal modes are plotted as functions of the non--commutative parameter $\Theta$ for $M = 1$, $ l = 1$ $( m = \pm 1 $), overtone number $n = 0$ and variation of Kalb--Ramond parameter $\ell = 0$ to $0.08$.
}
	\label{fig:L1M1}
    \end{figure}
\begin{figure}
	\centering
	\includegraphics[width=81mm]{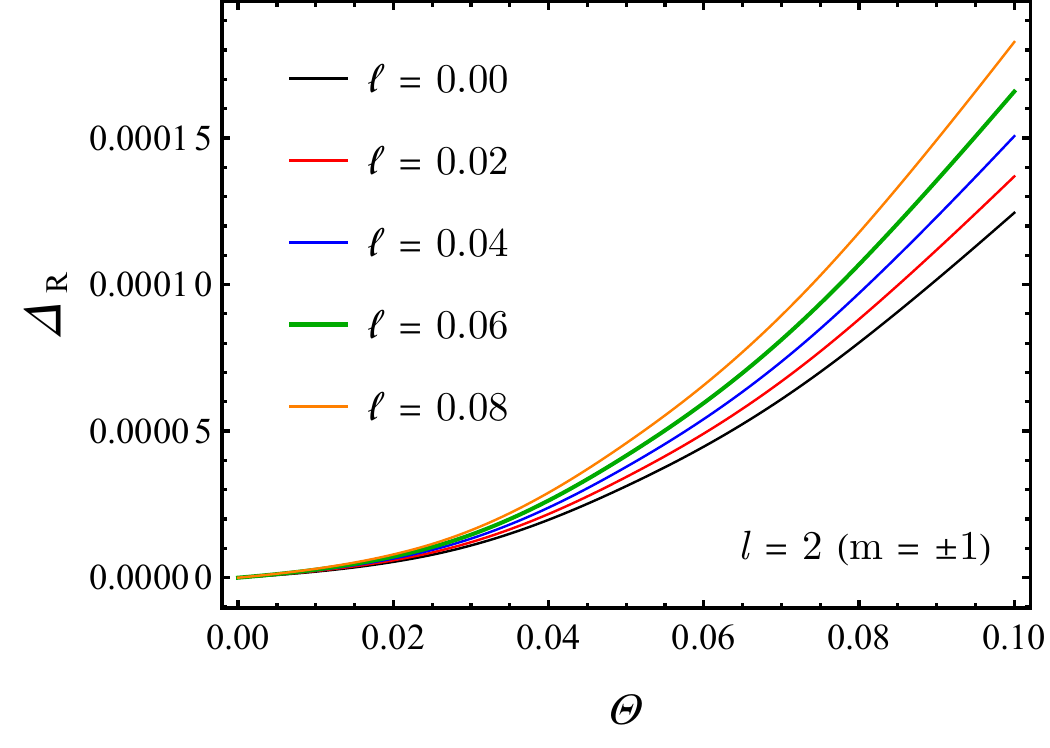}
	\hfil
    \includegraphics[width=81mm]{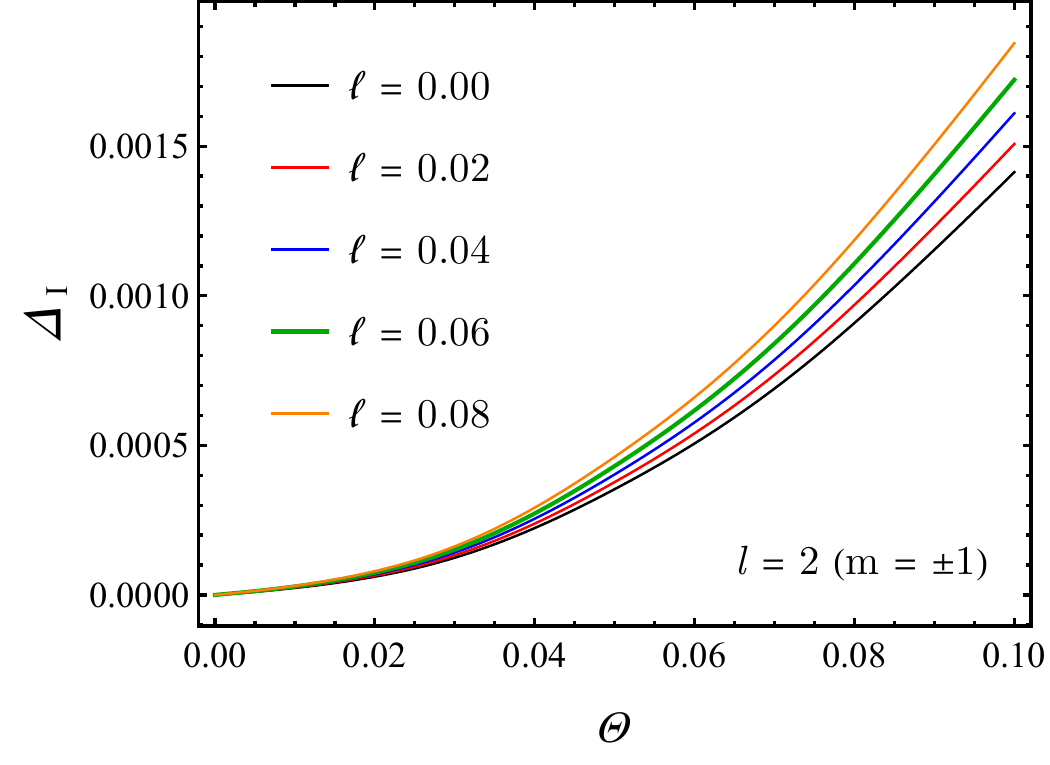}
	
	\caption{The variation in both real and imaginary components of the quasinormal mode frequencies, derived using the WKB approximation, is investigated as a function of the non--commutative parameter $\Theta$, for $M = 1$, $l = 2$ $( m = \pm 1 $), and the fundamental overtone $n = 0$.
}
	\label{fig:L2M1}
    \end{figure}
\begin{figure}
	\centering
	\includegraphics[width=80mm]{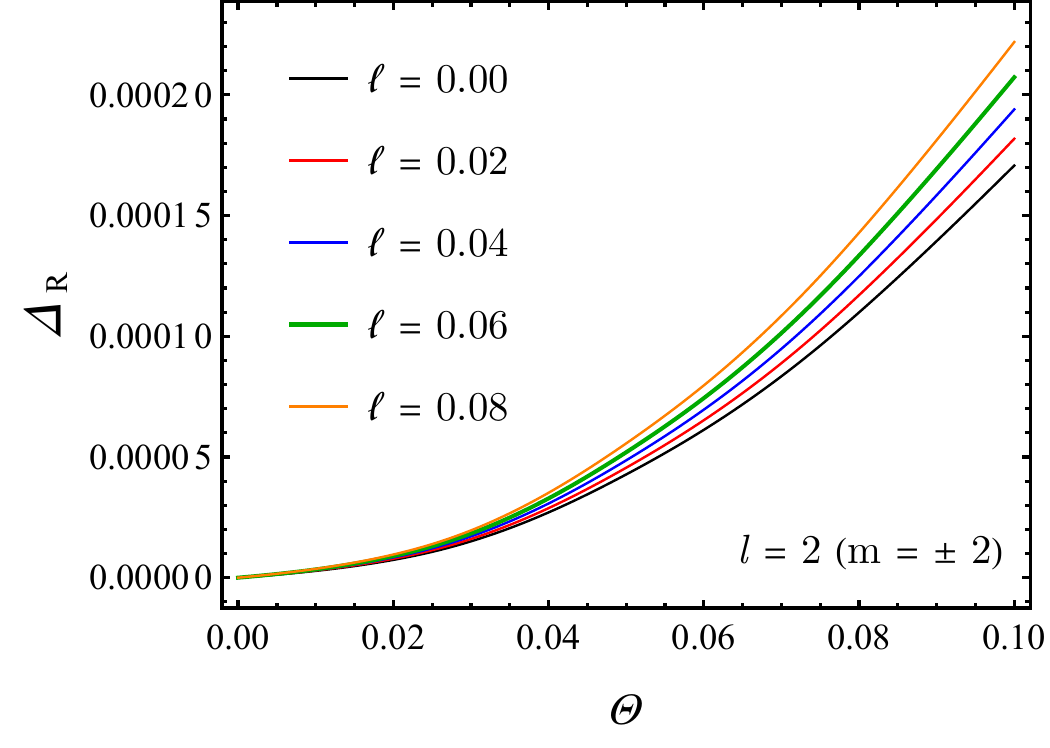}
	\hfil
    \includegraphics[width=80mm]{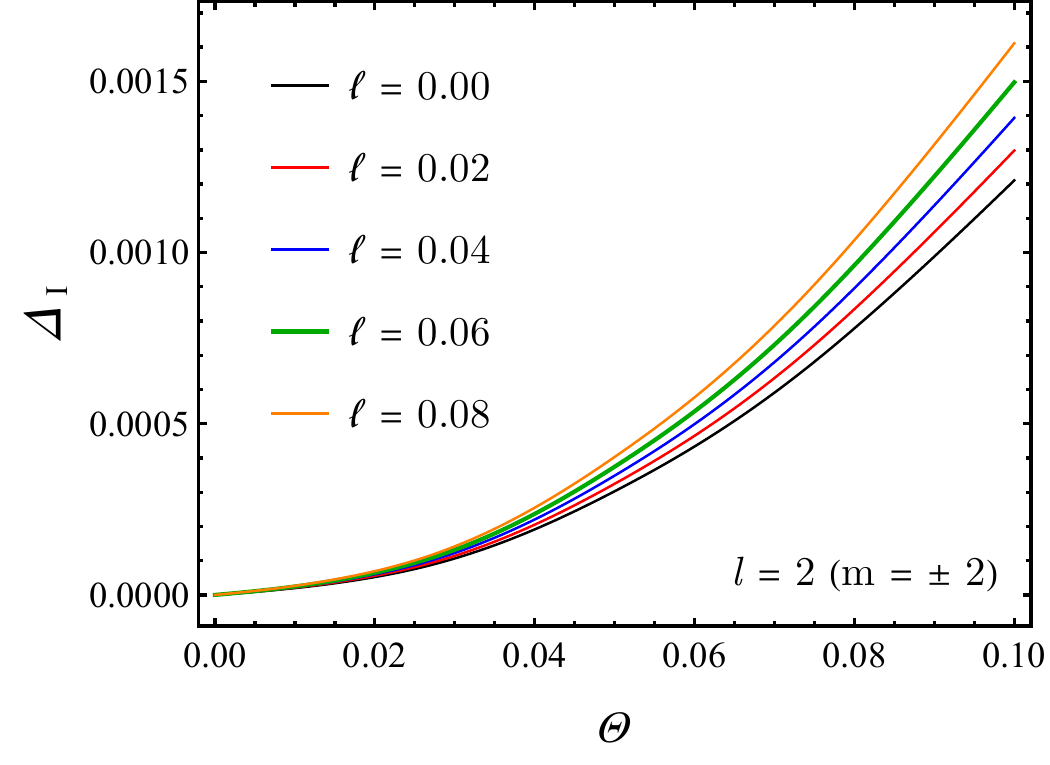}\\
	
	\caption{The real and imaginary deviations of quasinormal modes, calculated using the WKB method, are examined with respect to variations in the non--commutative parameter $\Theta$ for the case $M = 1$, $l = 2$ ($m = \pm 2$), and overtone number $n = 0$.}
	\label{fig:L2M2}
    \end{figure}
\begin{figure}
	\centering
	\includegraphics[width=80mm]{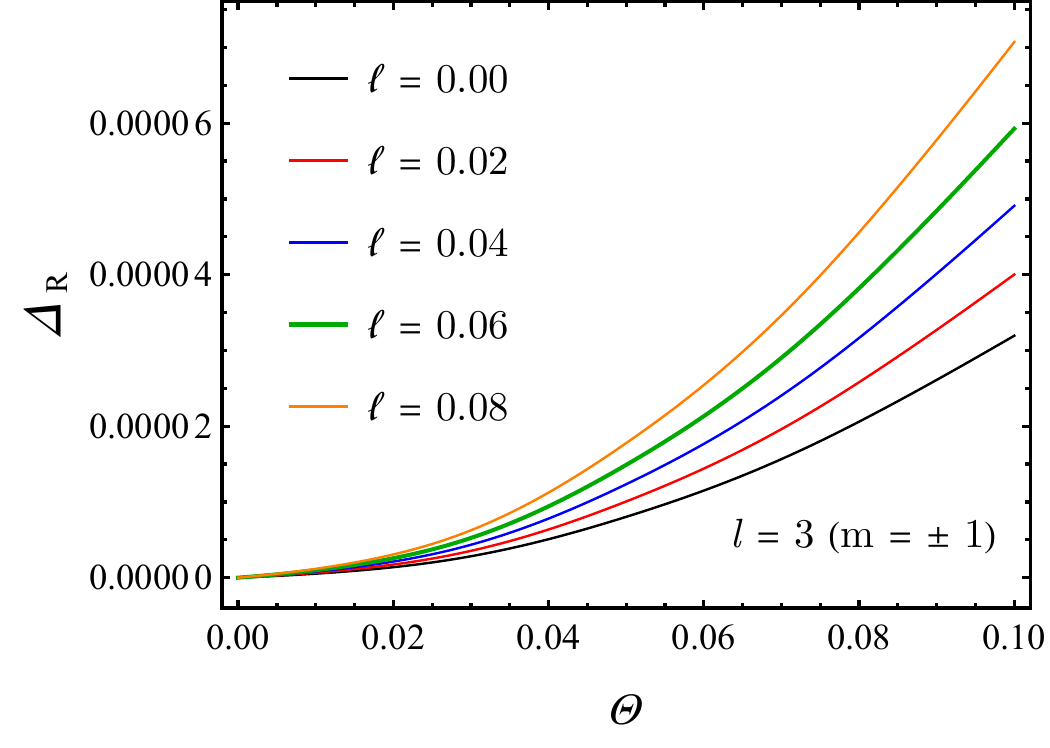}
	\hfil
    \includegraphics[width=80mm]{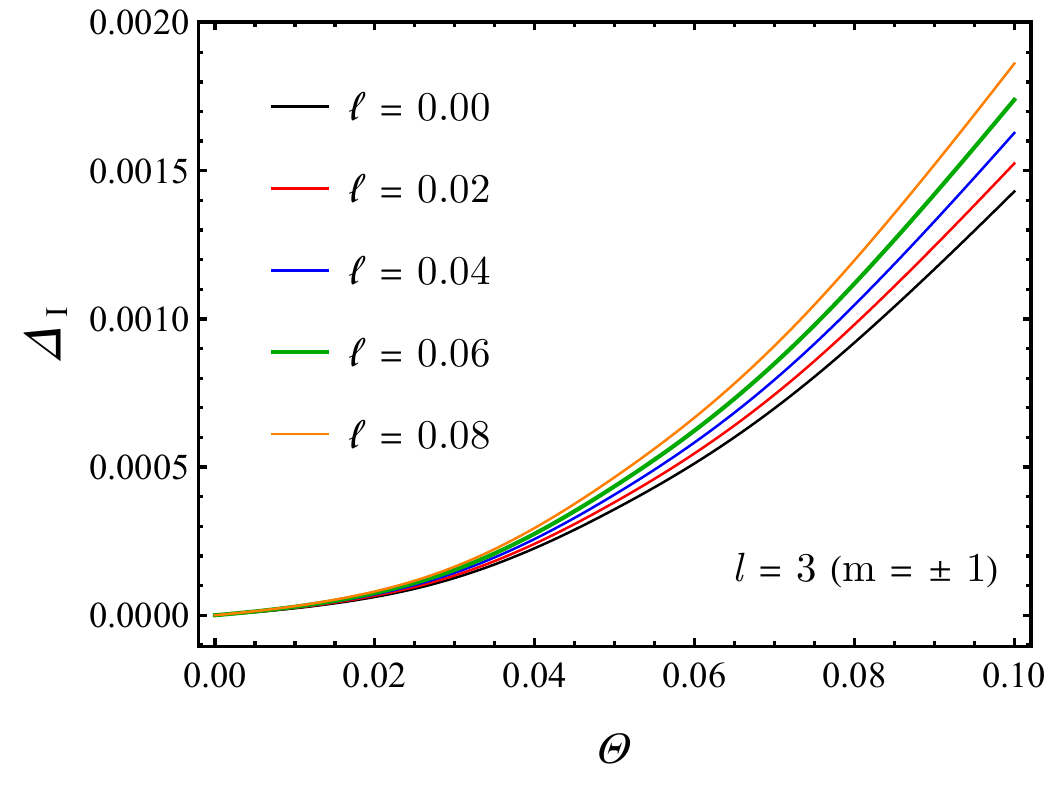}
	
	\caption{Effect of the non--commutative parameter  $\Theta$ on the QNM frequencies for $M = 1$, $l = 3$, $m = \pm 1$, and $n = 0$, for variation of Kalb--Ramond parameter.}
	\label{fig:L3M1}
    \end{figure}
\begin{figure}
	\centering
	\includegraphics[width=80mm]{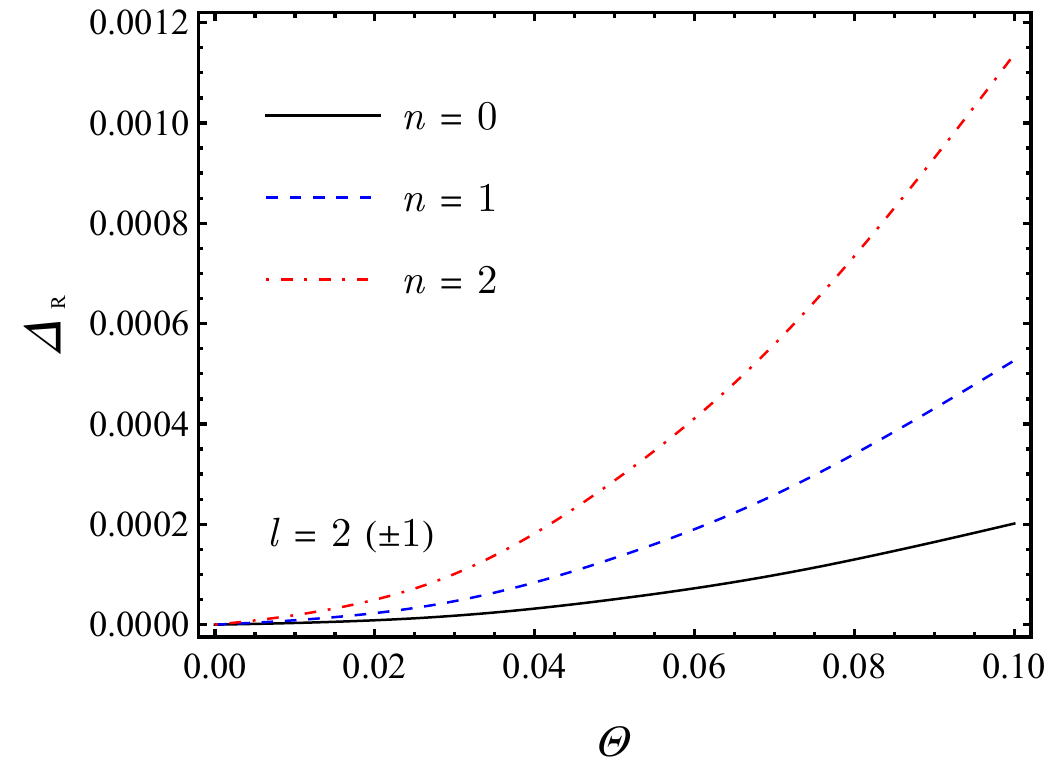}
	\hfil
    \includegraphics[width=80mm]{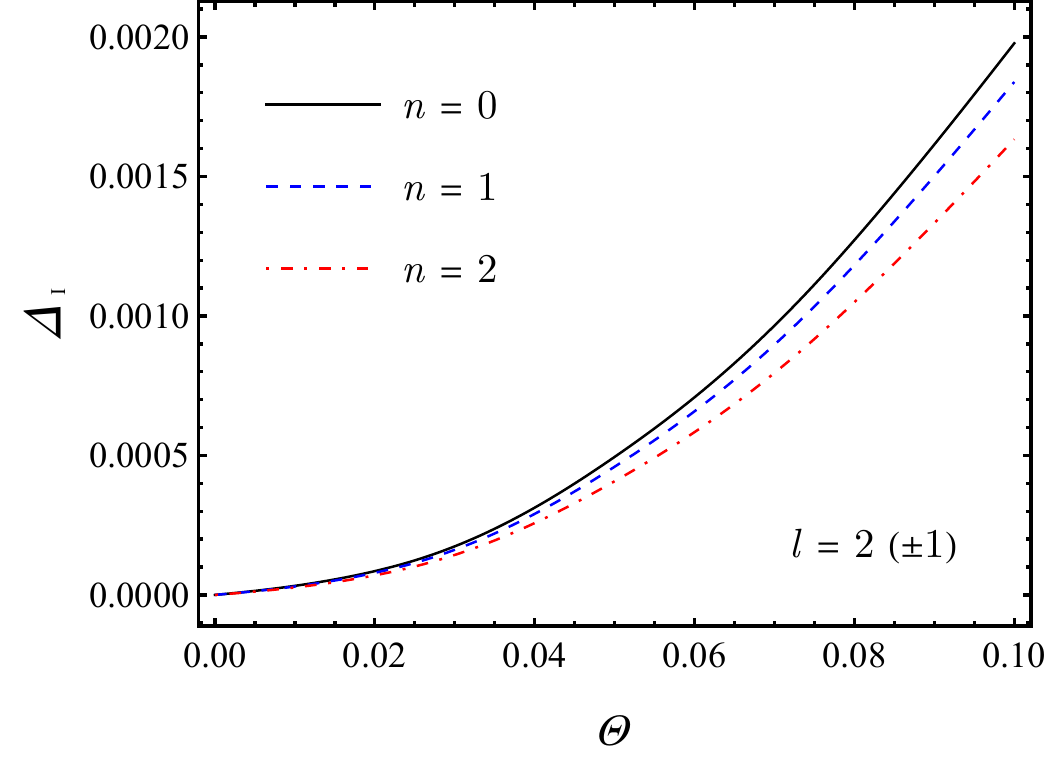}
	
	\caption{A comparison of the deviations in QNMs for $M = 1$ and $\ell = 0.1$, considering  $l = 2$ $( m = \pm 1)$, is presented for different overtone numbers $n = 0, 1, 2$ as the non--commutative parameter $\Theta$ varies.}
	\label{fig:L2M1N}
\end{figure}

Table.~\ref{tab:allQNM} presents the QNM for the Kalb--Ramond black hole under the influence of non--commutative geometry, computed using the WKB approximation with $M = 1$ and $l =1$. Both the real and imaginary parts of the QNMs exhibit a slight increase with $\Theta$. This trend holds across various values $l$, $m$, and overtone numbers $n$. Higher overtones and multipole modes demonstrate more pronounced damping (i.e., larger negative imaginary parts), reflecting stronger decay rates. The azimuthal number, as one should expect, also has a modification in the corresponding QNMs. For a fixed $l$, increasing $m$ leads to a very slight decrease in both the real and imaginary parts of the QNM frequencies. Overall, the data suggest that non--commutative corrections have a subtle but systematic effect on the oscillation frequencies and damping behavior of perturbations in this modified black hole background. Since these changes in the QNM frequencies are quite subtle—especially for smaller values of $\Theta$, where variations are almost indistinguishable—to facilitate a clearer interpretation of these dependencies, we introduce a normalized deviation parameter defined as
\begin{equation}
    \Delta = \frac{\omega^{\rm{NC}} - \omega^{\rm{KR}}}{\omega^{\rm{KR}}},
\end{equation}
which quantifies the relative difference between the QNM frequencies of the non--commutative ($\omega^{\text{NC}}$) black hole and those of the classical Kalb--Ramond black hole ($\omega^{\text{KR}}$). Both the real and imaginary components of this deviation are examined, with the results summarized in Figs.~\ref{fig:L1M1}--\ref{fig:L2M1N}.

The study considers three primary cases with multipole indices  $l = 1$, $l = 2$, and $l = 3$, along with their respective overtones satisfying $ n \leq l$, under the fixed Kalb--Ramond parameter $\ell = 0.1$ and mass condition $M = 1$. Fig.~\ref{fig:L1M1} illustrates that the real part of the QNM frequencies for $l = 1$, $m = \pm 1$, and $n = 0$ increases steadily as $\Theta$ grows, indicating a higher oscillation frequency of the scalar perturbation. Similarly, the absolute value of the imaginary component also increases with  $\Theta$, signifying faster decay and stronger damping. These trends persist across the other configurations shown in Figs.~\ref{fig:L2M1}--\ref{fig:L3M1}, demonstrating a consistent behavior for the different $(l, m)$ modes.

\begin{figure}
	\centering
	\includegraphics[width=80mm]{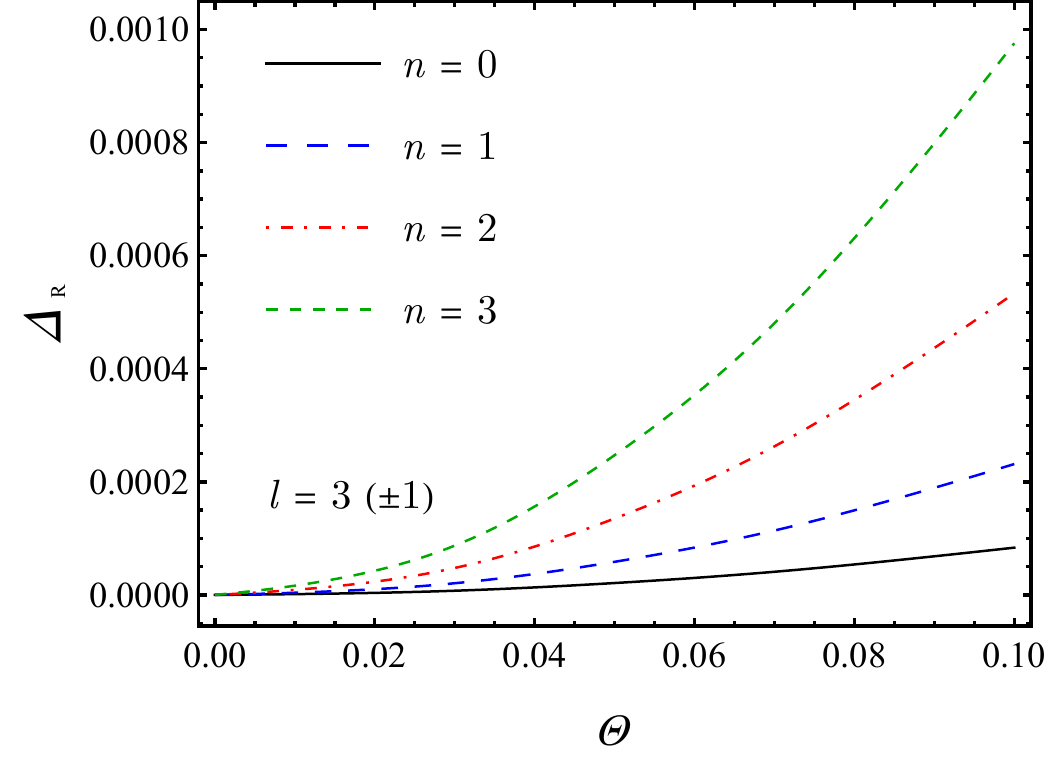}
	\hfil
    \includegraphics[width=80mm]{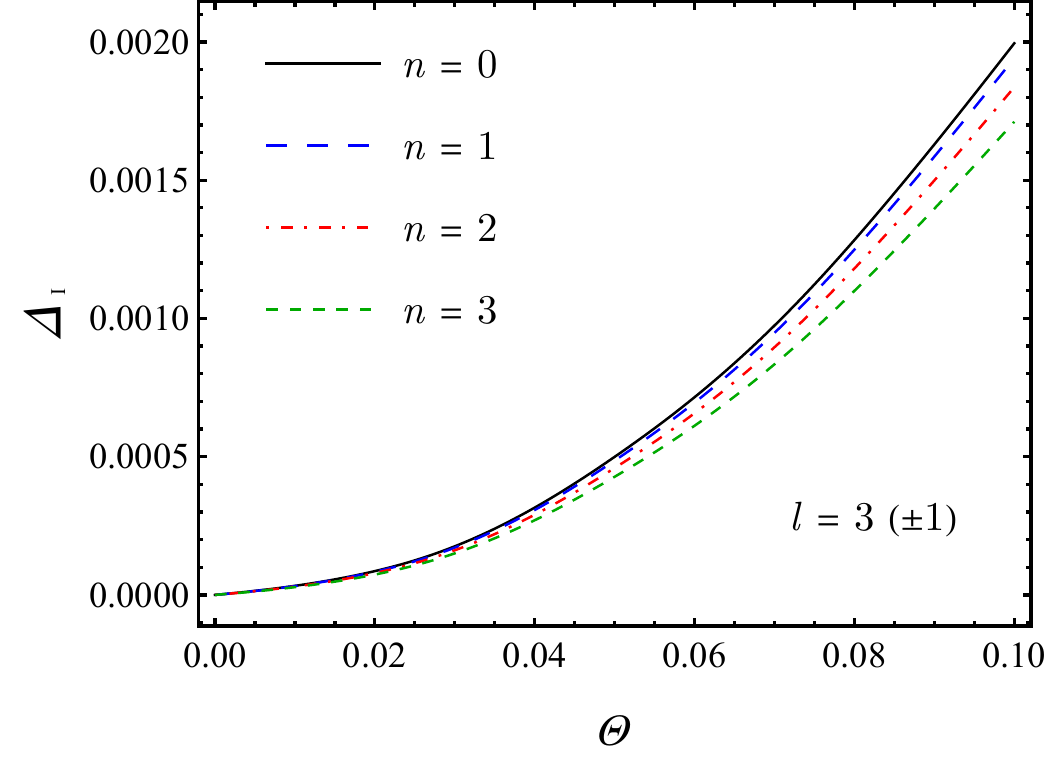}
	
	\caption{Deviation of QNMs as a function of the non--commutative parameter $\Theta$, for $M = 1$, $\ell = 0.1$, and multipole number $l = 3 (m = \pm 1)$, shown across overtone numbers $n = 0, 1, 2, 3$.}
	\label{fig:L3M1N}
    \end{figure}

Additional perspective is got by evaluating the dependence of the deviations on the overtone number $n$, as depicted in Figs.~\ref{fig:L2M1N}--\ref{fig:L3M1N}. The real parts of the frequencies, which correspond to oscillation rates, increase with $\Theta$ more significantly for higher overtones. The imaginary parts—which govern damping—also show a rising trend with $\Theta$, but the most prominent changes occur for the fundamental mode $(n = 0 $). This suggests that lower overtones are more sensitive to non--commutative corrections in terms of energy dissipation.



\section{\label{Sec7}Time--domain solution}

Understanding how scalar perturbations evolve over time is essential for analyzing the influence of quasinormal modes on dynamical scattering. Due to the complicated nature of the effective potential, a straightforward analytical approach is inadequate. To overcome this, the characteristic integration technique introduced by Gundlach et al. \cite{Gundlach:1993tp} is utilized, allowing for the accurate numerical modeling of wave evolution in time-dependent black hole backgrounds.

Several studies \cite{Baruah:2023rhd, Bolokhov:2024ixe, Guo:2023nkd, Yang:2024rms, Gundlach:1993tp, Skvortsova:2024wly, Shao:2023qlt} adopt a framework based on light-cone variables, where the transformations $u = t - r^{*}$ and $v = t + r^{*}$ are introduced. These variables significantly simplify the structure of the wave equation, making the analysis more tractable. When expressed in terms of $u$ and $v$, the wave equation takes on the following form:
\ie
\left(4 \frac{\partial^{2}}{\partial u \partial v} + V(u,v)\right) \psi (u,v) = 0 \label{timedomain}.
\fe

A practical way to numerically approach the equation consists in discretizing the system through a scheme that blends finite--difference techniques with additional computational refinements. In this manner, such an approach keeps both the precision of the results and the overall stability of the numerical evolution
\ie
\psi(N) = -\psi(S) + \psi(W) + \psi(E) - \frac{h^{2}}{8}V(S)[\psi(W) + \psi(E)] + \mathcal{O}(h^{4}).
\fe

The grid is constructed by assigning the points $S = (u, v)$, $E = (u, v + h)$, $W = (u + h, v)$, and $N = (u + h, v + h)$, where $h$ denotes the lattice step size. The initial setup is anchored on the null hypersurfaces $u = u_0$ and $v = v_0$, which serve as the foundational lines for launching the numerical evolution. Along the $u = u_0$ axis, the initial data are modeled using a Gaussian pulse centered at $v = v_c$, with its width governed by the parameter $\sigma$, as shown below
\ie
\psi(u=u_{0},v) = A e^{-(v-v_{0})^{2}}/2\sigma^{2}, \,\,\,\,\,\, \psi(u,v_{0}) = \psi_{0}.
\fe

Here, the evolution begins by assigning the scalar field a vanishing value along the initial surface $v = v_0$, that is, $\psi(u, v_0) = 0$. This choice simplifies implementation without restricting the general applicability of the method. The numerical integration proceeds by holding $u$ constant while incrementally advancing in the $v$ direction, making use of the null boundary data. For the sake of simplicity, the scalar perturbation is treated in a background where $M = 1$. The initial waveform is introduced as a Gaussian pulse centered at $v = 0$, with a fixed width $\sigma = 1$ and zero amplitude at the origin. The computational grid is defined over the square domain $[0, 1000] \times [0, 1000]$ in the $(u,v)$ plane, using a uniform spacing of $h = 0.1$.

\begin{figure}
    \centering
    \includegraphics[scale=0.51]{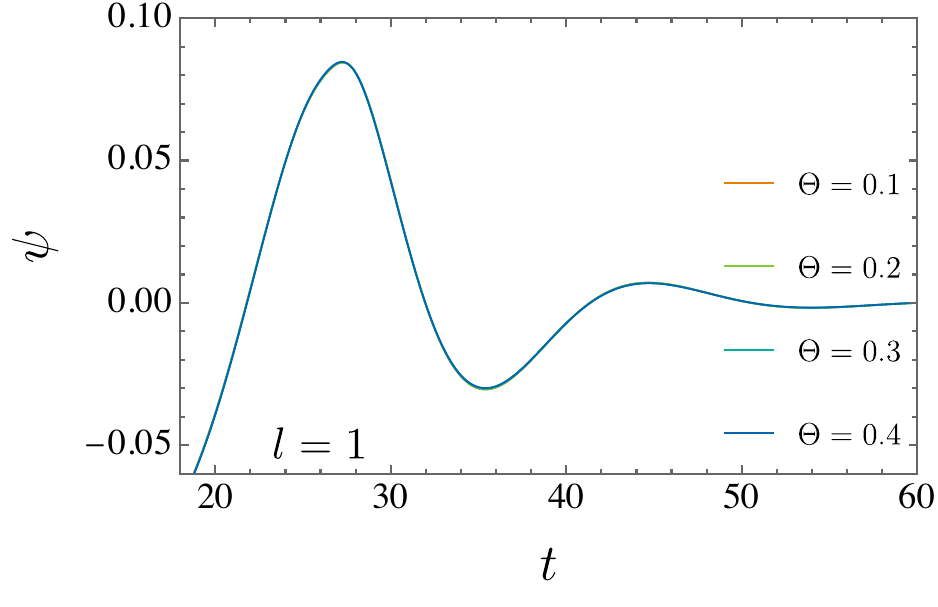}
     \includegraphics[scale=0.51]{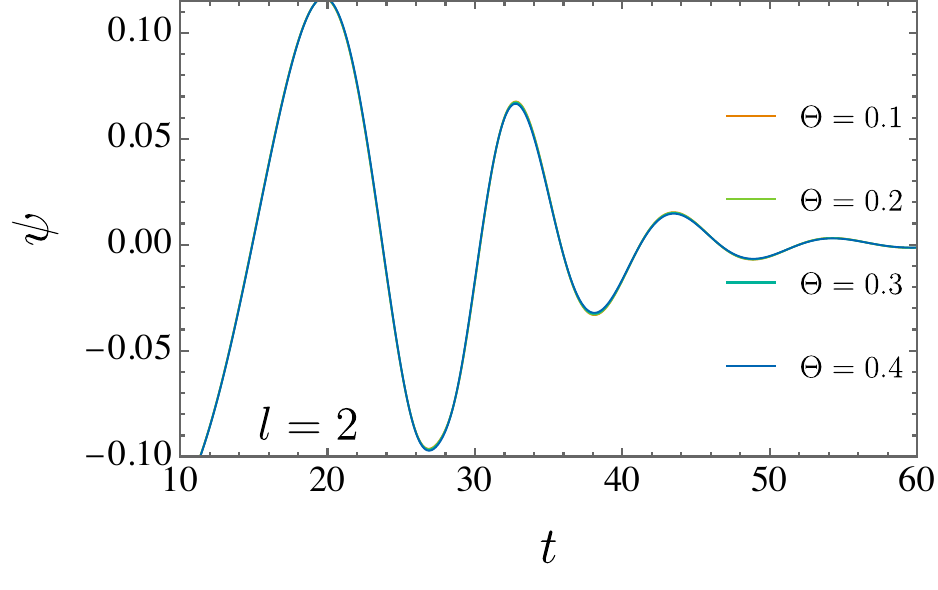}
      \includegraphics[scale=0.51]{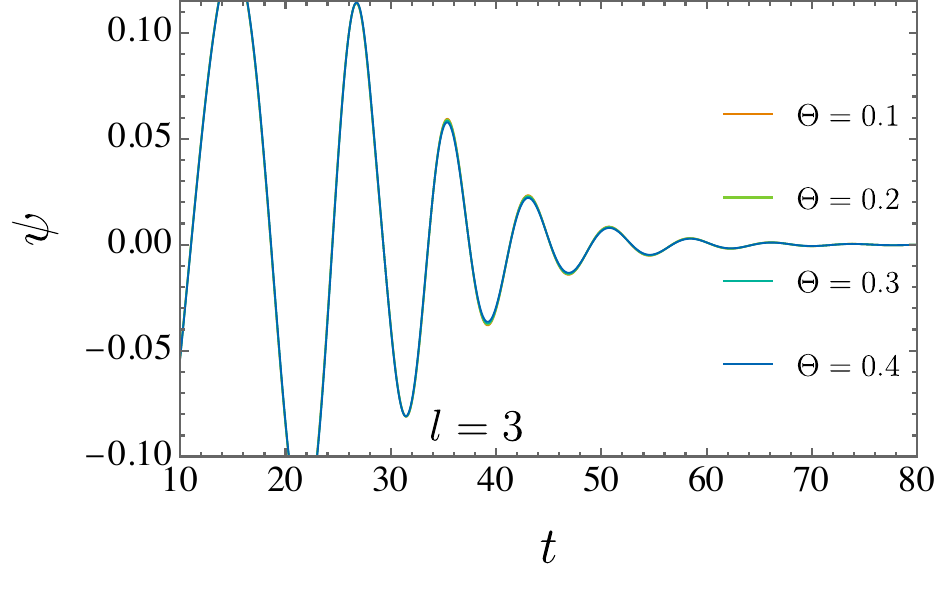}
    \caption{The time--dependent behavior of the field $\psi$ is illustrated for different angular indices, with panels corresponding to $l = 1$ (top left), $l = 2$ (top right), and $l = 3$ (bottom), under the fixed choice of $m = +1$. The parameter $\Theta$ is varied across the set ${0.1, 0.2, 0.3, 0.4}$ to examine its influence on the wave evolution, while the value of $\ell$ is kept unchanged at $0.1$ throughout the analysis.}
    \label{time-domain-wave}
\end{figure}

The progression of the scalar field $\psi$ over time is illustrated in Fig. \ref{time-domain-wave}, where distinct values of the non--commutative parameter $\Theta$—ranging from $0.1$ to $0.4$—are examined under the constraint $\ell = 0.1$. The behavior is evaluated across different angular modes: $l = 1$ (upper left), $l = 2$ (upper right), and $l = 3$ (bottom), with the azimuthal number fixed at $m = +1$. The evolution profiles indicate that the field tends to stabilize as time increases, eventually flattening out to small values.

\begin{figure}
    \centering
    \includegraphics[scale=0.51]{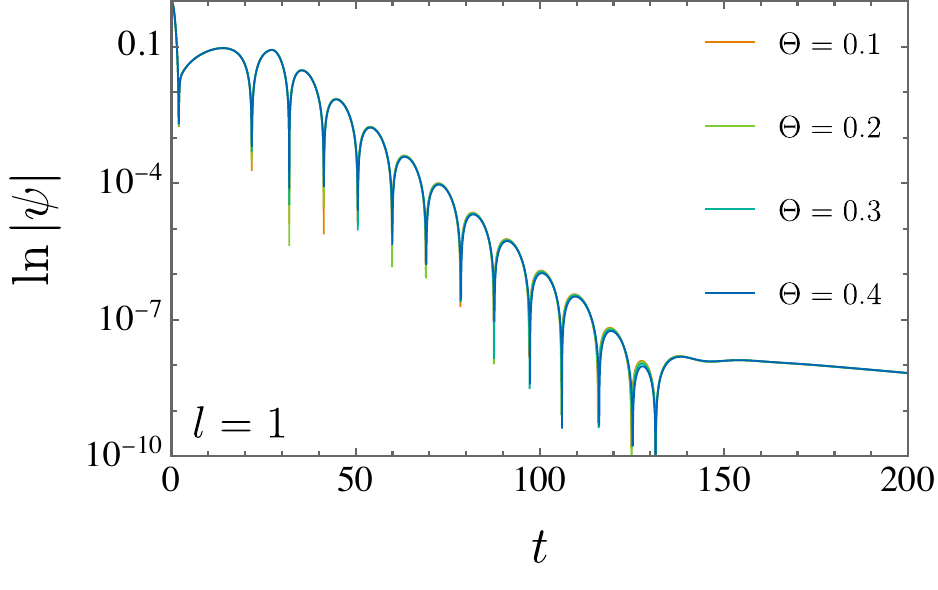}
     \includegraphics[scale=0.51]{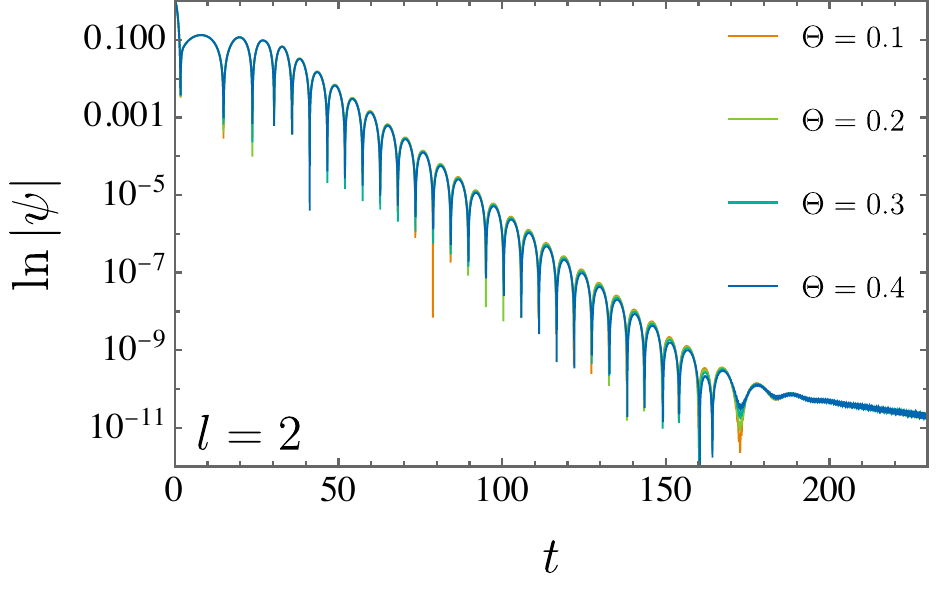}
      \includegraphics[scale=0.51]{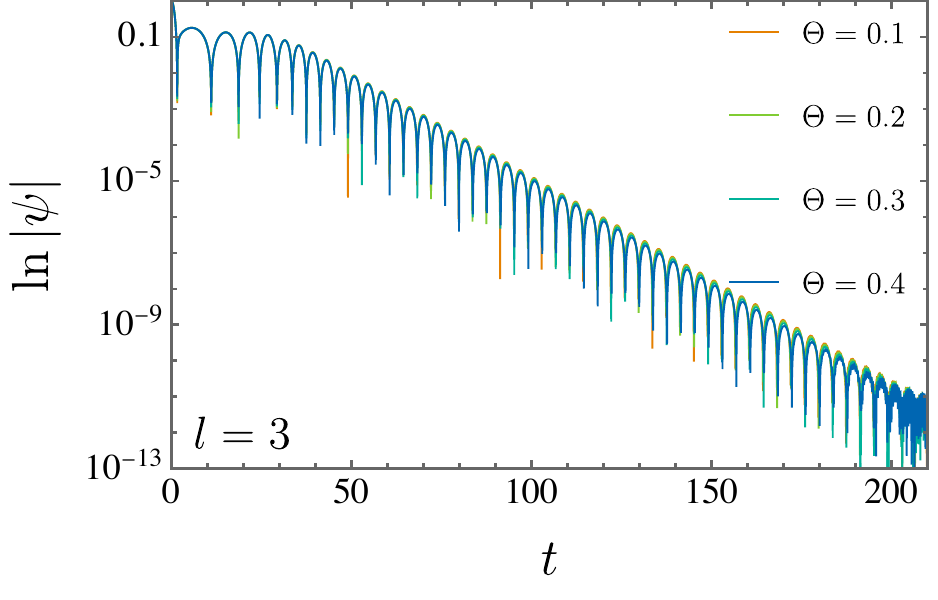}
    \caption{ The time evolution of the logarithmic amplitude $\ln|\psi|$ is shown for three angular modes: $l = 1$ (upper left), $l = 2$ (upper right), and $l = 3$ (bottom), all computed with $m = +1$ held constant. In this setting, the non--commutative parameter $\Theta$ is varied across the values $0.1$, $0.2$, $0.3$, and $0.4$, while the parameter $\ell$ remains fixed at $0.1$ throughout the entire analysis.}
    \label{time-domain-wave-abs}
\end{figure}

Complementing this analysis, Fig. \ref{time-domain-wave-abs} exhibits the natural logarithm of $|\psi|$ plotted against time using identical input parameters. In these ln plots, linear decay segments clearly reveal the exponential suppression of the signal, which aligns with the expected behavior for quasinormal damping.

\begin{figure}
    \centering
    \includegraphics[scale=0.51]{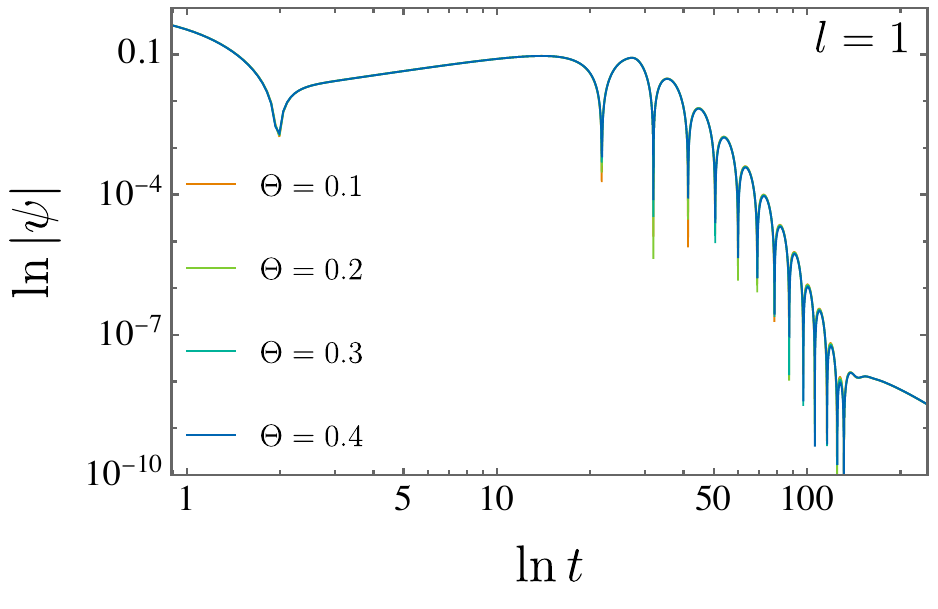}
     \includegraphics[scale=0.51]{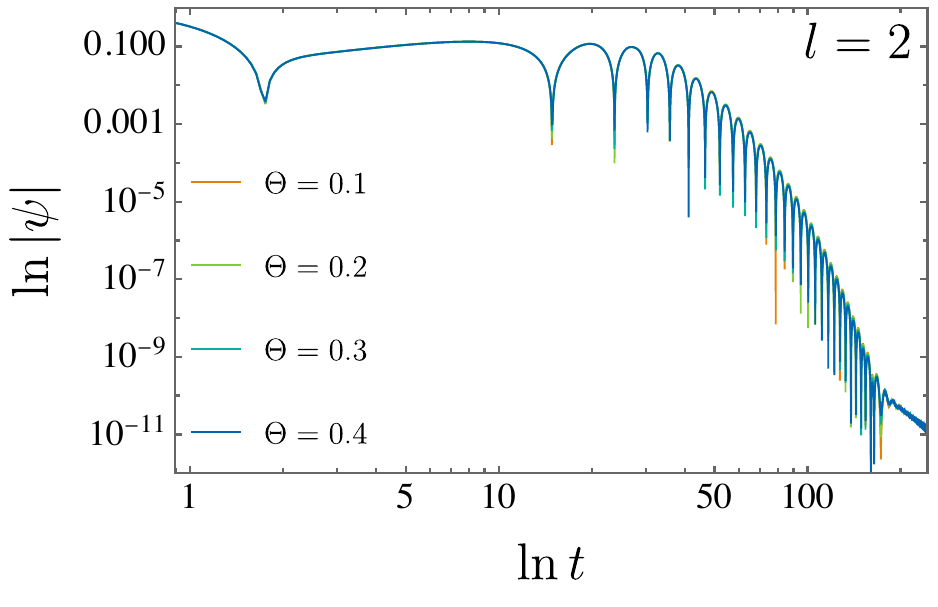}
     \includegraphics[scale=0.51]{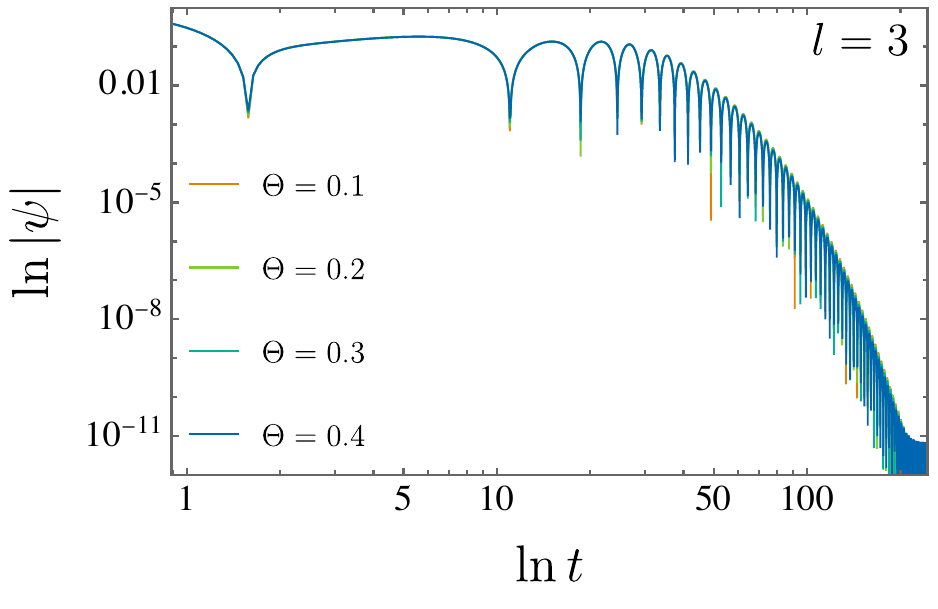}
    \caption{ A ln–-ln representation of the scalar field amplitude, $\ln|\psi|$ versus $\ln t$, is displayed for multiple angular momentum modes: $l = 1$ (upper left), $l = 2$ (upper right), and $l = 3$ (bottom), with the azimuthal index held at $m = +1$. This investigation is carried out under a fixed value of $\ell = 0.1$, while the parameter $\Theta$ is varied across the values $0.1$, $0.2$, $0.3$, and $0.4$ to assess its influence on the time--domain profile.}
    \label{time-domain-wave-abs-loglog}
\end{figure}

Additionally, Fig. \ref{time-domain-wave-abs-loglog} presents the same quantity in a ln–ln format, maintaining the same parameter choices for $\Theta$, $\ell$, $l$, and $m$. The results reinforce the conclusion that non--commutativity introduces stronger damping in the system’s response. This increased attenuation of the waveform, evident from the time profiles, is also supported by the quasinormal spectra provided in Table \ref{tab:allQNM}.

\section{\label{Sec12}Estimated constraints from the solar system tests}

General Relativity was initially confirmed based on tests performed in the environment of the Solar System, in which the mass of the sun played the role of the mass of the spherically symmetric body in Schwarzschild solution. As every measurement in sciences, they present error bars which are regions in which modifications of general relativity can be accommodated. Based on this, we shall perform an analysis of the Kalb--Ramond and non--commutative parameters to estimate the size of these contributions such that their effect do not contradict recent observations.

To analyze the motion of particles confined to the equatorial plane, where $\theta = \pi/2$, one considers the corresponding Lagrangian that governs their trajectories:
\begin{equation}\label{eq:lagr1}
{A}(r){{\dot t}^2} - B^{(\Theta,\ell)}{{\dot r}^2} - D^{(\Theta,\ell)}{{\dot \varphi }^2} = \eta\, .
\end{equation}

The velocity normalization condition is imposed through the relation $L(x, \dot{x}) = -\eta/2$, where the parameter $\eta$ distinguishes between different types of particles: $\eta = 0$ characterizes massless particles, whereas $\eta = 1$ corresponds to massive ones, whose motion is parametrized by their proper time $\lambda$.

Moreover, the expressions for the conserved quantities associated with the system—namely, the energy $E$ and the angular momentum $L$—can be written as follows
\begin{equation}\label{constant}
E = A^{(\Theta,\ell)}\dot t \quad\mathrm{and}\quad L = D^{(\Theta,\ell)}\dot \varphi.
\end{equation}

By combining Eq.~\eqref{eq:lagr1} with the expressions for the conserved quantities in \eqref{constant}, one obtains the expression shown below
\begin{equation}\label{massive}
    \left[\frac{\mathrm{d}}{\mathrm{d}\varphi}\left(\frac{1}{r}\right)\right]^2=r^{-4}D^2(r)\left[\frac{E^2}{A^{(\Theta,\ell)}B^{(\Theta,\ell)}\ell^2}-\frac{1}{B^{(\Theta,\ell)}\ell^2}\left(\eta+\frac{\ell^2}{D^{(\Theta,\ell)}}\right)\right]\, .
\end{equation}

Now, introducing the variable $u = L^2/(M r)$ and differentiating Eq.~\eqref{massive} with respect to $\varphi$, the dominant terms involving the parameters $\Theta$ and $\ell$ emerge in the following form
\begin{align}\label{eq:u-massive}
&\frac{\mathrm{d}^2 u}{\mathrm{d}\varphi^2} = \eta - u + \frac{3 M^2 u^2}{L^2}-\ell u\nonumber\\
&+\frac{\Theta^2}{2 L^8 (L^2 - 2 M^2 u)^2} \Bigg(L^{10} E^2 u-L^{10} \eta  u-10 L^8 E^2 M^2 u^2+7 L^8 \eta  M^2 u^2-L^8 M^2 u^3\nonumber\\
&+32 L^6 E^2 M^4 u^3-14 L^6 \eta  M^4 u^3+4 L^6 M^4 u^4-30 L^4 E^2 M^6 u^4+4 L^4 \eta  M^6 u^4+2 L^4 M^6 u^5\nonumber\\
&+8 L^2 \eta  M^8 u^5-24 L^2 M^8 u^6+24 M^{10} u^7 \Bigg).
\end{align}

Here, notice that the leading Newtonian effects appear in the first two terms on the right--hand side. A relativistic correction arises from the nonlinear term involving $M^2/L^2$, which, in SI units, is expressed as $G^2M^2/(c^4L^2)$—a quantity much smaller than the Newtonian components. The Kalb--Ramond background introduces its leading--order modification through the linear term $\ell u$, while non--commutative geometry contributes a correction that begins at second order in $\Theta$, as already shown in the other previous analyses in the paper. Given these considerations, a perturbative framework is adopted to extract constraints on the theory’s parameters. Accordingly, the analysis accounts for small perturbations in $\Theta$, $\ell$, and $M$. Within this regime, the dominant structure of Eq.~\eqref{eq:u-massive} becomes:
\begin{align}\label{eq:u}
     u''(\varphi)= \, \eta -\frac{u \left(2 L^4 (\ell+1)+L^2 \Theta ^2 \left(-E^2+\eta\right)\right)}{2 L^4}+\frac{3 M^2 u^2 \left(2 L^2+\Theta ^2 \left(-2E^2+\eta \right)\right)}{2 L^4}\, .
\end{align}

Several qualitative features emerge from the structure of this equation. Notably, parameter $\Theta$ appears in combination with the energy of the test particle, a behavior frequently encountered in quantum gravity models \cite{Addazi:2021xuf,AlvesBatista:2023wqm}. Additionally, both $\Theta$ and $\ell$ contribute terms that are independent of the mass of the central object, implying that these corrections modify the Newtonian limit directly. This characteristic aligns with expectations from various quantum gravity scenarios, where Planck--scale effects alter the Galilean and Minkowski limits of classical theories \cite{Amelino-Camelia:2008aez}.


\subsection{Perihelion precession of Mercury}

On the other hand, one of the classical benchmarks for testing General Relativity is the accurate prediction of Mercury’s perihelion advance, typically measured as angular displacement per century. In this context, Mercury is modeled as a massive test body traversing the static gravitational field produced by the Sun, corresponding to the case $\eta = 1$ discussed earlier. When either $\Theta = 0$ or $\ell = 0$, the leading deviation from Newtonian gravity in Eq.~\eqref{eq:u} arises from the term linear in $u$. For computational convenience, we introduce a reparametrization of the mass by defining $m = M/L$. Under this transformation, the primary correction to Newtonian dynamics becomes:
\begin{equation}\label{eq:merc}
     u''(\varphi)= \, \eta -u+\epsilon u+3 m^2 u^2\, .
\end{equation}
In this context, the parameter $\epsilon$ is defined as $\epsilon = -\ell - \frac{\Theta^2}{2L^2}(1 - E^2)$. To proceed further, we express the solution perturbatively in terms of $m$ and $\epsilon$, adopting the expansion $u = u_0 + m^2 u_m + \epsilon u_{\epsilon}$. Here, the term $u_0$ represents the Newtonian approximation, explicitly written as:
\begin{equation}
    u_0=1+e \cos(\varphi)\, ,
\end{equation}
{where $e$ is the eccentricity of the orbit}.
Here, by inserting the expanded form of $u$ into Eq.~\eqref{eq:merc}, one arrives at the following differential equation structure:
\begin{equation}\label{eq:pert_hay_merc}
    m^2 \left(3 (e \cos (\varphi )+1)^2-u_m''-u_m\right)-\epsilon \left(e \cos (\varphi )+1+u_{\epsilon}''+u_{\epsilon}\right)=0\, .
\end{equation}
Discarding mixed contributions of order $m^2\epsilon$, the remaining term involving $m^2$ represents the correction predicted by General Relativity, which is expressed as:
\begin{equation}
    u_m=3m^2\left[\left(1+\frac{e^2}{2}\right)-\frac{e^2}{6}\cos\left(2\varphi\right)+e\varphi\sin\left(\varphi\right)\right]\, .
\end{equation}

It is important to notice that the constant and oscillatory components enclosed in the brackets do not result in any net perihelion advancement, as the former remains fixed and the latter fluctuates symmetrically about zero. In other words, these terms are excluded from the final approximation. In contrast, the contribution linear in $\varphi$ grows with each orbital cycle, giving rise to a cumulative deviation that becomes observable over time.

When considering the correction associated with the $\epsilon$ parameter in Eq.~\eqref{eq:pert_hay_merc}, its influence must vanish to preserve the perturbative structure, yielding new corrections. Among the various sinusoidal terms that arise—most of which average to zero over a full orbit—only one term, $-\frac{1}{2}e\varphi \sin(\varphi)$, contributes meaningfully by producing a secular variation. After accounting for the Newtonian background, the relativistic correction, and this new contribution, and by reintroducing the original theoretical parameters, the resulting expression for $u(\varphi)$ becomes:
\begin{equation}
    u(\varphi)=1+e \cos (\varphi )+\frac{3M^2}{L^2}\left(1-\epsilon\frac{L^2}{6M^2}\right)e\varphi\sin(\varphi)\, .
\end{equation}

Given that the $\varphi\sin(\varphi)$ term has a negligible impact, it can be disregarded in the final approximation. This simplification allows the remaining two terms to be merged through appropriate trigonometric identities, resulting in:
\begin{equation}
    u(\varphi)\approx 1+e\cos\left[\left(1-\frac{3M^2}{L^2}\left(1-\epsilon\frac{L^2}{6M^2}\right)\right)\varphi\right]\doteq 1+e\cos\left[\left(1-\frac{3\widetilde{M}^2}{L^2}\right)\varphi\right]\, .
\end{equation}

The outcome may be viewed as an effective modification of the mass parameter, taking the form $\widetilde{M}^2 = M^2 \left(1 - \frac{\epsilon L^2}{6M^2} \right)$, thereby altering the conventional prediction from General Relativity. Based on this revised mass expression, the additional contribution to the perihelion precession is determined by:
\begin{equation}
    \Delta\Phi=6\pi\frac{\widetilde{M}^2}{L^2}=6\pi\frac{M^2}{L^2}\left(1-\epsilon\frac{L^2}{6M^2}\right)\, .
\end{equation}

As a result, a dimensionless deviation from the General Relativity prediction emerges, expressed as $\delta_{\text{Perih}} = -\frac{\epsilon L^2}{6M^2}$. Given that Mercury completes one orbit approximately every 88 days, the total number of revolutions over a century is roughly $100 \times 365.25 / 88 \approx 415$. Multiplying this orbital count by the angular shift per orbit yields the total perihelion precession per century. According to General Relativity, this shift amounts to $\Delta \Phi_{\text{GR}} = 42.9814''$ per century, closely matching the experimental measurement $\Delta \Phi_{\text{Exp}} = (42.9794 \pm 0.0030)''$/century \cite{Casana:2017jkc,Yang:2023wtu}, thereby validating the theoretical prediction. However, this high level of agreement can also be used to establish observational bounds on the parameter $\epsilon$, which in turn allows one to constrain the individual contributions from $\ell$ and $\Theta$.

The angular momentum $L$ of a planetary orbit can be expressed in terms of the semi-major axis $a$ and eccentricity $e$ through the relation $L^2 = M a (1 - e^2)$, while the specific orbital energy per unit mass is given by $E = -M/(2a)$ \cite{Goldstein:2002}. In natural units, the physical parameters for Mercury's orbit and the Sun are set as follows: the solar mass is $M = M_{\odot} = 9.138 \times 10^{37}$, the semi--major axis of Mercury’s orbit is $a = 3.583 \times 10^{45}$, and its eccentricity is $e = 0.2056$. Using these values, one obtains $L = 5.600 \times 10^{41}$, confirming the validity of a perturbative approach due to the smallness of $M^2/L^2$ contributions. The squared energy, calculated as $E^2 = 1.627 \times 10^{-16}$, is small enough that energy--dependent corrections can safely be disregarded.

With the parameter $\epsilon$ defined by $\epsilon = -\ell - \frac{\Theta^2}{2L^2}(1 - E^2)$, one can derive independent constraints on $\ell$ and $\Theta$. Setting $\Theta = 0$ isolates the Kalb--Ramond contribution, leading to the constraint $-1.817 \times 10^{-11} \leq \ell \leq 3.634 \times 10^{-12}$. Alternatively, taking $\ell = 0$ allows for bounding the non--commutative parameter, resulting in the range $-2976.57\,\text{m}^2 \leq \Theta^2 \leq 595.315\,\text{m}^2$. For completeness, both positive and negative branches of $\Theta^2$ are retained in the analysis.


\subsection{Deflection of light}

As a light ray passes close to a massive body, its trajectory bends, altering the apparent position of the source in the observer’s sky—a phenomenon commonly interpreted as a gravitational parallax. This bending is analyzed by studying null geodesics, which corresponds to setting $\eta = 0$ in Eq.~\eqref{eq:u}. To facilitate the analysis, the function $u$ is redefined as $u = 1/r$, leading to the following form:
\begin{align}\label{eq:light}
    u''(\varphi)=&\frac{-2L^2(1+\ell)+E^2\Theta^2}{2L^2}u+\frac{3 M \left(-2E^2 \Theta ^2+ 2L^2\right)}{2L^2}u^2 -\frac{\Theta^2}{2}u^3\, .
\end{align}

In this context, the ratio $L/E$ corresponds to the impact parameter $b$. To isolate the effects ascribed to $\Theta$, we first set $\ell = 0$. Notably, the resulting corrections involving the parameter $\Theta$ are universal—they appear independently of the central mass $M$—and thus modify the Newtonian trajectory from the outset. By appropriately redefining the mass term, the leading--order contributions can be expressed as follows:
\begin{equation}\label{eq:redef_light_theta}
     u''(\varphi)+\left(1-\frac{\Theta^2}{2b^2}\right)u=3\widetilde{M}u^2-\frac{\Theta^2}{2}u^3\, .
\end{equation}
Here, the effective mass is redefined as $\widetilde{M} = M\left(1 - \frac{\Theta^2}{b^2}\right)$. Setting the left-hand side of the equation to zero retrieves the Newtonian limit, from which a correction to the classical result naturally emerges \cite{Yang:2023wtu}
\begin{equation}
    u_0=b^{-1}\sin\left(\left(1-\frac{\Theta^2}{4b^2}\right)\varphi \right)\, ,
\end{equation}
with the initial angle as $\varphi_0 = 0$, the solution represents a linear, undeflected trajectory. By substituting this form into Eq.~\eqref{eq:redef_light_theta} and assuming the regime of small angular deviations ($\varphi \ll 1$), the resulting perturbed solution becomes:
\begin{equation}
    u(\varphi)=\frac{1}{b}\sin\left(\left(1-\frac{\Theta^2}{4b^2}\right)\varphi\right)+\frac{\widetilde{M}}{b^2(1-\Theta^2/(2b^2))}\left[1+\cos^2\left(\left(1-\frac{\Theta^2}{4b^2}\right)\varphi\right)\right]\, .
\end{equation}

As the light ray moves toward spatial infinity, its trajectory approaches $u \to 0$ (or equivalently, $r \to \infty$). To determine the angles at which the ray enters and exits the scattering region, one solves the condition $u = 0$. Introducing perturbations in both the angular variable $\varphi$ and the relevant model parameters yields the approximate angles: $\varphi_{\text{in}} = -2\bar{M}/b$ and $\varphi_{\text{ex}} = \pi + 2\bar{M}/b$, where the effective mass is defined as $\bar{M} = M\left(1 - \frac{\Theta^2}{4b^2}\right)$. The resulting deflection angle is then given by $\delta = -2\varphi_{u \to 0}$
\begin{equation}
    \delta_{\Theta}=\frac{4\bar{M}}{b}=4\frac{M}{b}\left(1-\frac{\Theta^2}{4b^2}\right)\, .
\end{equation}

When a light ray passes just above the Sun’s surface, the impact parameter can be approximated by the solar radius, $b \approx R_{\odot} = 4.305 \times 10^{43}$. The Sun’s mass is taken as $M = M_{\odot} = 9.138 \times 10^{37}$. The non--commutative correction to the deflection is captured by the dimensionless factor $1 - \Theta^2/(4b^2)$. Importantly, this modification is independent of the stellar mass and instead depends solely on the impact parameter—effectively linking it to the star’s radius.

General Relativity predicts a deflection angle of $\delta_{\text{GR}} = 4M/b = 1.7516687''$. Observationally, the measured deflection is expressed as $\delta_{\text{Exp}} = \frac{1}{2}(1 + \gamma) \times 1.7516687''$, with $\gamma = 0.99992 \pm 0.00012$ \cite{dsasdas}. To examine the influence of $\Theta$, the dimensionless factor $1 - \Theta^2/(4b^2)$ is compared with the empirical ratio $(1 + \gamma)/2$. This comparison yields the following constraint: $-3.872\times 10^{13}\, \text{m}^2\leq\Theta^2\leq 1.936\times 10^{14}\, \text{m}^2$.
 
To isolate the effects of the Kalb--Ramond field, one sets $\Theta = 0$ in Eq.~\eqref{eq:light}. Under this condition, the radial equation governing the trajectory simplifies to:
\begin{equation}\label{eq:light-hay}
    u''(\varphi)+(1+\ell) u=3 M u^2 \, .
\end{equation}

In this manner, the corresponding perturbed solution of this equation is therefore
\begin{equation}
    u(\varphi)=\frac{1}{b}\sin\left(\left(1+\frac{\ell}{2}\right)\varphi\right)+\frac{M}{b^2(1+\ell)}\left[1+\cos^2\left(\left(1+\frac{\ell}{2}\right)\varphi\right)\right]\, .
\end{equation}

By assuming the regime of small angular deviations, $\varphi \ll 1$, and following the same reasoning outlined previously, the resulting deflection angle in the presence of the Kalb--Ramond background becomes $\delta_{\ell} = \frac{4M}{b}(1 - \frac{3}{2}\ell)$. This expression allows us to infer the constraint on $\ell$, yielding the range $-1.333 \times 10^{-5} \leq \ell \leq 6.667 \times 10^{-5}$.


\subsection{Time delay of light}

The phenomenon known as Shapiro time delay \cite{Shapiro:1964uw} refers to the extra time required for radar signals to propagate to planets and return to Earth, due to the influence of the Sun’s gravitational field on the curvature of spacetime. To compute this effect, one examines the null geodesics governed by Eq.\eqref{massive}. Employing the condition for lightlike trajectories, together with the conserved quantities for energy and angular momentum provided in Eq.\eqref{constant}, it becomes possible to express the radial coordinate as a function of time
\begin{equation}
  \left(  \frac{\mathrm{d}r}{\mathrm{d}t}\right)^2=\frac{A^{(\Theta,\ell)}D^{(\Theta,\ell)}-\frac{\ell^2}{E^2}{(A^{(\Theta,\ell)})}^2}{B^{(\Theta,\ell)}D^{(\Theta,\ell)}}\, .
\end{equation}

According to Ref.~\cite{Wang:2024fiz}, the conserved quantities—angular momentum and energy—can be re-expressed in terms of the light ray’s minimum distance from the Sun, denoted by $b$. This turning point is determined by setting $\dot{r} = 0$, which leads to the condition $L^2/E^2 = D^{(\Theta,\ell)}(r_{\text{min}})/A^{(\Theta,\ell)}(r_{\text{min}})$. As a result, the propagation time can be reformulated as a function of the radial coordinate, taking the form:
\begin{equation}\label{eq:shapiro_main}
    \mathrm{d} t=\pm \frac{1}{A^{(\Theta,\ell)}}\frac{1}{\sqrt{\frac{1}{A^{(\Theta,\ell)}B^{(\Theta,\ell)}}-\frac{D^{(\Theta,\ell)}(r_{\text{min}})/A^{(\Theta,\ell)}(r_{\text{min}})}{B^{(\Theta,\ell)}D^{(\Theta,\ell)}}}}\, .
\end{equation}

The analysis proceeds by treating the non--commutative and Kalb--Ramond contributions independently, beginning with the non--commutative framework where $\ell$ is set to zero. To isolate the deviations from flat spacetime, only the dominant terms involving $M$ and $\Theta^2$ are preserved. Within this approximation scheme, the integration of Eq.~\eqref{eq:shapiro_main} leads to the following expression:
\begin{align}
    t=\sqrt{r^2-r_{\text{min}}^2}+M\left(\sqrt{\frac{r-r_{\text{min}}}{r+r_{\text{min}}}}+2\ln\left(\frac{r+\sqrt{r^2-r_{\text{min}}^2}}{r_{\text{min}}}\right)\right)\\
    +\frac{\Theta ^2 \left(2 (r_{\text{min}}-M) \arctan\left(\frac{\sqrt{r^2-r_{\text{min}}^2}-r}{r_{\text{min}}}\right)-\frac{M (19 r_{\text{min}}+26 r) \sqrt{r^2-r_{\text{min}}^2}}{r (r_{\text{min}}+r)}\right)}{8 r_{\text{min}}^2}\, .\nonumber
\end{align}

In the limit where $r \gg r_{\text{min}}$, the dominant contributions arise from both the General Relativity term and the leading--order effects associated with $\Theta$, and can be expressed as:
\begin{equation}\label{eq:sh_t_nc}
    t(r)=r+M+2M\ln\left(\frac{2r}{r_{\text{min}}}\right)-\frac{13M}{4r_{\text{min}}^2}\Theta^2\, .
\end{equation}

Let $t(r_E)$ represent the time taken for the signal to travel from the emitter to the Sun, and $t(r_R)$ the time from the Sun to the receiver. These travel times are computed using Eq.~\eqref{eq:sh_t_nc}, with $r_E$ and $r_R$ denoting the radial positions of the emitter and receiver, respectively. The full round--trip duration—covering the path from emission to reception and back—is then expressed as $T = 2t(r_E) + 2t(r_R)$. Under these conditions, the total propagation time becomes:
\begin{equation}
T_{\Theta}=2(r_E+r_R)+4M\left[1+\ln\left(\frac{4r_Rr_E}{r_{\text{min}}^2}\right)-\frac{13\Theta^2}{4r_{\text{min}}^2}\right]=T_{\text{flat}}+\delta T\, .
\end{equation}

In this manner, the Shapiro time delay quantifies the additional travel time experienced by a signal due to spacetime curvature induced by a massive object, relative to its travel time in flat spacetime. In the absence of gravitational effects, this baseline time is given by $T_{\text{flat}} = 2(r_E + r_R)$. Within the parametrized post--Newtonian (PPN) formalism, the relativistic contribution to this delay takes the form:
\begin{equation}
    \delta T = 4M\left(1+\frac{1+\gamma}{2}\ln \left(\frac{4r_Rr_E}{r_{\text{min}}^2}\right)\right)\, .
\end{equation}

High--precision data from the Cassini spacecraft \cite{Bertotti:2003rm,Will:2014kxa} have led to the most stringent bound on the PPN parameter $\gamma$, constraining it to $|\gamma - 1| < 2.3 \times 10^{-5}$. In natural units, the average Earth–Sun separation is defined as one astronomical unit (AU), corresponding to $r_E = 1\, \text{AU} = 2.457 \times 10^{45}$. During the measurement, the Cassini probe was positioned at $r_R = 8.46\, \text{AU}$ from the Sun, while the light signal’s minimum radial distance from the Sun was $r_{\text{min}} = 1.6, R_{\odot}$, with the solar radius given by $R_{\odot} = 4.305 \times 10^{43}$. Using these parameters, one obtains an upper bound on the non--commutative parameter: $|\Theta^2| \leq 5.001 \times 10^{14}\, \text{m}^2$.

In the case of the Kalb--Ramond background, we eliminate the non--commutative effects by setting $\Theta = 0$. Under this condition, the leading contributions resulting from the integration of Eq.~\eqref{eq:shapiro_main} are given by:
\begin{align}
   t=\sqrt{r^2-r_{\text{min}}^2}+M\left(\sqrt{\frac{r-r_{\text{min}}}{r+r_{\text{min}}}}+2\ln\left(\frac{r+\sqrt{r^2-r_{\text{min}}^2}}{r_{\text{min}}}\right)\right)\nonumber\\
  -\ell \left[\sqrt{r^2-r_{\text{min}}^2}+2M\left(\sqrt{\frac{r-r_{\text{min}}}{r+r_{\text{min}}}}+2\ln\left(\frac{r+\sqrt{r^2-r_{\text{min}}^2}}{r_{\text{min}}}\right)\right)\right]\, .
\end{align}

In the regime where $b \ll r$, the leading--order behavior is governed by the following terms:
\begin{equation}
     t=r+M+2M\ln\left(\frac{2r}{r_{\text{min}}}\right)-\ell \left[r+2M+4M\ln\left(\frac{2r}{r_{\text{min}}}\right)\right]\, .
\end{equation}

The Shapiro time delay arising from the Kalb–Ramond contribution can thus be expressed as:
\begin{equation} T_l=2(1-\ell)(r_E+r_R)+4M(1-2\ell)\left[1+\ln\left(\frac{4r_Rr_E}{r_{\text{min}}^2}\right)\right]\, .
\end{equation}

Based on the analysis of Cassini tracking data, the Kalb--Ramond parameter is constrained to $|\ell| \leq 5.700 \times 10^{-6}$, with the Sun’s mass $M$ employed in the computation. For completeness, both positive and negative values of the squared parameters $\Theta^2$ and $\ell$ were considered, reflecting the substitution $(\Theta^2, \ell) \rightarrow -(\Theta^2, \ell)$ in the formalism. A comprehensive overview of the resulting parameter bounds is provided in Table~\ref{tab:constr}.

\begin{table}[h!]
\centering
\caption{Bounds for $\Theta^2$ and $\ell$  derived from Solar System tests.}
\label{tab:constr}
\begin{tabular}{lc}
\hline\hline
\textbf{Solar System Test} & Constraints  \\
\hline
{\bf{Mercury precession}}   & \makecell{$-2976.57 \,\text{m}^2\leq\Theta^2\leq 595.315 \,\text{m}^2$ \\ $-1.817 \times 10^{-11} \leq \ell \leq 3.634 \times 10^{-12} $} \\
{\bf{Light deflection}}     & \makecell{$-3.872\times 10^{13}\, \text{m}^2\leq\Theta^2\leq 1.936\times 10^{14}\, \text{m}^2$ \\ $-1.333\times 10^{-5}\leq \ell\leq 6.667\times 10^{-5}$}  \\
{\bf{Shapiro time delay}}   & \makecell{$-5.001 \times 10^{14} \,\text{m}^2 \leq \Theta^2 \leq 5.001 \times 10^{14} \,\text{m}^2$ \\ $-5.700 \times 10^{-6}\leq \ell \leq 5.700 \times 10^{-6}$}  \\
\hline\hline
\end{tabular}
\end{table}


\section{\label{Sec13}Conclusion}

This study focused on constructing a novel black hole solution within the framework of a non--commutative gauge theory. To achieve this, we started from the black hole solution developed in Ref.~\cite{Yang:2023wtu} and adopted the Moyal deformation generated by the twist $\partial_r \wedge \partial_\theta$, in line with the method recently introduced in Ref.~\cite{Juric:2025kjl}. As a result, a novel black hole geometry emerged, shaped by non--commutative corrections encoded in the parameter $\Theta$ and by the Lorentz--violating contribution $\ell$ arising from Kalb--Ramond gravity.

We then examined the structure of the event horizon and found it remains unaffected by the presence of non--commutativity in this setting. The horizon radius was preserved as $r_h = 2(1 - \ell)M$. To investigate regularity, we computed the Kretschmann scalar. The results indicated that, provided geometric quantities such as the Christoffel symbols are not expanded to second order in $\Theta$, the black hole turned out to be regular. In particular, such a scalar approached $\tilde{\mathcal{K}}_{r \to 0} = \frac{1552}{3 \Theta^4}$ in the limit $r \to 0$.

{We also examined the thermodynamic properties of the system. Unlike the result reported in Ref.~\cite{Juric:2025kjl}, where the Schwarzschild case under the same Moyal twist led to an ill--defined surface gravity—preventing a consistent thermodynamic analysis—our Kalb--Ramond framework yielded a well--defined surface gravity. Consequently, the thermodynamic quantities could be consistently determined. In particular, we computed the Hawking temperature $T^{(\Theta,\ell)}$ and the heat capacity $C_V^{(\Theta,\ell)}$. Since the entropy remained $\pi r_h^2$ and the horizon was not modified by $\Theta$, a separate discussion of the entropy was not included.}

Additionally, by setting $T^{(\Theta,\ell)} \to 0$, we derived two candidate expressions for the remnant mass: $M^{(1)}_{\text{rem}} = \frac{i \sqrt{3} \Theta}{(4 - 4\ell) \sqrt{\ell + 1}}, \quad
M^{(2)}_{\text{rem}} = \frac{i \sqrt{3} \Theta}{4 (\ell - 1) \sqrt{\ell + 1}}$. However, both solutions were non--real. This outcome was interpreted as evidence that the black hole would undergo complete evaporation, leaving no remnant behind.

Furthermore, we examined the quantum radiation emitted by the black hole under study. To enhance the robustness of our analysis, both bosonic and fermionic particle modes were taken into account. In each case, the emission process was modeled via the quantum tunneling mechanism. The divergent integrals associated with the particle production rate were treated using the residue theorem, which allowed us to estimate the particle number densities $n(\Theta, \ell, \omega)$ for bosons and $n_{\psi}(\Theta, \ell)$ for fermions. A comparison between the two revealed that, in the low--frequency regime, bosonic emission is more pronounced than its fermionic counterpart—at least within the framework adopted in this work.

The effective potential for the massless scalar field was obtained by employing a perturbative expansion of the metric, expressed as $g_{\mu\nu}^{\text{NC}} = g_{\mu\nu} + \Theta^2 h_{\mu\nu}^{\text{NC}}$, which allowed for the application of the WKB method to compute the quasinormal modes. Overall, we observed that for a fixed value of $\ell$, increasing $\Theta$ led to more strongly damped oscillations. A similar trend was found when varying $\ell$ while keeping $\Theta$ fixed. These findings were further supported by a time--domain analysis, where the decay behavior of the perturbations was monitored as a function of time.

We further analyzed the evaporation lifetime of the black hole, denoted by $t_{\text{evap}}$. This was achieved using the Stefan--Boltzmann law, along with the high--frequency approximation for the absorption cross--section, where $\sigma \approx \pi R_{sh}^2$. Since we assumed complete evaporation, i.e., $M_f \to 0$, the corresponding evaporation time was found to be $t_{\text{evap}} = \frac{4096 \pi^3 M_i^3}{81 \xi} - \frac{20480 \pi^3 \ell M_i^3}{81 \xi} + \frac{35840 \pi^3 \Theta^2 \ell M_i}{81 \xi} - \frac{26624 \pi^3 \Theta^2 M_i}{243 \xi}.$ We also compared the final stages of the evaporation process for different black hole configurations. The hierarchy we found,
$t_{\text{evap}}^{\text{Schwarzschild}} > t_{\text{evap}}^{\text{NC Schwarzschild}} > t_{\text{evap}}^{\text{Kalb–Ramond}} > t_{\text{evap}}^{\text{NC Kalb–Ramond}},$
clearly indicated that both the non--commutative deformation and the Kalb--Ramond field contribute to a faster evaporation relative to the classical Schwarzschild case. In addition, we computed the energy and particle emission rates. For fixed values of $\ell$, an increase in $\Theta$ was shown to enhance both emission channels, reinforcing the conclusion that non--commutativity amplifies the evaporation process.

Finally, the constraints obtained from solar system observations establish the following bounds: from Mercury’s perihelion precession, we got $-2976.57\,\text{m}^2 \leq \Theta^2 \leq 595.315\,\text{m}^2$ and $-1.817 \times 10^{-11} \leq \ell \leq 3.634 \times 10^{-12}$; from light deflection, the allowed range was $-3.872 \times 10^{13}\,\text{m}^2 \leq \Theta^2 \leq 1.936 \times 10^{14}\,\text{m}^2$ and $-1.333 \times 10^{-5} \leq \ell \leq 6.667 \times 10^{-5}$; and from the Shapiro time delay, we had $-5.001 \times 10^{14} \,\text{m}^2 \leq \Theta^2 \leq 5.001 \times 10^{14} \,\text{m}^2$ and $-5.700 \times 10^{-6} \leq \ell \leq 5.700 \times 10^{-6}$.

As a prospective development, several promising directions deserve further attention. A natural continuation of the present study involves examining greybody factors and scattering processes under non--commutative corrections—both expected to significantly influence particle emission and wave propagation in this background. Furthermore, examining the behavior of light—through shadows, photon spheres, and lensing effects in both weak and strong gravity regimes—proves to be a valuable aspect to consider. These and other ideas are now under development.

\section*{Acknowledgments}
\hspace{0.5cm} A. A. Araújo Filho is supported by Conselho Nacional de Desenvolvimento Cient\'{\i}fico e Tecnol\'{o}gico (CNPq) and Fundação de Apoio à Pesquisa do Estado da Paraíba (FAPESQ), project No. 150891/2023-7. I. P. L. was partially supported by the National Council for Scientific and Technological Development - CNPq grant 312547/2023-4.  I. P .L. would like to acknowledge networking support by the COST Action BridgeQG (CA23130) and the COST Action RQI (CA23115), supported by COST (European Cooperation in Science and Technology). N. H. would like to acknowledge the contribution of the COST Action CA21106 - COSMIC WISPers in the Dark Universe: Theory, astrophysics and experiments (CosmicWISPers), the COST Action CA21136 - Addressing observational tensions in cosmology with systematics and fundamental physics (CosmoVerse), the COST Action CA23130 - Bridging high and low energies in search of quantum gravity (BridgeQG).


	\bibliography{main}
	\bibliographystyle{unsrt}
	
\end{document}